\documentclass[aps,prd,twocolumn,groupedaddress,nofootinbib,longbibliography,balancelastpage,superscriptaddress]{revtex4-1}
\usepackage{graphicx,color,amsmath,amssymb}
\usepackage[export]{adjustbox}
\usepackage{slashed}
\usepackage{booktabs} 
\usepackage{braket}
\usepackage[colorlinks=true
,urlcolor=blue
,anchorcolor=blue
,citecolor=blue
,filecolor=blue
,linkcolor=red
,menucolor=blue
,linktocpage=true
,pdfproducer=medialab
,pdfa=true
]{hyperref}
\usepackage{mathtools} 

\newcommand{\ga}{g_{a\gamma\gamma}}
\newcommand{\DM}{\scriptscriptstyle {\rm DM}}
\newcommand{\SM}{\scriptscriptstyle {\rm SM}}
\newcommand{\MW}{\scriptscriptstyle {\rm MW}}
\newcommand{\EG}{\scriptscriptstyle {\rm EG}}
\newcommand{\bE}{\mathbf{E}}
\newcommand{\bbe}{\mathbf{e}}
\newcommand{\bB}{\mathbf{B}}
\newcommand{\bJ}{\mathbf{J}}
\newcommand{\bx}{\mathbf{x}}
\newcommand{\bk}{\mathbf{k}}
\newcommand{\bv}{\mathbf{v}}
\newcommand{\hn}{\hat{\mathbf{n}}}
\newcommand{\hz}{\hat{\mathbf{z}}}

\begin{document}

\title{The Cosmic Axion Background}

\author{Jeff A. Dror}
\email{jdror1@ucsc.edu}
\affiliation{Department of Physics and Santa Cruz Institute for Particle Physics, University of California, Santa Cruz, CA 95064, USA}
\affiliation{Berkeley Center for Theoretical Physics, University of California, Berkeley, CA 94720, USA}
\affiliation{Theory Group, Lawrence Berkeley National Laboratory, Berkeley, CA 94720, USA}

\author{Hitoshi Murayama}
\email{hitoshi@berkeley.edu}
\email{hitoshi.murayama@ipmu.jp}
\affiliation{Berkeley Center for Theoretical Physics, University of California, Berkeley, CA 94720, USA}
\affiliation{Theory Group, Lawrence Berkeley National Laboratory, Berkeley, CA 94720, USA}
\affiliation{Kavli Institute for the Physics and Mathematics of the Universe (WPI), University of Tokyo, Kashiwa 277-8583, Japan}

\author{Nicholas L. Rodd}
\email{nrodd@berkeley.edu}
\affiliation{Berkeley Center for Theoretical Physics, University of California, Berkeley, CA 94720, USA}
\affiliation{Theory Group, Lawrence Berkeley National Laboratory, Berkeley, CA 94720, USA}


\begin{abstract}
Existing searches for cosmic axions relics have relied heavily on the axion being non-relativistic and making up dark matter. 
However, light axions can be copiously produced in the early Universe and remain relativistic today, thereby constituting a Cosmic {\em axion} Background (C$a$B). 
As prototypical examples of axion sources, we consider thermal production, dark-matter decay, parametric resonance, and topological defect decay. 
Each of these has a characteristic frequency spectrum that can be searched for in axion direct detection experiments.
We focus on the axion-photon coupling and study the sensitivity of current and future versions of ADMX, HAYSTAC, DMRadio, and ABRACADABRA to a C$a$B, finding that the data collected in search of dark matter can be repurposed to detect axion energy densities well below limits set by measurements of the energy budget of the Universe.
In this way, direct detection of relativistic relics offers a powerful new opportunity to learn about the early Universe and, potentially, discover the axion.
\end{abstract}

\maketitle

\section{Introduction}

\begin{figure*}
\begin{center} 
\includegraphics[width=.93\textwidth]{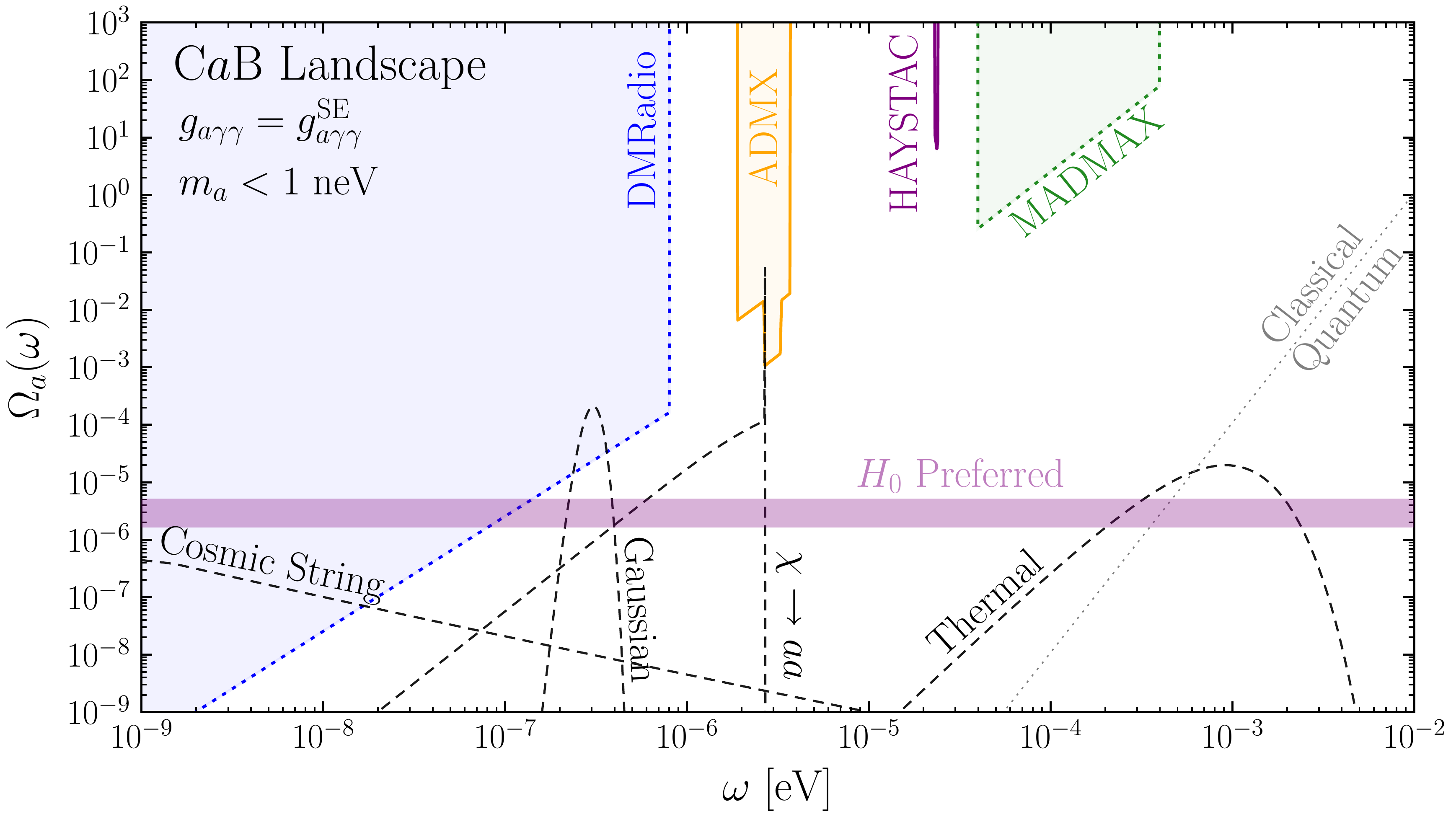}
\end{center}
\vspace{-0.3cm}
\caption{A representative depiction of the landscape of the cosmic axion background (C$a$B), showing the differential axion energy density, given in \eqref{eq:Omegaomega}, as a function of energy.
The black dashed curves show four different realizations of the C$a$B, corresponding to thermal production (with $T_a = T_0$, the CMB temperature), a Gaussian distribution representative of parametric-resonance production (with $\rho_a = \rho_\gamma$, $\bar{\omega}=0.3~\mu{\rm eV}$, and $\sigma/\bar{\omega}=0.1$), dark-matter decay ($\chi \to aa$), and cosmic-string production ($f_a = 10^{15}$ GeV, $T_d = 10^{12}$ GeV).
For the dark-matter decay distribution the parameters are set to parameters already accessible to ADMX, as justified later in this work.
In particular, we take $m_{\DM} \simeq 5.4~\mu{\rm eV}$ and $\tau \simeq 2 \times 10^3\, t_U$, with $t_U$ the age of the universe.
While the thermal distribution will always peak roughly where shown and the cosmic-string production is dominant at lower frequencies, the parametric resonance and dark-matter decay signals can populate the full energy energy range shown.
In all cases we set the axion photon coupling to the largest allowed value consistent with star-emission bounds over this energy range, $\ga^{\rm SE} = 0.66 \times 10^{-10}~{\rm GeV}^{-1}$.
The colored regions denote the sensitivity in this same space that could be obtained by reanalyzing existing ADMX and HAYSTAC data, or with the future sensitivities of DMRadio and MADMAX.
In determining the sensitivities, we have assumed that the C$a$B axion-photon coupling saturates star-emission bounds.
We show the region of parameter space where the C$a$B could partially alleviate the Hubble tension, labelled $H_0$ Preferred.
Finally, the gray dotted line depicts the approximate boundary, to the left of which the C$a$B has sufficient number densities to be treated as a classical wave.}
\label{fig:CaB}
\end{figure*}

The existence of an axion with mass well below the electroweak scale could resolve the strong CP puzzle~\cite{Peccei:1977hh,Peccei:1977ur,Weinberg:1977ma,Wilczek:1977pj}, and is entirely in line with UV expectations given the ubiquity of axions in string theory, where they arise from the deconstruction of extra-dimensional gauge fields~\cite{Svrcek:2006yi,Arvanitaki:2009fg,Halverson:2019cmy}.
The discovery of cosmologies where such a particle produced in the early Universe could constitute dark matter~\cite{Preskill:1982cy,Abbott:1982af,Dine:1982ah} has motivated a broad program to detect non-relativistic axions, and the development of instruments that will cover enormous swaths of unexplored parameter space in the coming decades.
Yet the axion need not be dark matter, and the mere existence of an axion in the spectrum implies the possibility that a relic population of these states was produced in the early history of the Universe.
Generically, such a population could be relativistic --- a characteristic feature of the axion is an approximate shift symmetry, leading to a potential suppressed by powers of the axion decay constant, $f_a$, and correspondingly the axion mass, $m_a$, is expected to be small.
Accordingly, the Universe may be awash in a sea of relativistic axions, a Cosmic {\em axion} Background (C$a$B).

In this work we will broadly discuss the production and detection of such a C$a$B.
The possibility of a relativistic axion population is not new, and has been discussed in several contexts, including axion contributions to $\Delta N_{\rm eff} $~\cite{Baumann:2016wac}, axions with keV energies motivated by the prospect of moduli decaying into axions through Planck-suppressed higher dimensional operators~\cite{Conlon:2013isa,Conlon:2013txa,Cicoli:2014bfa,Cui:2017ytb}, or constraints on the conversion of relativistic axions off primordial magnetic fields in the early Universe~\cite{Higaki:2013qka,Evoli:2016zhj}.
However, our focus here is to systematize the study of the C$a$B and demonstrate the terrestrial detection prospects, thereby opening new paths to discovery.
In addition to outlining a number of distinct scenarios where relativistic axions could be produced in the early, or even late, Universe, we will demonstrate that such a population can leave a detectable fingerprint in instruments designed to search for dark matter. Studies of an additional relativistic component to the Universe are particularly relevant in light of recent discrepancies in measurements of the Hubble constant between the early ($H_0 = 67.4 \pm 0.5~{\rm km}/{\rm s}/{\rm Mpc}$) and late ($H_0 = 73.3 \pm 0.8~{\rm km}/{\rm s}/{\rm Mpc}$) Universe~\cite{Verde:2019ivm}. An additional contribution to $N_{\rm eff}$ of around $0.4$ -- which relativistic axions could provide -- may play a role in resolving the discrepancy, as they can relax the uncertainties in the early Universe measurement, giving a value of $66.3 \pm 1.4$, which would reduce, although far from resolve, the tension~\cite{Aghanim:2018eyx}. This provides an experimental target which we will denote by ``$H_0$ Preferred'' throughout.

A simplified representation of the C$a$B landscape discussed in this work, is provided in Fig.~\ref{fig:CaB}.
The black dashed curves show the differential axion energy density, $\Omega_a(\omega)$ (a precise definition is provided below), as a function of the energy, $\omega$, for the C$a$B variants discussed in this work.
The colored and shaded regions show the reach of two existing (solid curves) and future (dotted curves) instruments in this same space.
We will explain this figure in more detail later in the introduction, but already we emphasize that dark-matter searches will probe interesting C$a$B parameters, particularly at lower frequencies.
In Fig.~\ref{fig:CaB}, and throughout this work, we will focus on the axion-photon coupling,
\begin{equation} 
{\cal L} \supset - \frac{\ga}{4} a \tilde{F}_{\mu\nu} F^{\mu\nu} = \ga a \bE \cdot \bB \,.
\label{eq:Lga}
\end{equation} 
In general, the coupling of the axion to the Standard Model (SM) is highly uncertain and there exist experiments targeting a number of different axion-SM couplings (for a review, see e.g. \cite{Graham:2015ouw,Irastorza:2018dyq}). 
While we restrict our discussion to $\ga$, many aspects of the C$a$B extend to more general couplings.

At present, there are two primary classes of searches for axion backgrounds using the coupling in \eqref{eq:Lga}.
The two strategies are broadly distinguished by where the axions are produced: a relativistic population produced in the cores of compact astrophysical objects or a non-relativistic dark-matter population.
For existing relativistic searches, the axions are produced in compact objects, such as stars like the Sun, which act as a bright source of axions with energies in the $\sim$keV range.
Avoiding excess cooling of these objects from axion emission already puts a stringent bound on $\ga$~\cite{Raffelt:1996wa} with comparable limits obtained by directly searching for the emitted axions in helioscopes~\cite{Sikivie:1983ip} or absorption in direct detection experiments~\cite{Moriyama:1995bz}.
Together these searches, which we collectively refer to as ``star-emission'' bounds, are able to set strong bounds on axions with $m_a \lesssim 1~{\rm keV}$, with the strongest limits across this full range given by $\ga \lesssim 0.66 \times 10^{-10}~{\rm GeV}^{-1}$ as determined by the CAST helioscope~\cite{Anastassopoulos:2017ftl} and observations of Horizontal Branch stars~\cite{Ayala:2014pea,Carenza:2020zil}.
For $m_a \lesssim 10^{-10}~{\rm eV}$, these bounds can be strengthened by X-ray searches from conversion of axions emitted by SN-1987A~\cite{Payez:2014xsa} (assuming the supernova remnant is a proto-neutron star~\cite{Bar:2019ifz}), NGC~1275~\cite{Reynolds:2019uqt}, and super star clusters~\cite{Dessert:2020lil}, reaching $\ga \lesssim 3.6 \times 10^{-12}~{\rm GeV}^{-1}$.
We collectively denote these existing star-emission bounds by $\ga^{\rm SE}$.
These will be relevant as in the current work we will only consider axions with $m_a \ll 1~{\rm keV}$, and therefore the same axions constituting the cosmic background could also be produced in these astrophysical objects, and must therefore satisfy $\ga \leq \ga^{\rm SE}$.

Dark-matter searches instead look for non-relativistic axions with a much larger local number density~\cite{Sikivie:1983ip}.
Traditionally, axion dark matter has been searched for in microwave cavity haloscopes~\cite{Krauss:1985ub,Sikivie:1985yu}.
In the presence of a large magnetic field, axions in the $1-50~\mu$eV mass range can resonantly excite the modes of an $\mathcal{O}({\rm m})$ sized cavity (as $m_a^{-1} \sim \mu{\rm eV}^{-1} \sim {\rm m}$).
This detection principle underlies many of the strongest existing bounds on axion dark matter, as determined by the ADMX~\cite{Asztalos:2003px,Du:2018uak,Braine:2019fqb} and HAYSTAC~\cite{Zhong:2018rsr} collaborations (see also Ref.~\cite{Lee:2020cfj}), which already require dark-matter axions in this mass range to have $\ga$ orders of magnitude below $\ga^{\rm SE}$.
Ideas are currently being developed to extend the accessible axion dark-matter mass window to both higher and lower values.
For $m_a \leq 1~\mu{\rm eV}$, resonant conversion can still be obtained when the axion power is read out through a high quality-factor lumped-element resonator~\cite{Sikivie:2013laa,Chaudhuri:2014dla,Kahn:2016aff,Silva-Feaver:2016qhh}.
A broadband readout of the signal in this mass range has already been used to set limits comparable to $\ga^{\rm SE}$ by the ABRACADABRA~\cite{Ouellet:2018beu,Ouellet:2019tlz} and SHAFT~\cite{Gramolin:2020ict} instruments, and in the future DMRadio will aim to significantly improve on these pathfinding results~\cite{Chaudhuri:2014dla,Silva-Feaver:2016qhh,SnowmassOuellet,SnowmassChaudhuri}.
At higher masses the MADMAX Collaboration will search for dark matter using a dielectric haloscope, which searches for the electromagnetic emission that an axion generates at dielectric boundaries in the presence of a magnetic field~\cite{TheMADMAXWorkingGroup:2016hpc,Millar:2016cjp,Ioannisian:2017srr}.
Other proposed instruments searching for dark matter through the axion-photon coupling include resonant frequency conversion in superconducting cavities~\cite{Berlin:2019ahk,Lasenby:2019prg}, looking for a phase difference in locked lasers~\cite{Liu:2018icu,Obata:2018vvr}, exciting quasi-degenerate modes in a superconducting cavity \cite{Berlin:2020vrk}, detection of small energy deposits in crystals~\cite{Marsh:2018dlj,Trickle:2019ovy}, and matching the axion mass to a plasma frequency~\cite{Lawson:2019brd,Gelmini:2020kcu}, although this list is far from exhaustive.
In summary, it is likely that in the coming decades the axion dark matter hypothesis will either be confirmed, or required to satisfy $\ga \ll \ga^{\rm SE}$ in the mass range $1~{\rm neV} \lesssim m_a \lesssim 1~{\rm meV}$.

Let us now sketch how this progress in the search for axion dark matter can be repurposed to search for the C$a$B.
The detectable power deposited by an axion population via \eqref{eq:Lga} is naively $\propto \ga^2 \rho_a$ up to the details of the experimental readout.
Taking the experimental factors as constant between the two scenarios, we can obtain an estimate of the sensitivity for a dark-matter instrument to the C$a$B by matching the power between the two, i.e. $(\ga^2 \rho_a)_{{\rm C}a{\rm B}} = (\ga^2 \rho_a)_{\DM}$.
Assuming axions fully constitute dark matter, astrophysical observations determine that $\rho_a = \rho_{\DM} \simeq 0.4~{\rm GeV/cm}^3$, see e.g.~\cite{deSalas:2020hbh}.\footnote{We take $\rho_{\DM} = 0.4~{\rm GeV/cm}^3$ throughout.}
The unknown parameter being searched for is then $\ga$, which beyond $\ga \leq \ga^{\rm SE}$ is a free parameter, although in specific scenarios like the QCD axion sharper predictions are possible.
Regardless, for a given instrument we can project the reach in $\ga$ to determine the associated reach in deposited axion power.
For the C$a$B both $\ga \leq \ga^{\rm SE}$ and $\rho_a$ are free parameters.
If the C$a$B is a relic of the early Universe, measurements of $\Delta N_{\rm eff}$ further require the energy density to be less than that of the Cosmic Microwave Background (CMB)~\cite{Aghanim:2018eyx}, $\rho_a \lesssim \rho_{\gamma}$, although the density may be predicted in certain scenarios.
This poses an immediate challenge: for equal $\ga$, the power deposited by the C$a$B will be at least a factor of $\rho_{\DM}/\rho_{\gamma} \simeq 10^9$ smaller.
The situation is even more dire.
The detectability of power deposited by axion dark matter is enhanced by the exceptionally long coherence time of the signal, originating from the narrow energy distribution associated with non-relativistic dark matter in the Milky Way.
Indeed, for dark matter we expect $\Delta \omega/\omega \sim 10^{-6}$, whereas generically the C$a$B will have a broad distribution in energy, $\Delta \omega/\omega \sim 1$.
As we will review, this typically enhances sensitivity to the dark matter signal by a further three orders of magnitude relative to the C$a$B.
Accordingly, for equal $\ga$, a relativistic axion that is a relic of the early Universe is at most a trillionth as detectable as dark matter.

As we will show, the challenge is not insurmountable. Upcoming axion dark-matter instruments will have a sensitivity that such a C$a$B will be detectable.
This is demonstrated in Fig.~\ref{fig:CaB}, where we recast the existing results of ADMX and HAYSTAC and the expected future reach of DMRadio and MADMAX onto the equivalent C$a$B parameter space, assuming $\ga^{{\rm C}a{\rm B}} = \ga^{\rm SE}$.
In detail, we define $\Omega_a(\omega)$ as the relic density per unit log (angular) frequency of the axion
\begin{equation}
\Omega_a(\omega) = \frac{1}{ \rho_c}\frac{d\rho_a}{d \ln \omega}\,,
\label{eq:Omegaomega}
\end{equation}
with $\rho_c = 3 M_{\rm Pl}^2 H_0^2$ the critical density.\footnote{We take $M_{\rm Pl} \simeq 2.4 \times 10^{18}~{\rm GeV}$, the reduced Planck constant.}
Fixing the coupling, we can recast the stated dark-matter sensitivity to a sensitivity on $\rho_a$ and hence $\Omega_a(\omega)$.
The figure demonstrates that DMRadio will be sensitive to scenarios where $\Omega_a(\omega) \lesssim 5 \times 10^{-5}$, roughly corresponding to $\rho_a < \rho_{\gamma}$ -- a target cavity instruments may also reach in the future -- and further the $H_0$ preferred parameter space discussed earlier.

The sensitivity to such small energy densities suggests that the data collected by axion direct detection experiments can be repurposed to probe a range of cosmic sources of axions beyond non-relativistic dark matter.
The axion distribution can be narrow or broad and have a peak frequency over a wide range of energies, depending on how and when they were produced, which motivates the discussion in this work on mechanisms for generating the C$a$B.
In particular, we discuss a thermal axion background, emission from cosmic strings, and production from a parametric resonance in the early Universe, which is expected to produce a roughly Gaussian distribution, all of which are shown in Fig.~\ref{fig:CaB}.
For such cosmic relics, the axions will free-stream over long distances and their spectrum ultimately depends on the cosmic history of the Universe.
In this sense, axion experiments looking for a stochastic axion background are in close analogy with searches for a stochastic gravitational wave background (only axions may have a much larger coupling).\footnote{For reference, current pulsar timing arrays and laser interferometers have sensitivity to gravitational wave backgrounds of relic densities, $\Omega_{{\rm GW}}$, of ${\cal O}(10^{-10})$~\cite{Lentati:2015qwp,Shannon:2015ect,Arzoumanian:2018saf} and ${\cal O}(10^{-7})$~\cite{LIGOScientific:2019vic}, respectively.}

While ADMX is close, no existing instrument is currently sensitive to cosmological relics.
This motivates scenarios where the C$a$B is produced in the late Universe, where $\rho_a $ can be larger than $\rho_{\gamma}$, and in particular we discuss dark matter decaying to two relativistic axions, $\chi \to aa$, with $m_a \ll m_{\chi}/2$.
The resulting spectrum of axions receives two contributions: one from the decay of dark matter within the Milky Way, which generates a sharp $\Delta \omega/\omega \sim 10^{-3}$ spectrum, and the broader spectrum resulting from dark-matter decays throughout the Universe.
Both contributions can be seen in the spectrum shown in Fig.~\ref{fig:CaB}.
As we will show, dark-matter instruments can be repurposed into axion telescopes to search for this dark-matter indirect-detection channel.
Such searches can further exploit the fact that the Milky Way signal will undergo a daily modulation in microwave cavity instruments, as the relative direction of the signal, primarily from the Galactic Center, and the experimental magnetic field vary throughout the day.
Indeed, we will show that ADMX is currently sensitive to unexplored parameter space -- a reanalysis of their existing data may already reveal a signal of the C$a$B.

In the remainder of this work we will expand the above discussion as follows.
To begin with, in Sec.~\ref{sec:prod}, we introduce different possible C$a$B sources focusing on thermal production, dark-matter decay, parametric-resonance production, and  emission from topological defects.
Then, in Sec.~\ref{sec:det} we study the viability of detecting a relativistic axion background with instruments designed to search for dark-matter axions through the axion-photon coupling.
As already mentioned, we focus on the axion-photon coupling, and further will restrict our attention to the sensitivity with resonant cavity instruments such as ADMX and HAYSTAC, and lumped-circuit readout approaches such as DMRadio.
Our analysis will justify the sensitivities shown for these instruments in Fig.~\ref{fig:CaB}.
We will not, however, return to carefully consider the sensitivity of instruments focused on higher mass axion dark matter $m_a \sim \omega > 100~\mu{\rm eV}$, such as MADMAX.
Detecting a relic C$a$B requires sensitivity to $\ga$ many orders of magnitude below $\ga^{\rm SE}$.
This simply will not be achieved in any proposed high mass instrument.\footnote{That the arguably most well motivated C$a$B candidate -- a relativistic thermal relic -- is expected to peak in this energy range justifies considering dedicated experimental efforts, although we will not pursue this in the present work.}
In Sec.~\ref{sec:lim} we then combine the results to determine projected limits on various C$a$B scenarios, and finally present our outlook in Sec.~\ref{sec:conc}.

\section{C$a$B Sources}
\label{sec:prod}

We now turn to a discussion of specific production mechanisms for the C$a$B.
As mentioned already, axions can be produced in the early and late Universe, and we will consider examples of both.
In each scenario, our goal will be to characterize the associated axion energy spectrum, which will be a central ingredient when we come to detection.
For this purpose we will again use $\Omega_a(\omega)$ as defined in \eqref{eq:Omegaomega}.
We emphasize that the present discussion is not intended to be an exhaustive consideration of all scenarios from which a C$a$B could emerge, rather, we simply demonstrate that there are many possibilities.
Nonetheless, the analysis will reveal a common theme that emerges across production mechanisms, in particular that the C$a$B will generically be a broad distribution, $\Delta \omega/\omega \sim 1$.
When contrasted with the highly coherent signal predicted for dark matter, this expectation will represent a fundamental difference when approaching searches for relativistic axions.

\subsection{Thermal Relic}
\label{sec:thermal}

We begin by studying the simplest example of a C$a$B source, thermal production during the early Universe.
Early studies of thermal axion production can be found in~\cite{Turner:1986tb,Chang:1993gm,Masso:2002np,Hannestad:2005df,Graf:2010tv} with a more detailed analysis performed in~\cite{Salvio:2013iaa,Ferreira:2018vjj,Arias-Aragon:2020shv}. Fundamentally, if an axion was ever in thermal contact with the SM bath at high temperatures, then a residual thermal population is expected to exist to the present day, generating a C$a$B with the closest resemblance to the CMB. 
Indeed, a thermal axion relic will also be described by a blackbody spectrum, so that
\begin{equation}
\Omega_a(\omega) = \frac{1}{2\pi^2 \rho_c} \frac{\omega^4}{e^{\omega/T_a} - 1}\,,
\end{equation}
with a total energy density comparable to that of the CMB.

\begin{figure} 
\vspace{-0.1cm}
\begin{center} 
\includegraphics[width=8cm]{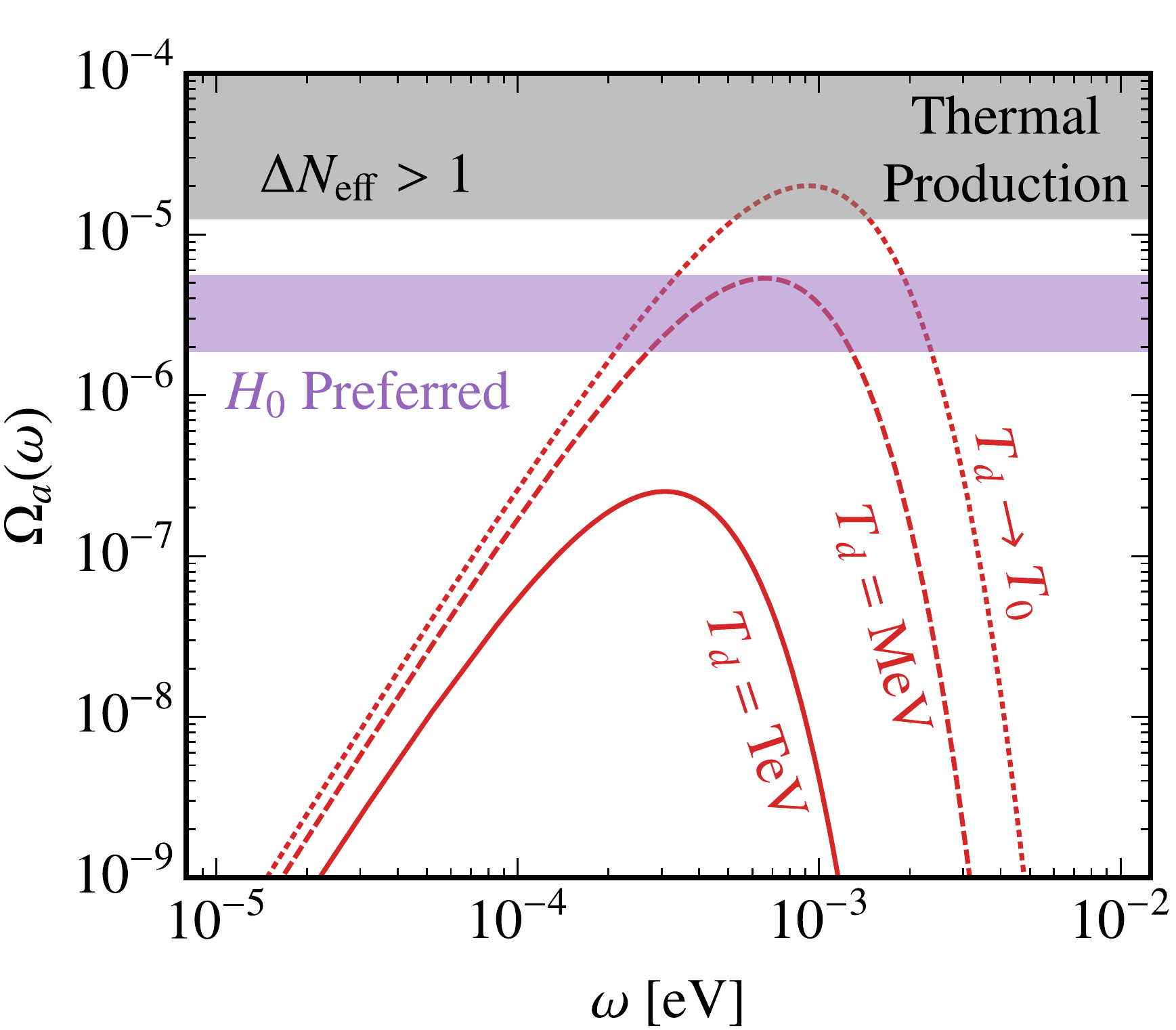} 
\end{center}
\vspace{-0.3cm}
\caption{The spectrum of thermal axions for different decoupling temperatures, $T_d$.
The region excluded by measurements of $\Delta N_{\rm eff}$ is shown, as well as the range where a contribution to $\Delta N_{\rm eff}$ can partially alleviate the $H_0$ tension.}
\label{fig:thermal}
\end{figure}

The above distribution is defined by a single parameter, the present day axion temperature, $T_a$.\footnote{We note that $T_a$ is not a true temperature since the axion is expected to have feeble self-interactions. 
Nevertheless, as for the CMB, frequencies above the horizon size at thermal decoupling redshift uniformly with the expansion of the Universe, implying that treating $T_a$ as an actual temperature is an excellent approximation.}
Remaining agnostic as to the exact axion-SM interaction that brought the axion into thermal equilibrium initially, at some temperature, $T_d$, the two will decouple.
If we assume that since axion freeze-out there has not been any entropy dilutions beyond those in the SM, and further that there was not an early period of matter domination, then as entropy is approximately conserved, we can relate the present and decoupling axion temperatures as follows,
\begin{equation}
T_a  \simeq T_0 \left( \frac{g_{\ast,S} (T_0) }{g_{\ast,S} (T_d)} \right)^{1/3}\,.
\label{eq:tratio}
\end{equation}
Here $T_0 \simeq 2.7~{\rm K}$ is the present day CMB temperature and $g_{\ast,S}(T)$ is the number of entropic degrees as a function of temperature.
Accordingly, we can specify the thermal axion in terms of $T_a$ or $T_d$.
The spectrum for different decoupling temperatures is shown in Fig.~\ref{fig:thermal}.
Generically, a thermal distribution is associated with a number density of axions of ${\cal O}(1-100~{\rm cm}^{-3})$ and a peak energy of $\sim 10^{-4}~{\rm eV}$.
Again, both are comparable to the CMB.
As the figure shows, $T_d \lesssim 1~{\rm MeV} $ is excluded by $\Delta N_{\rm eff}$ measurements, which would include the case where $T_a = T_0$ as shown in Fig.~\ref{fig:CaB}.
Nevertheless, the range $1~{\rm MeV} \lesssim T _d \lesssim 1~{\rm GeV} $ is somewhat favored as a solution to the present $H _0$ tension, a possibility that was studied in detail in Ref.~\cite{DEramo:2018vss}.

Ultimately, $T_d$ is determined by the microphysics responsible for the axion coming into thermal contact.
Axions coupled to photons have a maximum possible decoupling temperature since processes such as, $ \gamma e \rightarrow a e $, will keep it in equilibrium. Equating this rate with Hubble leads to an estimate of the decoupling temperature, $ T _d  \sim ~{\rm TeV} \left( g  ^{ {\rm SE }}_{ a \gamma \gamma }/g_{ a \gamma \gamma } \right) ^2  $. We conclude that axions saturating the star-emission bounds would have a decoupling temperature of around a TeV, while additional interactions can keep the axion thermally coupled at lower temperatures. This motivates a range of decoupling temperatures. Lastly, we emphasize that a thermal abundance of axions will always form as long as the temperature of the Universe was ever above the decoupling temperature, making this population a robust prediction for any theory without a low reheating temperature after inflation.

\subsection{Dark-Matter Decay}
\label{sec:DMDecay}

Dark matter need not be absolutely stable, and axions offer one possible decay channel.
Provided that the dark-matter mass is significantly larger than $m_a$, then the axions produced through this process will be relativistic.
If these same axions have a sizable photon coupling then they are in principle detectable in terrestrial experiments, opening up a new channel for the indirect detection program. 
An important aspect of this scenario, is that since the dark-matter abundance is considerably larger than the CMB ($\rho_{\DM}/\rho_{\gamma} \simeq 10^9$), decaying dark matter can result in a C$a$B energy density that is significantly larger than allowed for a relic population, given bounds from $\Delta N_{\rm eff}$.\footnote{This possibility was also noted in Ref.~\cite{Cui:2017ytb} as an opportunity for generating keV scale relativistic axions which would be detectable in axion helioscopes.}
Accordingly, in the short term decaying dark matter represents the most accessible C$a$B candidate.

Currently, this scenario is only constrained indirectly as a consequence of the fact that a significant fraction of dark matter decaying into radiation would modify the expansion history of the Universe~\cite{Gong:2008gi,Poulin:2016nat}. 
Qualitatively, these bounds require the decay rate to be less than the current Hubble rate $\Gamma \lesssim H_0$, so that the dark-matter lifetime is longer than the age of the Universe, $\tau = 1/\Gamma \gtrsim t_U$.
As dark matter decaying to a relativistic species would modify the expansion history, this possibility has been suggested as a potential resolution to the Hubble tension~\cite{Vattis:2019efj} (though this was later refuted \cite{Haridasu:2020xaa}).
Given the existing tension in $H_0$ measurements between the early and late Universe, the current bounds depend noticeably on which data set is used.
Recently, using only local measurements, the Dark Energy Survey constrained the dark-matter lifetime to $\tau \gtrsim 50~{\rm Gyr} \sim 3.6\, t_U$~\cite{Chen:2020iwm} and we consider this as our nominal bound.
Although this value is likely to be revised with developments on the Hubble tension, this will not qualitatively impact our discussion.

The first goal of this section will be to describe the C$a$B that results from decaying dark matter. We will then outline an explicit decaying dark-matter model, and lastly, we discuss the feasibility of detecting axions arising from the related mechanism of neutrino decays.

\subsubsection{The Axion Spectrum from Decaying Dark Matter}

Axions produced from dark-matter decay will have a spectrum that results from two distinct sources: the decay of galactic dark matter within the Milky Way and contribution from decays of the extragalactic dark matter throughout the Universe. While in both cases the fundamental process will be dark matter, which we denote $\chi$, decaying to axions, the resulting spectra will be significantly different. Nonetheless, the contributions produce similar axion abundances.

Consider first the extragalactic contribution resulting from dark matter decaying to axions throughout the isotropic, homogeneous, and expanding Universe.
The number density of axions observed today produced per unit time and per unit energy is given by the product of several factors.
The first of these is the number density of dark-matter particles at a given time $t$, $\rho_{\DM}(t)/m_\chi$.
We must also weight this by the rate at which dark matter decays at this time, which is $\Gamma e^{-\Gamma t}$ (we assume that the decay rate $\Gamma$ is constant through cosmic history).
Each decay is associated with a differential energy spectrum of the emitted axions, $dN/d\omega'$, normalized such that its integral over all $\omega'$ gives the number of emitted axions.
As the emitted axions are assumed to be relativistic, the axion energy as observed today will be suppressed by a ratio of scale factors, $\omega = \omega' a$, where $a$ is the scale factor at time $t$ and we take $a_0=1$.
Finally, the present number density will be diluted as compared to the density emitted at $t$, as the Universe is now larger by a factor of $1/a^3$.
Combining these factors and then integrating over all time from $t=0$ to the present $t=t_0$, we obtain the total extragalactic differential number density as
\begin{equation} 
\frac{dn_a}{d\omega} = \int_0^{t_0} dt\,a^3 \Gamma e^{- \Gamma t} \frac{\rho_{\DM}(t)}{m_\chi} \frac{dN}{d\omega'} \bigg|_{\omega' = \omega/a}\,.
\label{eq:dndw-egldecay}
\end{equation}
Changing integration variables to the scale factor, we can write this as
\begin{equation} 
\Omega_a(\omega) \simeq \frac{\Omega_{\DM}\,\omega^2 }{m_{\chi}} \int_0^1 \frac{da}{a}\, \frac{\Gamma e^{-\Gamma t(a)}}{H(a)} \frac{dN}{d\omega'} \bigg|_{\omega' = \omega/a}\,,
\end{equation} 
with $\Omega_{\DM} \simeq 0.27$ the cosmological dark-matter density and $t(a)$ the age of the Universe as a function of scale factor, so that $t(1)=t_U$.

\begin{figure}
\vspace{-0.1cm}
\begin{center} 
\includegraphics[width=8cm]{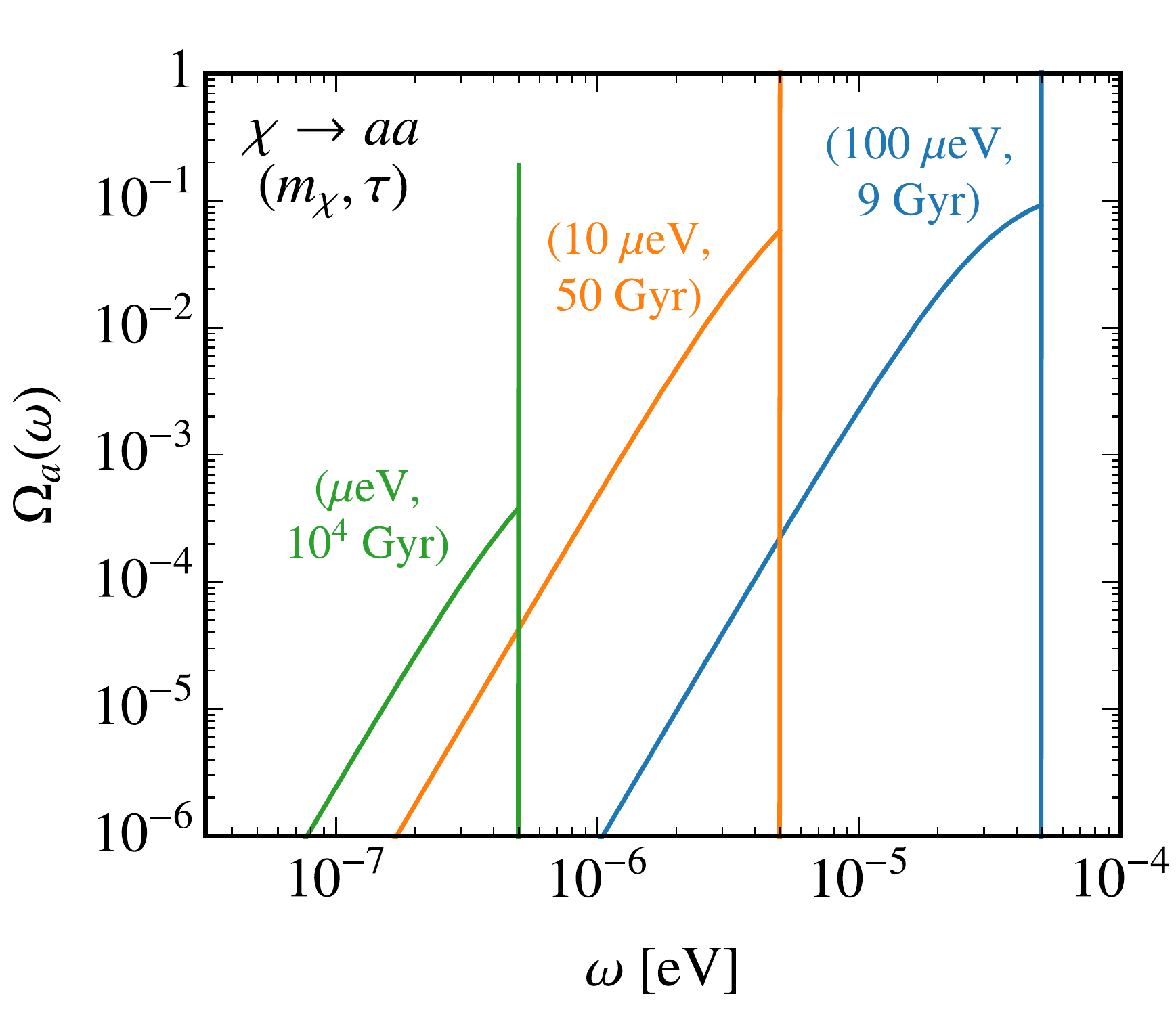} 
\end{center}
\vspace{-0.3cm}
\caption{The spectrum of axions arising from a component of dark matter decaying into axions through, $\chi \to a a$, for several dark-matter masses and lifetimes. The extragalactic component gives a broad spectrum of axions due to cosmological redshift while decays within the Milky Way produce a narrow spectrum at half the dark-matter mass.}
\label{fig:DMDecay}
\end{figure}

The above result is appropriate for a general dark-matter decay axion spectrum.
Throughout this work, however, we will specialize to the simple example of a two-body decay, $\chi \to aa$, again assuming $m_a \ll m_{\chi}/2$, so that the produced axions are relativistic.
In this case, the spectrum takes the following particularly simple form,
\begin{equation} 
\frac{dN}{d\omega'} = 2\delta(\omega'-m_\chi/2)\,. 
\label{eq:2bodyspec}
\end{equation} 
Inserting this into the above gives,
\begin{equation}
\Omega_a(\omega) \simeq \Omega_{\DM} \left( \frac{2\omega}{m_{\chi}} \right)^2 \frac{e^{-t(2\omega/m_{\chi})/\tau}}{\tau\,H(2\omega/m_{\chi})} \Theta(m_{\chi}/2-\omega)\,,
\label{eq:Omega-egdecay}
\end{equation}
where $\Theta$ is the Heaviside step-function and we have exchanged the decay rate for the lifetime.
The axion energy spectrum as observed today is shown in Fig.~\ref{fig:DMDecay} for different dark matter lifetimes and masses.
The sharp peak is associated with galactic decays, described shortly, but the broad continuum arises from the above expression. 
The redshifting of the axions produced throughout the Universe smooths the sharp two-body spectrum into a continuum.

We now compute the axion abundance created within the Milky Way. 
For this, the number density of axions per unit energy around Earth is determined by the conventional indirect detection expression for decaying dark matter, and given by\footnote{For $\tau \gg t_U$, we have $e^{-t_U/\tau} \simeq 1$, and so this factor is commonly neglected in indirect detection analyses.}
\begin{equation} 
\frac{dn_a}{d\omega} = \frac{e^{-t_U/\tau}}{4\pi m_\chi \tau} \frac{dN}{d\omega}  \int ds\,d\Omega\, \rho_{\DM}(s,\Omega)\,.
\label{eq:dndw-localdecay}
\end{equation}
Note we have assumed the observable decays within the Milky Way occur at $t = t_U$. 
The integral at the end of this is commonly referred to as the $D$-factor in the indirect detection literature (for details see e.g.~\cite{Lisanti:2017qoz}).
For a canonical Milky Way dark-matter profile, the $D$-factor has a full sky integrated value of $D_{\MW}\simeq 2.7 \times 10 ^{ 32 }~{\rm eV} / {\rm cm} ^2 \cdot {\rm sr} $.\footnote{To obtain this value we assumed a canonical Navarro-Frenk-White profile~\cite{Navarro:1995iw,Navarro:1996gj}, took the Earth-Galactic Center distance as 8.127 kpc~\cite{Abuter:2018drb}, and a local dark-matter density of 0.4 GeV/cm$^3$.}

The expression in \eqref{eq:dndw-localdecay} again holds for a general spectrum.
If we specialize to the case of $\chi \to aa$, then the local axion energy density per unit log frequency is approximately given by,
\begin{equation} 
\Omega_a (\omega) \simeq \frac{\omega^2 e^{-t_U/\tau}}{2\pi m_{\chi} \tau \rho_c} \delta (\omega - m_\chi/2 ) D_{\MW}\,.
\label{eq:Omega-localdecay}
\end{equation} 
Integrating \eqref{eq:Omega-egdecay} and \eqref{eq:Omega-localdecay}, we can determine the total energy density for the two contributions.
This is maximized for $\tau \sim t_U$, where we have $\rho_a^{\MW} \simeq 2 \rho_a^{\EG} \simeq 10^3 \rho_{\gamma}$, so that, as claimed, the energy densities from the two contributions are comparable, and a combined density larger than the CMB can be obtained for a range of lifetimes ($\rho_a \lesssim \rho_{\gamma}$ for $\tau \gtrsim 10^4 \,t_U$).

The reason \eqref{eq:Omega-localdecay} is an approximation is because it assumes the observed axion spectrum is the same as that in the dark-matter rest frame.
While this is often a reasonable approximation, axion experiments are often sensitive to extremely narrow energy distributions --- recall, for dark matter, $\Delta \omega/\omega \sim 10^{-6}$.
This motivates a more detailed consideration of the axion energy distribution.
There are two contributions that will resolve the distribution in \eqref{eq:Omega-localdecay} to have a finite width: the velocity dispersion of dark matter in the Milky Way and the finite velocity of the Earth through the dark-matter halo.
Both velocities result in a net motion between the observer and source of axions, and therefore the axion energies will be Doppler shifted by a factor of $v \sim 10^{-3}$, which is the magnitude of both velocity components.
In this work, we will simply replace $\delta(\omega - m_\chi/2)$ in \eqref{eq:Omega-localdecay} with a Gaussian of width $10^{-3}$ centered at half the dark-matter mass. The actual distribution is more complex, indeed it depends on the dark-matter distribution and varies across the sky given the motion of the Earth in the halo frame (for further details, see Ref.~\cite{Speckhard:2015eva}).
Nevertheless, the main aspect of the distribution relevant for forecasting sensitivities is the width, and the Gaussian approximation adequately accounts for this.

\subsubsection{An Explicit Model: Decaying Scalar Dark Matter} 
\label{sec:DMdecaymodel}

Above we considered the axion abundance produced through dark-matter decay, with all model dependence in the axion spectrum, $dN/d\omega$, and the lifetime $\tau$.
We now study a decaying dark-matter model which predicts a detectable C$a$B, and generates the simple $\chi \to a a$ spectrum used above.

Consider a theory with a complex scalar field, $\Phi$, with potential,
\begin{equation} 
V(\Phi) = \lambda^2 \left( \left| \Phi \right|^2 - \frac{f_a^2}{2} \right)^2 \,.
\label{eq:Vphi}
\end{equation} 
The theory has a spontaneously broken U(1) and we identify the Goldstone boson with the axion and the radial mode with dark matter, decomposing the field as $\Phi = (\chi + f_a) e^{ia/f_a}/\sqrt{2}$.

In the broken phase, the relevant axion dark-matter couplings are,\footnote{The potential also contains terms which can mediate annihilation to axions, $\chi \chi \to a a$.
For the masses considered in this work, this annihilation is completely subdominant to the decay.}
\begin{equation}
V(a, \chi) \supset \frac{1}{2} (2 \lambda^2 f_a^2) \chi^2 + \frac{1}{2} (2 \lambda^2 f_a) \chi a^2 \,,
\end{equation}
from which we identify the dark-matter mass as $m_{\chi} = \sqrt{2} \lambda f_a$. Further, the axion dark-matter coupling allows us to compute the rate of dark-matter decay as,
\begin{equation} 
\Gamma_{\chi \to aa} = \frac{m_\chi^3}{16 \pi f_a^2}\,, 
\label{eq:chidecay}
\end{equation}
with corresponding axion spectrum as given in \eqref{eq:2bodyspec}. 
In order for $\Gamma_{\chi \to aa}$ to be, at least, comparable to the age of the Universe, we then require $f_a$ to be well below the weak scale.
This may seem hard to reconcile given the stringent bounds on the axion-photon coupling, $ \ga^{\rm SE} \ll 1~{\rm TeV}^{-1}$, however, this can be natural if the axion obtains its photon coupling through axion-axion or photon-dark-photon mixing, and as we demonstrate in App.~\ref{app:mixing} this does not require any elaborate model building. Nevertheless, this does require forbidding any significant terms in the scalar potential mixing between the SM Higgs and $ \Phi $. In generating $\ga$, $\chi$ may also obtain a coupling directly to photons.
Even though searches for $\chi \to \gamma \gamma$ are significantly more stringent than the axion searches discussed in this work, these constraints are not significant in the parameter space of interest, as we work in the limit of $f_a \ll \ga^{-1}$.

\subsubsection{Cosmic Neutrino Background Decay}
A C$a$B may also be produced as a byproduct of neutrino decays of the cosmic neutrino background (C$\nu$B). Flavor off-diagonal couplings of axions to neutrinos can be the result of a global lepton number broken by multiple scalar fields with the axion playing the role of a Majoron~\cite{Gelmini:1980re} or, more generally, the familon~\cite{Feng:1997tn}. A generic axion can have a coupling to neutrinos given by ($Q = Q^\dagger$),
\begin{equation} 
{\cal L} \supset Q_{ij}\frac{\partial_\mu a}{f_a} \nu^\dagger_i \bar{\sigma}^\mu  \nu_j \,.
\end{equation} 
Here we assume the neutrinos are Majorana and work with two-component fermion notation.
The neutrino decay rates were recently calculated in Ref.~\cite{Dror:2020fbh} for $L_i - L_j$ gauge bosons in the high-energy limit and the results can be translated to axions with the relation, $1 / f_a \leftrightarrow g_X / m_X $,
\begin{align} 
\Gamma_{\nu_i \rightarrow \nu_j a} & = \frac{1}{16 \pi m_i} \overline{\left| {\cal M} \right|^2} \left(1 - m_j^2 / m_i^2 \right)\,, \\ 
\overline{ \left| {\cal M} \right|^2} & = \frac{1}{f_a^2} \left( ( m_i^2 - m_j^2)^2 \text{Re} Q_{ij}^2 + ( m_i + m_j )^4 \text{Im} Q_{ij}^2 \right)\,.\nonumber
\end{align} 
Parametrically, $\Gamma_{\nu_i \to \nu_j a} \sim  m_\nu^3 Q^2 /f_a^2$ and for the decay to be comparable to Hubble while avoiding the star-emission bounds requires $f_a \ll 1~{\rm TeV}$ while keeping $\ga \ll 1~{\rm TeV} ^{-1} $. As mentioned previously, this can occur naturally for axions that inherit a photon interaction through axion-axion or photon-dark photon mixing, see also App.~\ref{app:mixing}.

The strongest constraints on the neutrino lifetime are from observations of neutrino free-streaming in the CMB~\cite{Hannestad:2004qu}.
Current limits allow for neutrino lifetimes well below the age of the Universe; indeed, a recent reanalysis of the bounds in Ref.~\cite{Barenboim:2020vrr} found a conservative limit that is on the order of several days. Accordingly there is a significant possibility that neutrino decays populate the C$a$B. For decays while the neutrinos are still relativistic, the axions are produced with an energy comparable with the neutrino temperature, and hence this results in a spectrum similar to the thermal background considered in Sec.~\ref{sec:thermal}. Axions produced from late-time neutrino decays -- after neutrinos have become non-relativistic -- will have a peaked spectrum around the neutrino mass. In either event, the resulting spectrum will be subject to the same challenge as the thermal background, in that it is located at high frequencies where it is unlikely to be observable in the near future due to the lack of sensitive experiments at these energies. As such, we will not evaluate this case in detail, but note should the thermal C$a$B become accessible, then likely so too would this scenario.

\subsection{Parametric Resonance}
\label{sec:parametric-resonance}

A C$a$B can also be produced through the process of parametric resonance in the early Universe~\cite{Dolgov:1989us,PhysRevD.42.2491,Kofman:1994rk,Kofman:1997yn}. In order for this process to occur, the axion must be coupled to a scalar field which is heavily displaced from its minimum after inflation. This will occur by quantum fluctuations for any scalar field, unless it is fixed to the origin by an effective mass larger than the Hubble scale at inflation. When such a scalar field begins to oscillate about its minimum it will produce axions with a bose-enhanced rate that will typically deplete its energy density within an e-fold.
The characteristic axion energy as observed today will have redshifted dramatically and can be much lower than the energy of axions produced by perturbative decay of a scalar field.
As the parametric-resonance phenomena is a non-perturbative process that occurs out of equilibrium, computing the spectrum in detail requires evolving multiple scalar fields on the lattice. We will not attempt such a calculation here but instead perform qualitative estimates. Earlier work on relativistic axion production considered potential modifications to $\Delta N_{\rm eff}$ and parallel production of gravitational waves~\cite{Ema:2017krp}. In this subsection we review the dynamics of axion production and explore the parameter space leading to a detectable axion background. We follow the notation and discussion of Refs.~\cite{Co:2017mop,Dror:2018pdh} which studied the prospect of ultralight bosonic dark matter produced through parametric resonance. 

We now focus on an explicit realization of the parametric-resonance phenomena, which can be achieved using the same model introduced in Sec.~\ref{sec:DMDecay}. Recall, there we had the axion arise from a global symmetry breaking complex scalar, $\Phi$, with a radial mode $\chi$ playing the role of dark matter.
Our starting point will again be the potential given in \eqref{eq:Vphi} and as our initial condition we take $\chi$ to have a large field value, $\chi_i \gg f_a$. At early times we assume the second derivative of the potential with respect to the field is greater than Hubble squared, $V''(\chi_i) \gtrsim H^2$, such that the scalar is stuck and the field redshifts as vacuum energy. When $V''(\chi_i) \sim H^2$ the field begins to oscillate, resulting in exponential production of both the radial and axion modes.
Since $\chi_i \gg f_a$, $m_\chi^2 \chi^2$ is small relative to $\lambda \chi^4$ and can be neglected during the oscillations. This leads to a broad resonance that rapidly depletes the energy density stored in the original scalar field. Furthermore, since the axion energy at the time of production is set by the effective mass of $\chi$, $m_{\chi}^{\rm eff}(\chi) \equiv \lambda \chi$, it is independent of the temperature of the SM bath and often considerably smaller. Subsequent redshift until today can lead to relativistic axions over a wide range of energies, well below the temperature of the CMB and potentially within reach of low-frequency axion haloscopes.

We now estimate the abundance and energy spectrum of the axion and radial mode.
Axions are emitted with energy $\omega_a \sim m_\chi^{\rm eff}(\chi_i)$ during radiation domination at a temperature of oscillation, $T_{\rm osc} \sim \sqrt{\Gamma M_{\rm Pl}}$, where $\Gamma$ is the oscillation timescale.
For perturbative production, $\Gamma$ cannot be arbitrarily large, in detail $\Gamma \lesssim m_\chi^3 / f_a^2$.
For parametric resonance, the particle production will occur within a few oscillations, so $\Gamma \sim \lambda \chi_i$.
The characteristic axion energy, as measured today, is redshifted using $T_{\rm osc}$ and given by,
\begin{equation}\begin{aligned}
\bar{ \omega}_a & \simeq m_\chi^{\rm eff}( \chi_i ) \left( \frac{s(T_0)}{s(T_{\rm osc})} \right)^{1/3}\\ 
& \simeq 10^{-15}~{\rm eV} \left( \frac{m_\chi^{\rm eff} (\chi_i)}{1~{\rm MeV}} \right)^{1/2} \,,
\end{aligned}\end{equation}
where $s(T)$ is the entropy of the SM bath.
Accordingly, provided $m_\chi^{\rm eff} (\chi_i) \ll M_{\rm Pl}$, we have $\bar{\omega}_a \ll T_0$, and the axion energy will be well below the cosmic photon temperature.

With an estimate for where the spectrum will peak, next we consider the $a$ and $\chi$ comoving number densities after the oscillations have concluded.
These can be parameterized as,
\begin{equation}
Y_a = f \frac{\rho_{\chi,{\rm osc}}}{\bar{ \omega}_a' s(T_{\rm osc})}\,,\hspace{0.3cm}
Y_{\chi} = (1-f) \frac{\rho_{\chi,{\rm osc}}}{\bar{ \omega}_{\chi}'s (T_{\rm osc})}\,,
\end{equation}
where $f$ denotes the fraction of energy transferred to axions, $\bar{\omega}_a'$ and $\bar{\omega}_{\chi}'$ are the mean energies of each particle at the time of production, and $\rho_{\chi,{\rm osc}} \simeq \frac{1}{4} \lambda^2 \chi_i^4$ is the total energy density in the radial direction prior to oscillations. Since the vacuum mass of radial mode can be neglected in this limit both $a$ and $\chi$ are produced with comparable energy densities, $f \simeq 1/2$ and comparable energies, $\bar{\omega}_a' \simeq \bar{\omega}_{\chi}' \simeq \lambda \chi_i$. As noted above, the energy is determined by the effective mass, which is driven by the quartic. These determine the comoving number densities to be
\begin{equation} 
Y_a \simeq Y _\chi \simeq \frac{0.01}{\lambda^{1/2}} \left( \frac{\chi_i}{M_{\rm Pl}} \right)^{3/2} \,,
\end{equation} 
so that the axion energy density today is fixed by the initial scalar field value,
\begin{equation} 
\frac{\rho_a}{\rho_c}  \simeq 3 \times 10^{-7} \left( \frac{\chi_i}{M_{\rm Pl}} \right)^2\,. 
\label{eq:PR-rho}
\end{equation} 
We conclude for $\chi_i < M_{\rm Pl}$, parametric resonance produces a maximum axion relic density a few orders of magnitude below that of the CMB. 

The characteristic frequency of the axions may span many orders of magnitude, and depends on the initial field value as well as the quartic. To explore the parameter space it is helpful to focus on a specific case where $\chi$ makes up dark matter.\footnote{Alternatively, if $\chi$ has significant interactions with the SM, it may transfer its entropy into the rest of the thermal bath and be cosmologically unobservable as considered in Ref.~\cite{Co:2017mop}.} 
Requiring $\chi$ to have the observed dark-matter abundance provides an additional constraint,
\begin{equation} 
m_\chi  Y_{\chi} \simeq m_\chi \frac{0.01}{\lambda^{1/2}} \left( \frac{\chi_i}{M_{\rm Pl}} \right)^{3/2} \sim 1~{\rm eV}\,. 
\label{eq:darkmatter}
\end{equation} 
To study the resulting model space, we take the free parameters to be the three parameters $\{ m_\chi,\, \chi_i,\, \lambda \}$, one combination of which is restricted by the requirement of \eqref{eq:darkmatter}.
Note, the value of the vacuum expectation value is a dependent parameter, $f_a = m_{\chi} / \sqrt{2} \lambda$.
This leads to a prediction for the axion energy today, 
\begin{equation} 
\bar{\omega}_a  \simeq 5 \times 10^{-7}~{\rm eV} \left( \frac{m_\chi}{1~{\rm eV}} \right) \left( \frac{\chi_i}{M_{\rm Pl}} \right)^2 \,.
\label{eq:PR-omega}
\end{equation} 
The relative spread in the axion spectrum must be determined using lattice simulations though we expect it to be ${\cal O}(1)$. Simulations for a similar theory have found the spectrum to be roughly a Gaussian with a relative width of order unity~\cite{Micha:2004bv}. Here we have only included the influence of redshift on the axion energy spectrum. It is known that there are additional number-changing processes that tend to move the axion spectrum toward a thermal distribution. These are slow and not expected to effectively thermalize axions on a cosmological timescale, but may shift the peak axion frequency by an order of magnitude~\cite{Micha:2004bv}.

\begin{figure}
\vspace{-0.1cm}
\begin{center} 
\includegraphics[width=8.5cm]{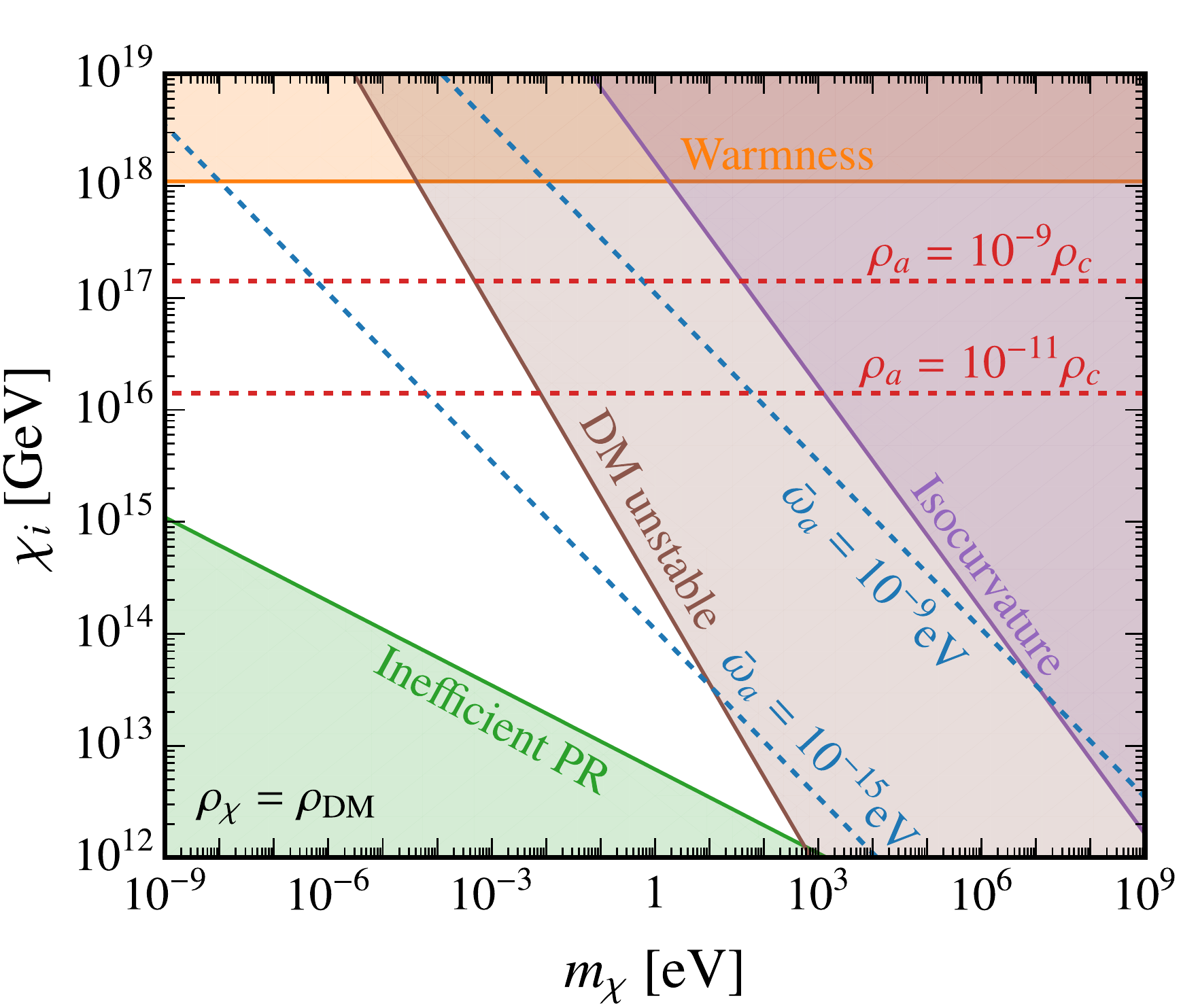} 
\end{center}
\vspace{-0.3cm}
\caption{The parameter space of parametric resonance producing axions and $\chi$, where everywhere in the plot $\chi$ constitutes all of dark matter.
The constraints shown arise from dark-matter stability, warmness, efficient parametric resonance, and isocurvature.
Further, we show two contours of axion energy density, and separately the mean C$a$B energy.
The higher mean energy of $\bar{\omega}_a = 10^{-9}$ eV falls entirely in the region excluded by DM stability, however this constraint is removed if we no longer assume $\chi$ constitutes the dark matter of the Universe.}
\label{fig:PR}
\end{figure}

The parameter space of PR production, assuming $\chi$ makes up dark matter, has several constraints summarized below:\footnote{Other bounds on this scenario include requiring $\chi_i$ to be sub-Planckian, avoiding an epoch of early matter domination (for consistency), and perturbativity of the quartic.
These bounds are subdominant to those we consider for the entire allowed region.}
\begin{enumerate} 
\item {\bf DM Unstable}: As discussed in section \ref{sec:DMDecay}, $\chi$ may also decay (perturbativity) into axions with a rate given by \eqref{eq:chidecay}. As discussed earlier in the context of dark-matter indirect detection (see Sec.~\ref{sec:DMDecay}), we use the nominal bound of $\tau \gtrsim 50~{\rm Gyr} \sim 3.6\, t_U$~\cite{Chen:2020iwm}.
\item {\bf Warmness}: This bound arises from the requirement that $\chi$ are cold enough to constitute dark matter today, roughly taken to be $p_\chi (T) / m_\chi \lesssim 10^{-3}$ at recombination. In detail, we require
\begin{equation} 
\frac{p_\chi(T_{\rm eq})}{m_\chi} \simeq \frac{10^{-11} ~{\rm eV}}{m_\chi} \left( \frac{m_\chi (\chi_i)  }{1~{\rm MeV}} \right)^{1/2} \lesssim 10^{- 3}\,.
\end{equation} 
\item {\bf Inefficient PR}: Parametric resonance is efficient at producing axions when the initial field value is much larger than its vacuum value. Otherwise the resonance is a narrow feature and is unable to convert all the energy density in $\chi$ into field excitations. For this condition we take the rough bound, $\chi_i \gtrsim 10 f_a$.
\item {\bf Isocurvature}: During inflation we assume $\chi$ has an effective mass below the Hubble scale at inflation, $\lambda \chi_i \lesssim H _{\rm inf}$ such that fluctuations during inflation displace the field away from the minimum. These isocurvature perturbations can be observed in the CMB, placing a bound $\chi_i/H _{\rm inf} \gtrsim (\pi^2 \beta {\cal P}_R (k_\ast))^{-1/2}$, where $\beta \leq 0.011$ is the isocurvature fraction and ${\cal P}_R(k_\ast) \simeq 2.1 \times 10^{-9}$ is the observed amplitude of the curvature power spectrum at the pivot scale~\cite{Akrami:2018odb}. Combining these results places a limit on the scalar quartic, $\lambda$.
\end{enumerate}

The parameter space of relativistic axions produced through parametric resonance in light of these bounds is shown in Fig.~\ref{fig:PR}, where we have fixed the abundance of $\chi$ to match that of dark matter.
We see that for $\chi$ to constitute dark matter, we require $m_\chi \lesssim 1~{\rm keV} $ and the condition of a detectable C$a$B further restricts $\chi$ to large initial field values and smaller masses.

These results demonstrate that a consistent C$a$B produced through parametric resonance can occur over an enormous range of frequencies.
From \eqref{eq:PR-rho}, detectability will be maximized for $\chi_i \sim M_{\rm Pl}$ (up to consistency of the warmness criteria).
The spectrum is then expected to be roughly an $\mathcal{O}(1)$ width Gaussian~\cite{Micha:2004bv}, with peak frequency determined from \eqref{eq:PR-omega} as $\bar{\omega}_a \simeq 5 \times 10^{-7}~{\rm eV}\, (m_{\chi}/1~{\rm eV})$.
In the scenario where $\chi$ constitutes dark matter, this allows a mean energy as low as $\bar{\omega}_a \sim 10^{-28}$ eV, when $m_{\chi} \sim 10^{-22}$ eV is in the fuzzy dark-matter regime, and as high as $5 \times 10^{-12}$ eV when saturating the dark-matter stability criteria, shown in Fig.~\ref{fig:PR}.
Removing the requirement that $\chi$ be dark matter, the frequency range can then be extended even further, particular to higher frequencies which may be accessible by DMRadio, or even ADMX and HAYSTAC.

\subsection{Topological Defect Decay}
\label{sec:cosmic-string}

\begin{figure*}
\begin{center} 
\includegraphics[width=8.5cm]{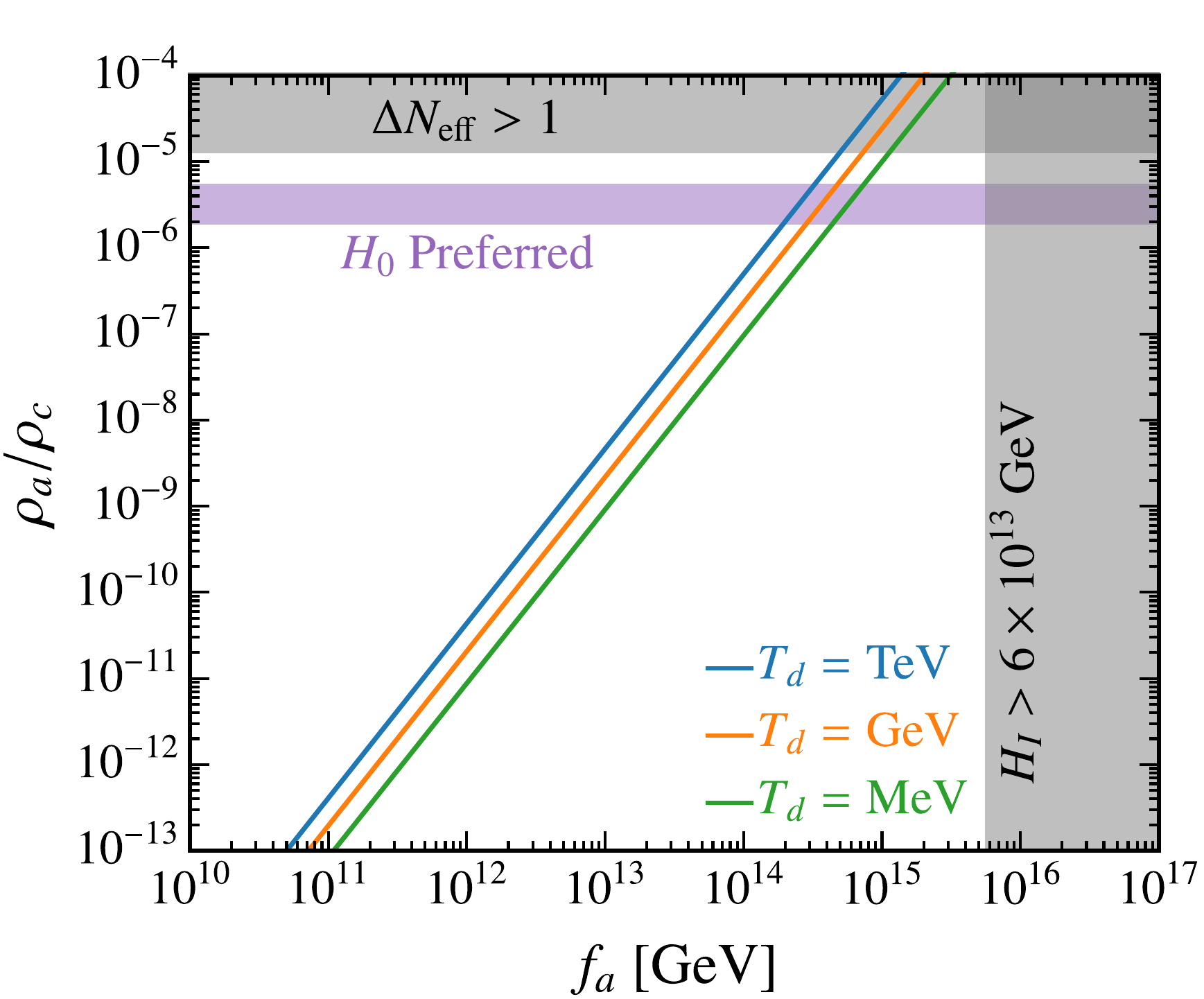} 
\includegraphics[width=8.5cm]{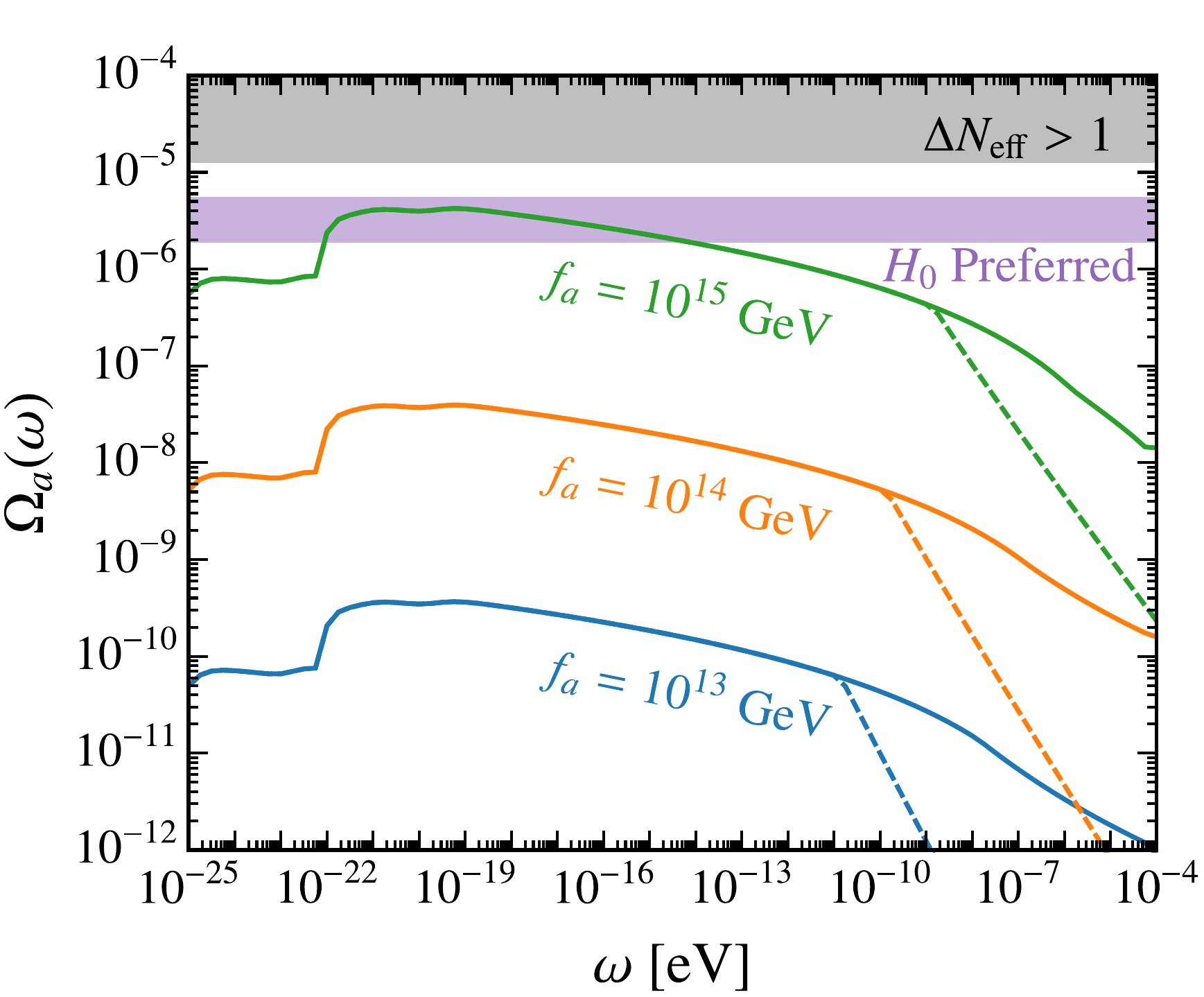} 
\end{center}
\vspace{-0.3cm}
\caption{The energy density (left) and spectrum (right) of a C$a$B produced from the emission of cosmic strings.
The relic density is shown as a function of the axion decay constant for different decoupling temperatures, $T_d$.
The spectrum is shown as a function of energy, however, the shape at high frequencies is sensitive to the decoupling temperature, which we take to be equal to $f_a$ (solid) or $f_a / 10^3$ (dashed). 
In both figures we also show bounds from $\Delta N_{\rm eff}$ and the region preferred to mildly alleviate the Hubble tension.
On the left, we further show the region excluded by the maximum possible reheating scale of the Universe, derived by assuming instantaneous reheating from the maximum scale of inflation.}
\label{fig:drhodomega_string}
\end{figure*}

A C$a$B may also be produced through the decay of topological defects.
In this section we study the abundance and energy spectrum of axions emitted from a network of cosmic strings formed during a thermal phase transition.
Axions produced during the phase transition itself are in thermal contact with the SM bath and will contribute to the thermal background discussed in Sec.~\ref{sec:thermal}. We focus on the axions produced after chemical decoupling, when the cosmic-string network has entered the time where the network has a constant number of strings per Hubble volume (up to log-violations, which we also take into account), commonly referred to as the ``scaling regime''. In computing the spectrum we work in the limit $m_a \to 0$ where strings remain until late times and there are no domain walls. If there is a finite mass, and the domain-wall number is equal to 1, the network will quickly collapse when $H \sim m _a$. This will produce an additional burst of axion production and a sharp drop in the axion spectrum at a characteristic frequency. We do not consider these effects but they may produce additional distinctive signals.

The spectrum of axions emitted by cosmic strings is still an active area of discussion in the literature with the debate centered on whether the typical axion energy emitted by a string is of order the inverse length or inverse thickness of the string (in particular, see Refs.~\cite{Gorghetto:2018myk,Gorghetto:2020qws} and~\cite{Buschmann:2019icd,Dine:2020pds}). We estimate the abundance and spectrum following numerical simulations done for the QCD axion in Refs.~\cite{Gorghetto:2018myk,Gorghetto:2020qws}, where the simulations suggest that the spectrum is dominated by low-energy axions.

To begin with, the energy density of cosmic strings can be parameterized using the average length of string within a Hubble length, $\xi$, and the energy of that string, given by the product of its tension, $\mu_{\rm eff}$, and a Hubble length, $1/H$. 
The total energy is then averaged over Hubble volume, $1/H^3$.
Following Ref.~\cite{Gorghetto:2018myk} we write this as,
\begin{equation} 
\rho_s = \frac{\xi(t) \mu_{\rm eff}(t)}{t^2} \,.
\label{eq:rhos}
\end{equation} 
This form is convenient since both $\xi$ and $\mu_{\rm eff}$ only evolve logarithmically with time.
Their evolution and can be parameterized as~\cite{Gorghetto:2018myk}
\begin{equation}\begin{aligned} 
\xi(t) & \simeq \alpha \ln  \frac{m_r}{H} + \beta \,, \label{eq:xxval}\\ 
\mu_{\rm eff}(t)  & \simeq \pi f_a^2 \ln \frac{m_r \gamma}{H \sqrt{\xi}} \,.
\end{aligned}\end{equation} 
Here $m_r \sim f_a $ is the string width, $\gamma$ is roughly a constant in time which we will approximate as unity, $\alpha \simeq 0. 24 \pm 0.02 $~\cite{Gorghetto:2020qws} is also a constant, and finally we take $\beta \simeq 0$ since we are interested in late times, where the log term is the dominant contribution.

The rate of axion energy emission during the scaling regime per unit volume, $\Gamma$, is given by the difference of the energy density of ``free'' strings (strings without inter-commutation and radiation) and the energy density stored in strings,
\begin{equation} 
\Gamma = \dot{\rho}^{\rm free}_s - \dot{\rho}_s \,.
\label{eq:Gamma}
\end{equation} 
We assume that both energy densities are equal at an initial time, $t _i$.
The free energy density at a later time $t$ is then,
\begin{equation} 
\rho_s^{\rm free} = \frac{\xi(t_i) \mu_{\rm eff}(t)}{t_i t} \,.
\label{eq:rhofree}
\end{equation} 
This follows as $\rho_s^{\rm free} \propto t^{-1}$ and we require $\rho_s^{\rm free}$ and $\rho_s$ to match at $t = t _i$.
Inserting \eqref{eq:rhos} and \eqref{eq:rhofree} into \eqref{eq:Gamma} and working in the large log limit gives,
\begin{equation} 
\Gamma \simeq 2 H (\rho_s / \rho_{\SM}) \rho_{\SM}\,,
\end{equation} 
where $\rho_{\SM} $ is the total energy density in the SM and the combination
\begin{equation} 
\frac{\rho_s}{\rho_{\SM}} = \frac{4 \xi \mu_{\rm eff}}{3 M_{\rm Pl}^2}\,,
\end{equation} 
has only a logarithmic time dependence.
The relic density in axions today  is then given by,
\begin{equation}\begin{aligned}
\frac{\rho_a}{\rho_c} & = \frac{1}{\rho_c} \int_{a_d}^1 \frac{da}{a} a^4 \frac{\Gamma(a)}{H}\\ 
&= \frac{8}{3 M_{\rm Pl}^2}  \int_{a_d}^1 \frac{da}{a} a^4 \xi \mu_{\rm eff} \frac{\rho_{\SM}}{\rho_c}\,,
\end{aligned}\end{equation}
where $a_d$ is the scale factor at the time the network enters the scaling regime. This expression applies during both radiation and matter domination, however we note that the simulations to estimate $\xi$ were only performed for radiation domination, and we focus on axions produced during this epoch. The relic density for different decoupling temperatures is shown on the left of Fig.~\ref{fig:drhodomega_string}. In addition to bounds on the energy density, cosmic strings have a constraint on the maximum value of the decay constant.
The energy scale of inflation is given in terms of the tensor-to-scalar ratio as~\cite{Baumann:2009ds},
\begin{equation} 
V^{1/4} \sim 10^{16}~{\rm GeV} \left( \frac{r}{0.01} \right)^{1/4} \,.
\end{equation} 
Using the upper bound on $r < 0.056$~\cite{Akrami:2018odb} and setting $V = 3 M_{\rm Pl}^2 H_I^2$ we can derive an upper bound on the Hubble scale of inflation, $H_I < 6 \times 10^{13}~{\rm GeV}$. Relating $H_I$ to the reheat temperature of the universe, $H_I \sim T^2_{\rm RH} / M_{\rm Pl}$, gives a maximum possible $T_{\rm RH}$. To have a thermal phase transition in the early universe requires $T_{\rm RH}$ to be above the critical temperature for a phase transition, $ \sim f _a $, putting an upper bound on the decay constant. Lastly, we note that if the cosmic string network remains until recombination there is an additional bound from the string energy density imprinted on the cosmic microwave background~\cite{Charnock:2016nzm}. The spectrum of axions produced after this epoch corresponds to frequencies, $\omega \lesssim 10^{-31}~{\rm eV} $, and are not observable with the experiments considered in this work. In presenting our axion spectrum and experimental projections, we assume the network collapses before this time. 

\begin{figure} 
\vspace{-0.1cm}
\begin{center} 
\includegraphics[width=8.5cm]{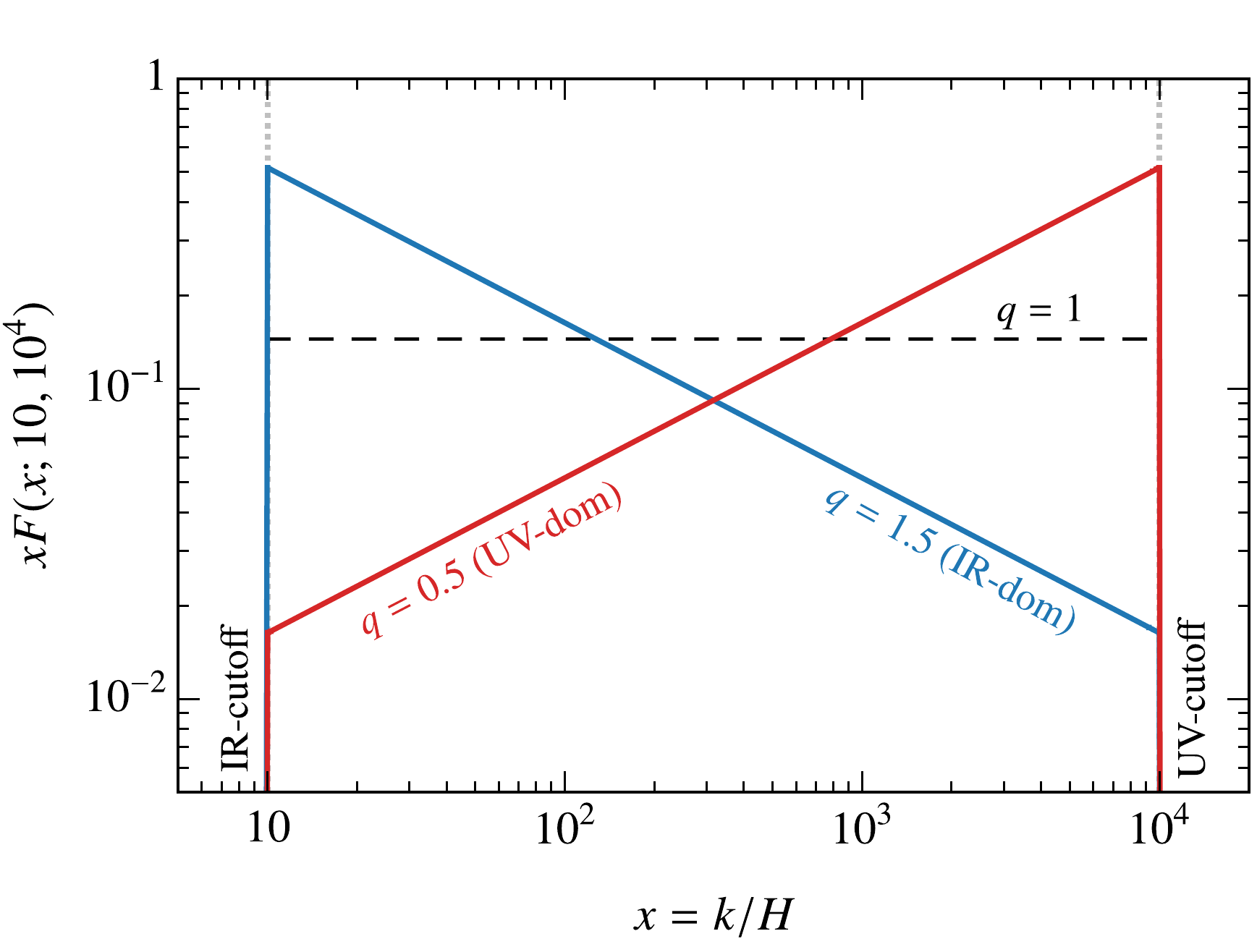} 
\end{center}
\vspace{-0.3cm}
\caption{The function in \eqref{eq:stringF}, which determines the spectrum of axion emission from cosmic strings for different spectral indices, $q$.
If $q < 1$ the emission spectrum is dominated by axions with a wavelength of order the string width while if $q > 1$ the spectrum is dominated by modes of order the string length. These two scenarios are referred to as UV and IR dominated, respectively. We assume an IR dominated spectrum in this work as suggested in Refs.~\cite{Gorghetto:2018myk,Gorghetto:2020qws}.}
\label{fig:spec}
\end{figure}

We now move on to calculate how this energy is distributed. The emission spectrum of axions from strings has been a source of uncertainty in the literature with the debate centered around whether axion emission is dominated by coherent motion of the string producing axions with wavelength of order the string length (``IR-dominated'') or by small loops and kinks along the string producing axions with wavelength of order the string width (``UV-dominated'').
This has profound consequences for QCD axion dark matter as it predicts a relic abundance produced from topological defect decay with an uncertainty of a few orders of magnitude (see Refs.~\cite{Gorghetto:2018myk,Gorghetto:2020qws} when compared to Ref.~\cite{Buschmann:2019icd}).
Fundamentally, the enormous separation of scales between the string length and its width make this a challenging problem to resolve.
In either case, the spectrum can likely be approximated by a power-law parameterized by a spectral index $q$ with a high and low energy cut-off \cite{Gorghetto:2018myk},
\begin{equation} 
\renewcommand{\arraystretch}{1.5}
F(x; x_1, x_2) = \left\{ \begin{array}{lc} {\cal N} x^{-q}, & x_1< x < x_2 \\ 0,& {\rm otherwise}
\end{array} \right.
\label{eq:stringF}
\end{equation}
where ${\cal N} \equiv (q-1) x_1^{q-1} / (1 - (x_1/x_2)^{q-1})$ normalizes $F$ such that the integral over all $x$ is unity. 
Here $x$ is the appropriately normalized energy, while $x_1$ and $x_2$ are IR and UV cutoffs.
In Fig.~\ref{fig:spec} we show $x F(x)$ for different values of $q$, demonstrating that for $q < 1$ the spectrum is UV-dominated while for $q > 1$ the spectrum is IR-dominated. 
While the debate is yet to be settled (in particular, see Ref.~\cite{Dine:2020pds}), a recent analysis~\cite{Gorghetto:2020qws} suggests that the spectrum is IR dominated during the scaling regime with estimates for the IR and UV cutoffs of $x_1 \simeq 10$ and $x_2 \simeq m_r / H$ respectively.
Furthermore, the authors of Ref.~\cite{Gorghetto:2020qws} find a best fit for the spectral index over time of,
\begin{equation} 
q(t) \simeq  0.51 + 0.053 \ln \frac{m_r}{H}\,,
\end{equation} 
and we assume this form in our results.

Given a spectral function $F(x)$, the axion spectrum as observed today is the appropriately weighted time-integral of this expression,
\begin{equation} 
\Omega_a(\omega) = \frac{8\omega}{3 M_{\rm Pl}^2}\int_{a_d}^1 \frac{da}{a}  \frac{\xi \mu_{\rm eff}}{H} \frac{\rho_{\SM}}{\rho_c} a^3 F \left( \omega / H a \right)\,,
\label{eq:OmegaString} 
\end{equation} 
where $\omega$ is the axion energy as measured today.
The spectrum for different $f_a$ is shown on the right of Fig.~\ref{fig:drhodomega_string}.
At low energies the spectrum is roughly a constant (a consequence of the network being in the scaling regime) while at high energies the spectrum falls off as it relies on producing axions with energies much larger then the Hubble scale from cosmic-string oscillations.
The frequency where the drop begins depends on the decoupling temperature of the axion with the SM bath, $T_d$, with solid lines denoting $T_d = f_a$ and dashed curves showing $T_d = f_a / 10^3$.
In all cases, the abrupt change just below $\omega = 10^{-22}~{\rm eV}$ is associated with a drop in the cosmic string energy density after the QCD phase transition, where the number of relativistic degrees of freedom in the SM drops considerably.

As for all C$a$B candidates, in addition to the dependence on the energy density and spectrum, axion detection from cosmic strings is sensitive to the axion-SM coupling.
For generic axions, the axion-photon coupling is $\ga \lesssim \alpha / 2\pi f_a$, which when combined with the densities above would be challenging to observe.
Accordingly, when we discuss the experimental prospects, we will again consider $\ga$ larger than this simplest expectation, which can be induced by mechanisms including the clockwork~\cite{Choi:2015fiu} (see Refs.~\cite{Agrawal:2017cmd} and~\cite{Dror:2020zru} for recent summaries of such mechanisms in the context of the QCD axion and ultralight axion dark matter, respectively).
Nonetheless, we note that if experiments could probe the scenario where $\ga \sim \alpha /2\pi f_a$, it would be possible to probe {\em all} value of $f_a$. This is because the detectable C$a$B power is $\propto \ga^2 \rho_a$, and for cosmic strings we have $\rho_a \propto \mu_{\rm eff} \propto f_a^2$, resulting in $\ga^2 \rho_a$ being independent of $f_a$.

\section{Detecting the C$a$B}
\label{sec:det}

Having motivated the possibility of a local C$a$B, we now turn to the question of how that population could be detected. 
We focus on detection at a few of the many instruments constituting the burgeoning program to detect ultralight dark matter. 
Our central conclusion will be that experiments designed with axion dark matter in mind are generally also sensitive to a relativistic population. 
Indeed, it is possible that ADMX has already collected a detectable signal that would have been missed by an analysis focused on the non-relativistic axion.

Qualitatively, detection of relativistic axions proceeds as for their non-relativistic counterparts. 
In both cases, the axion can be described as an oscillating classical wave,\footnote{The classical wave description holds in the limit of a large number of states per de Broglie volume, $n_a \lambda_{\rm dB}^3 \gg 1$. 
For dark-matter axions, using the mean expected dark-matter density and speed, this is satisfied for $m_a \lesssim 10~{\rm eV}$. 
For the C$a$B, we instead have $n_a \lambda_{\rm dB}^3 \sim (\rho_a/\rho_{\gamma})(\bar{\omega}/1\,{\rm meV})^{-4}$. 
In the present work we will consider detection exclusively in scenarios with $\bar{\omega} \ll 1\,{\rm meV}$ and sufficient densities that classicality applies. 
Nevertheless, our description will not apply for arbitrarily large mean energies or small densities. The approximate boundary between the two regimes is shown in Fig.~\ref{fig:CaB}.} which through a coupling to the SM induces a detectable time-varying signal in, for example, electromagnetic waves or nuclear spins. A central difference is the signal bandwidth. For dark matter, the expectation is that the signal power will be deposited in an extremely narrow range of frequencies centered around its unknown mass $m_a$. In general, the oscillation frequency is set by the axion energy. For a non-relativistic particle, the energy is $\omega \simeq m_a (1+v^2/2)$, and given our expectation for the local dark matter is that the speeds vary over a range $\Delta v \sim 10^{-3}$ and take a mean value $\bar{v} \sim 10^{-3}$, axion dark matter carries a large quality factor of $Q_a^{\DM} = \bar{\omega}/\Delta \omega \sim 10^6$, where $\bar{\omega}$ is the average energy. For a C$a$B the expectation is that the local axion field has a wide distribution of energies, such that generically $Q_a^{\text{C$a$B}} \sim 1$. An exception is dark matter decaying to axions within the Milky Way, where we expect $Q_a^{\text{C$a$B}} \sim 10^3$. Regardless, in either case $Q_a^{\text{C$a$B}} \ll Q_a^{\DM}$, and this will represent a challenge to detection. There are additional important differences between the relativistic and dark-matter cases -- for instance, the relativistic signal can exhibit a unique daily-modulation signal even at a single detector -- and we will explore these as well. As is the case for the bulk of this work, we restrict our attention to experiments focused on the axion coupling to electromagnetism, $\ga$, although much of our formalism can be lifted for other SM couplings.

We divide our discussion of the C$a$B detection into five parts. Firstly we outline several basic features of a relativistic axion population -- its expected amplitude and distribution in both time and frequency -- applicable to any detection strategy. We next use these results to sketch our expected sensitivity to the C$a$B by comparing the experimentally detectable power associated with the relativistic and dark-matter axion field. Having provided a general sensitivity estimate sufficient for understanding Fig.~\ref{fig:CaB}, we then focus specifically on the axion-photon coupling, detailing axion electromagnetism with a specific focus on the differences in the relativistic case. Finally, we apply these lessons to existing axion dark-matter detection strategies, and discuss representative examples of both broadband and resonant detection strategies approaches. In the following section we will use these results to set estimated limits on a number of different C$a$B scenarios discussed in Sec.~\ref{sec:prod}.

\subsection{Properties of the Relativistic Axion}
\label{sec:prop-rel-ax}

To start our discussion, we will outline general properties of the relativistic axion field relevant to its detection.
In particular, we treat the C$a$B as the superposition of many non-interacting axion particles with energies $\omega$ drawn from probability distribution $p(\omega)$.\footnote{Formally we can define $p(\omega) = (1/n_a) dn_a/d\omega$, as it is the differential number density that controls the probability of observing an axion at a given energy.}
Within this framework, we will derive the expected amplitude of the axion field $a$ -- more specifically of $a^2$ -- in both the time and frequency domain, and further quantify the fluctuations around the central value.
Any experimental detection will involve a coupling to the axion field, and therefore the measurements will inherit these average values and fluctuations.
We will consider a general energy distribution, and show that our results contain the non-relativistic limit as a special case.
Indeed, our results are a direct generalization of the non-relativistic field, which will allow us to bootstrap known dark-matter results to the C$a$B.

\begin{figure*}[htb]
\centering
\includegraphics[width=.32\textwidth]{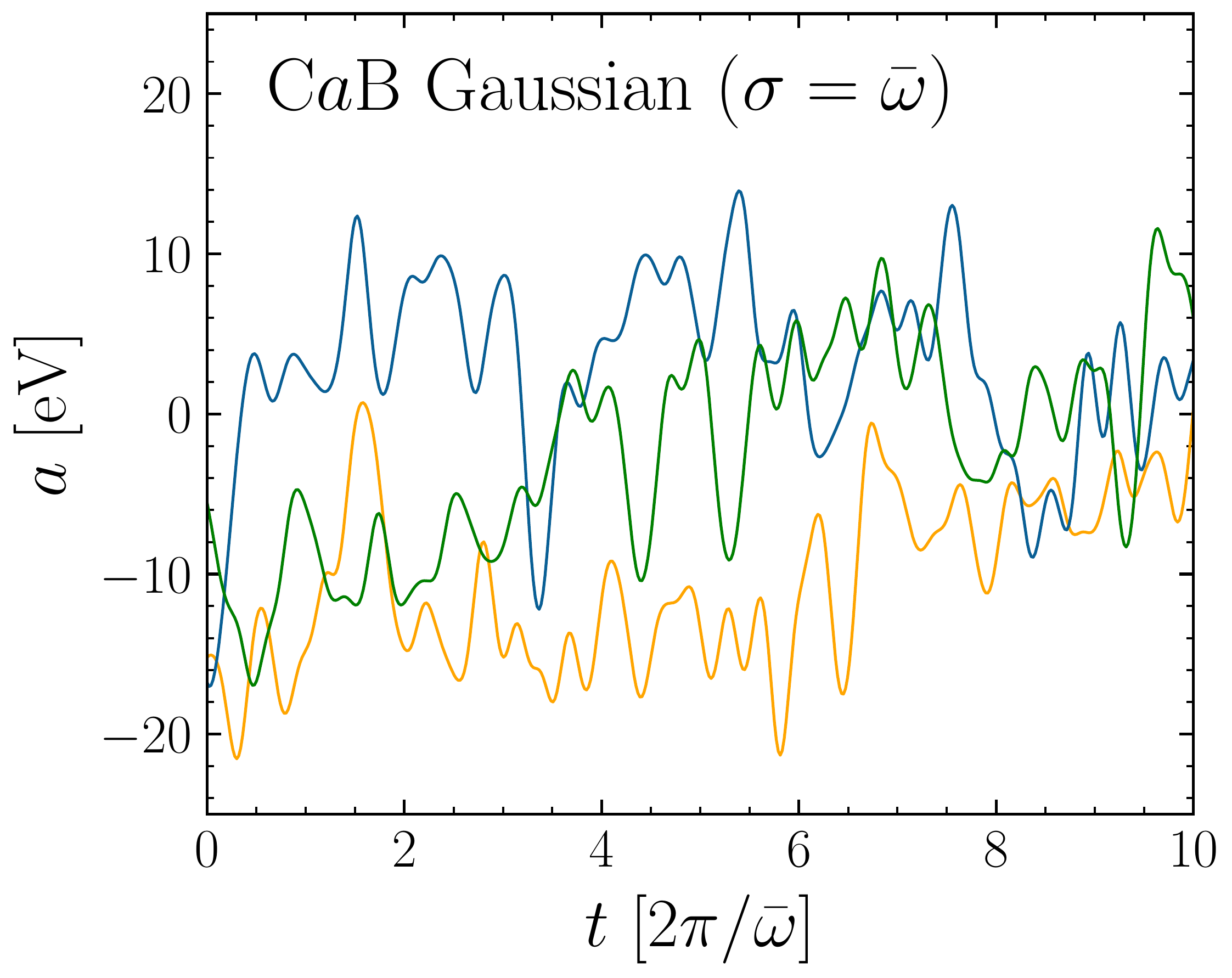}
\hspace{0.15cm}
\includegraphics[width=.32\textwidth]{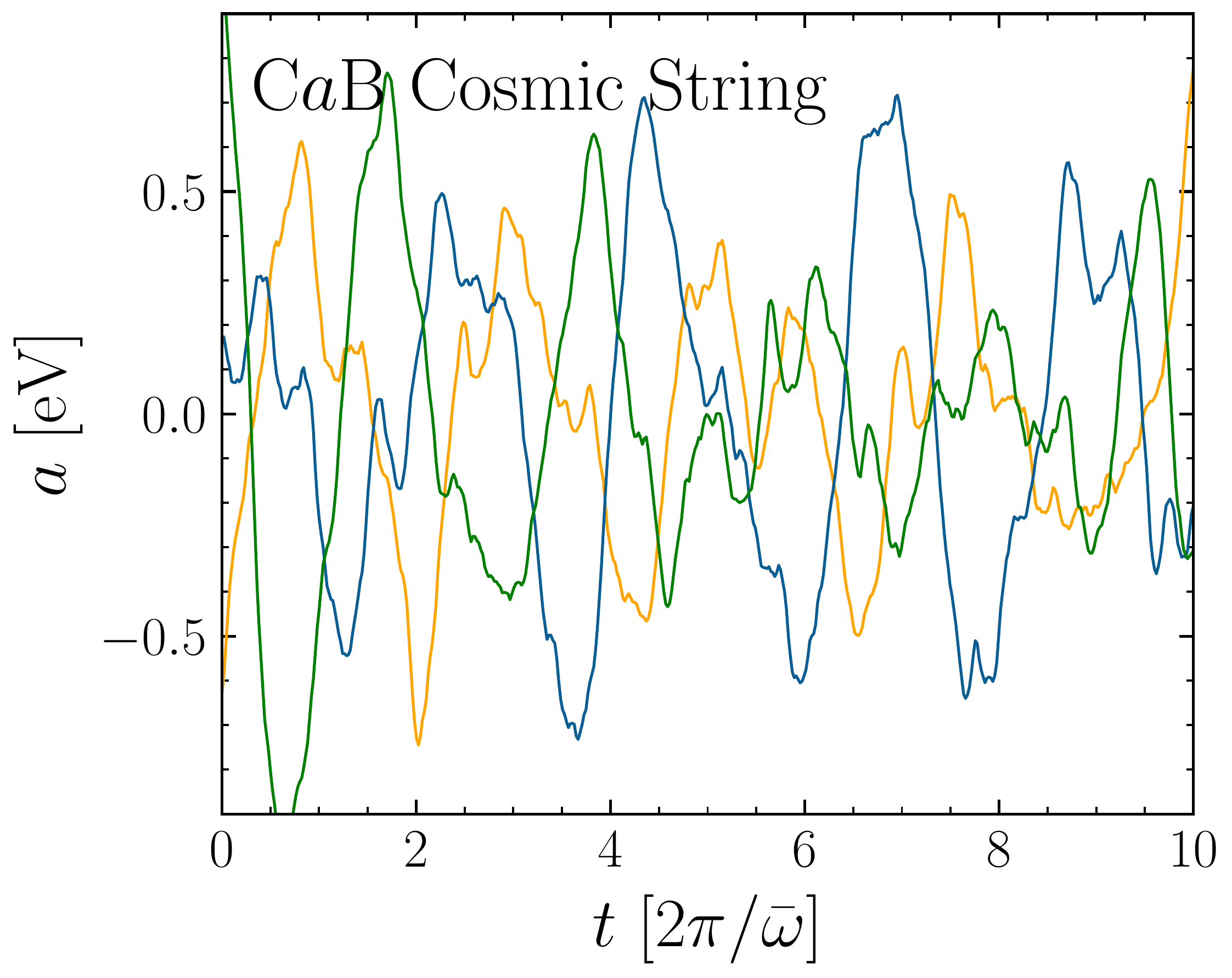}
\hspace{0.15cm}
\includegraphics[width=.32\textwidth]{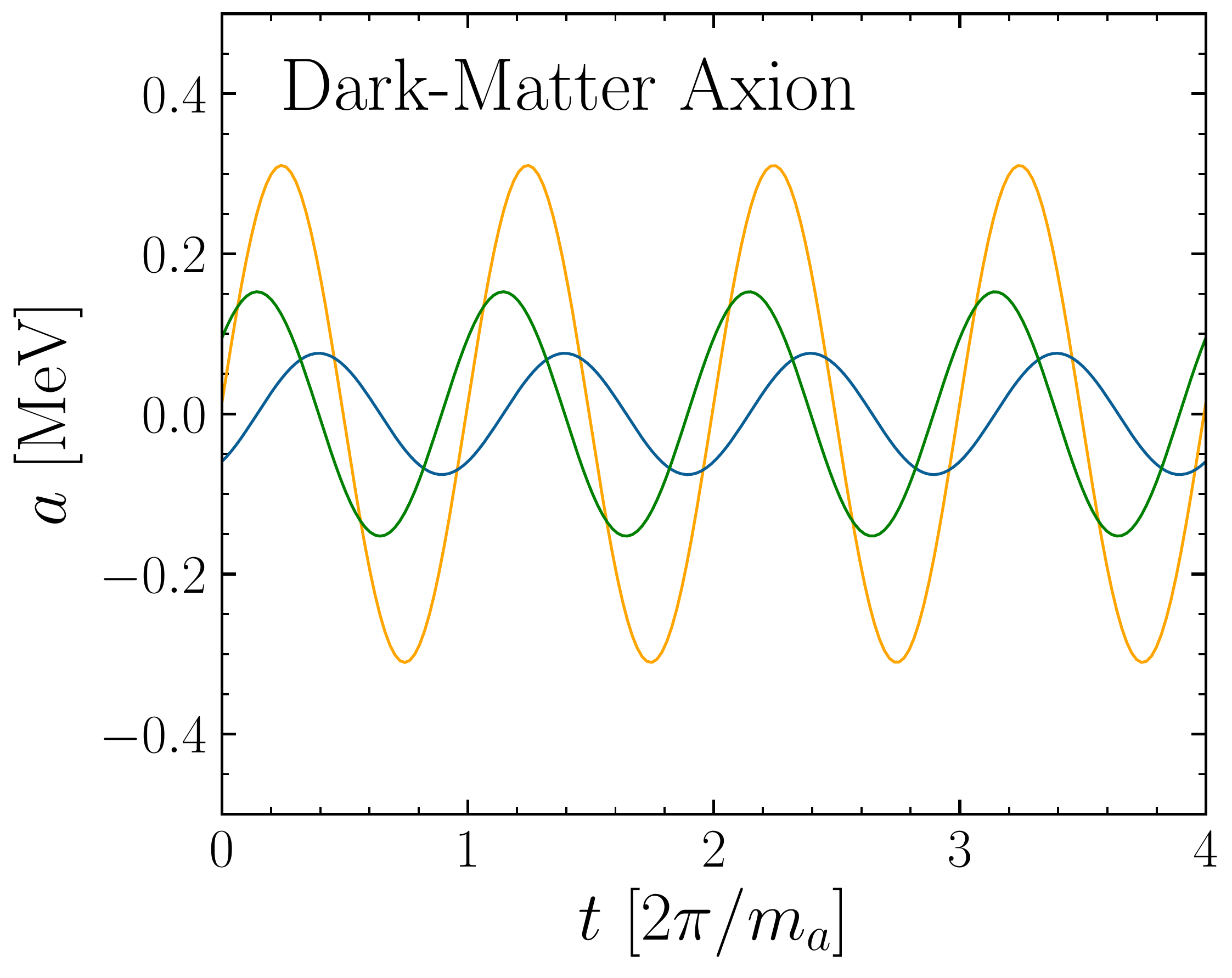}
\vspace{-0.3cm}
\caption{The local axion field for three different $p(\omega)$: a wide Gaussian (left), the cosmic-string distribution (center), and the expected dark-matter distribution (right).
The broader $p(\omega)$ expected for the C$a$B generates the additional structure seen for the relativistic axion field.
For each, the three curves represent distinct realizations of $a(t)$ computed directly from \eqref{eq:axion-sum}, with $N_a = 10^6$.
To aid the comparison, for each distribution we choose parameters such that $\bar{\omega} \simeq 10$ neV.
For the two relativistic examples, we have $\rho_a \sim \rho_{\gamma}$, whereas for dark matter we take $\rho_a = \rho_{\DM}$.
See text for additional details.
}
\label{fig:a-timedomain}
\end{figure*}

For the energies and densities considered in this work, the axion field will always contain an enormous number of particles per de Broglie volume.
Consequently, the field can be described in terms of an emergent classical wave.
In this respect, the C$a$B directly mirrors non-relativistic dark matter for $m_a \ll 10~{\rm eV}$, where the associated statistics were derived in Ref.~\cite{Foster:2017hbq}, and we will generalize a number of results from that reference.
We imagine the classical axion wave as constructed from a large number, $N_a$, of non-interacting waves,
\begin{equation} 
a(t) = \sqrt{\frac{2\rho_a}{N_a \bar{\omega}}}\sum_{i=1}^{N_a} \frac{1}{\sqrt{\omega_i}} \cos \left[ \omega_i t + \phi_i \right].
\label{eq:axion-sum}
\end{equation} 
Each element of this sum is associated with a random variable $\omega_i$, an energy drawn from $p(\omega)$. Beyond their energy, however, there is no reason to imagine the various states are phase coherent, and this is ensured by the uniform random variable $\phi_i \in [0,2\pi)$. The amplitude is fixed by ensuring the field carries energy density $\rho_a$, which would be equal to $\rho_{\DM}$ for non-relativistic dark matter.
In the discretized picture, $\bar{\omega} = N_a^{-1} \sum_i \omega_i$, but more generally we take $\bar{\omega} = \int d \omega\,\omega p(\omega)$.

In principle, there is an additional contribution to the phase neglected in \eqref{eq:axion-sum}: the spatial variation controlled by $- \bk_i \cdot \bx$, where $\bk_i$ is the particle momentum.
If we imagine measuring the axion field at a single point, this contribution is irrelevant at the level of the phase, as we can always center our coordinates such that $\bx=0$.\footnote{If the axion field is measured at multiple spatially separated locations, however, the $\bk \cdot \bx$ contribution to the phase is physical, and can be used to perform interferometry on the wave~\cite{Foster:2020fln}.}
Yet where it can be relevant is through effects sensitive to the spatial gradients of the axion field.
As we will discuss in Sec.~\ref{sec:ra-em}, whilst these gradients are usually neglected for a non-relativistic field, they are parametrically important for a relativistic population.
The amplitude of these effects is fixed by the massive dispersion relation, $|\bk_i| = \sqrt{\omega_i^2-m_a^2}$, and therefore also controlled by the energy distribution, $p(\omega)$.
The direction of $\bk$, however, is not, and will itself be drawn from a distribution on the celestial sphere.
For instance, if the dark matter is made of axions, the direction of $\bk$ may point towards a dark-matter stream incident on the Earth, or in the relativistic case the C$a$B would be biased towards the center of the Milky Way if it originates from dark-matter decay.
As shown in~\cite{Foster:2020fln}, the angular distribution can be fully incorporated into the description of the non-relativistic axion field, and the arguments there can be generalized to relativistic axions.\footnote{In the non-relativistic case, it is convenient to express the energy and momentum both in terms of the particle velocity, $\bv$.
In \cite{Foster:2020fln}, it was shown how the statistics of the axion field can be described in terms of $p(\bv)$ (often written $f(\bv)$), which includes directional information.
This approach can be generalized to the relativistic case by describing the field in terms of $\bk$ and $p(\bk)$, rather than the energy as we do in the text.}
We will not pursue this direction in the current work, however.
While the directional distribution will be relevant for the fine details of the relativistic signal -- in particular as it relates to daily-modulation effects unique to the relativistic axion, discussed in Sec.~\ref{sec:ra-em} -- it unnecessarily complicates an estimate of the experimental reach, which is our focus.

To make progress in our description of the axion field, we re-organize the sum in \eqref{eq:axion-sum} such that states with nearby energy are combined.
Specifically, we partition the particles into sets, indexed by $j$, containing all those with $\omega \in [\omega_j,\,\omega_j+\Delta \omega]$, within which the states are distinguished only by the random phase.
Combining the particles within a given energy cell then amounts to a random walk in the complex plane (see Ref.~\cite{Foster:2017hbq}), leaving
\begin{equation}
a(t) = \sqrt{\frac{\rho_a}{\bar{\omega}}} \sum_j \alpha_j \sqrt{\frac{p(\omega_j) \Delta \omega}{\omega_j}} \cos \left[ \omega_j t + \phi_j \right]\,.
\label{eq:axion-sum-E}
\end{equation}
The end stage of the walk is a new random phase $\phi_j$, with the distance traveled dictated by the Rayleigh random variable $\alpha_j$, drawn from $p(\alpha) = \alpha\, e^{-\alpha^2/2}$, and the density of states at that energy, controlled by $p(\omega_j)$.
Through its dependence on $\alpha_j$ and $\phi_j$, $a(t)$ is itself a random variable.
Although $\langle a \rangle = 0$, we expect $\langle a^2 \rangle > 0$; indeed, $a^2$ is an exponentially random variable, with mean (approximating $\Delta \omega$ as differential)
\begin{equation}
\langle a^2 \rangle = \frac{\rho_a}{\bar{\omega}} \int_0^{\infty} \frac{d\omega}{\omega}p(\omega) = \frac{\rho_a \langle 1/\omega \rangle}{\bar{\omega}}\,.
\label{eq:asq-mean}
\end{equation}
For a non-relativistic axion, $\langle 1/\omega \rangle^{-1} \simeq \bar{\omega} \simeq m_a$, and we recover the familiar dark-matter result, $\langle a^2 \rangle = \rho_a/m_a^2$.
For a general energy distribution, however, there is no such simplification (to be clear $\langle \bar{\omega}/\omega \rangle \neq 1$).
To exemplify this point, consider a $p(\omega)$ which is log flat over $[\omega_1,\omega_2]$.
If $\omega_1 \sim \omega_2$, we have $\langle \bar{\omega}/\omega \rangle \sim 1$.
However, if they are parametrically separated, $\omega_1 = \epsilon\, \omega_2$ for $\epsilon \ll 1$, then $\langle \bar{\omega}/\omega \rangle \sim (\epsilon \ln^2 \epsilon)^{-1} \gg 1$.

To complete our discussion of the relativistic axion in the time domain, in Fig.~\ref{fig:a-timedomain} we show three realizations of the axion field, determined directly from \eqref{eq:axion-sum}, for three different $p(\omega)$.
In the left two figures we take $m_a \ll \bar{\omega}$, in order to depict examples of the C$a$B, which are then contrasted with the expected dark-matter axion on the right.
On the left, we take $p(\omega)$ to be a positive-definite normal distribution, which can be considered an example of a C$a$B emerging from parametric-resonance production.
We take $\bar{\omega}=10$ neV, CMB energy density, $\rho_a = \rho_{\gamma}$, and further set the distribution to be wide, specifically $\sigma = \bar{\omega}$.
The fact that a number of frequencies are contributing is visible in the realizations.
In the middle, we take an even broader $p(\omega)$, corresponding to the C$a$B as predicted from cosmic-string production.
In detail, the distribution is determined by \eqref{eq:OmegaString} with $f_a = 10^{15}$ GeV and $T_d = 10^{12}$ GeV, such that $\rho_a \sim \rho_{\gamma}$.
Nevertheless, we only draw frequencies in a restricted range of $\omega \in [5~{\rm neV},\,1~\mu{\rm eV}]$, over which we have $\bar{\omega} \sim 10$ neV.
The presence of both high and low-frequency contributions in $p(\omega)$ can be seen in the realizations.
Finally, on the right, we show the conventional dark-matter axion scenario, with $\bar{\omega} \simeq m_a = 10$ neV, and small variations around this as predicted by the standard halo model.
The variations are not visible in the time domain, with the period of the realizations highly regular.
The statistical nature of the amplitude discussed above, can be seen.

While we can understand the time dependence of the C$a$B, its properties are more transparent in the frequency domain.
As such, consider the Fourier transform of \eqref{eq:axion-sum-E}.
We imagine making measurements of the axion field at a frequency $f=1/\Delta t$ for a total integration time $T$, thereby collecting a set of $N = T/\Delta t$ discrete measurements of the field, which we denote by $\{a_n = a(n\Delta t)\}$.
We then calculate the power spectral density (PSD), which quantifies the power in the field at a given frequency, as
\begin{equation}\begin{aligned}
S_{a}(\omega) = \frac{(\Delta t)^2}{T} \left| \sum_{n=0}^{N-1} a_n e^{-i\omega n \Delta t} \right|^2.
\end{aligned}\end{equation}
Technically $\omega$ is a discrete variable, given by $2 \pi k/T$, with $k = 0,1,\ldots,N-1$ the relevant Fourier mode, although we will often assume a sufficiently long integration time that we can approximate $\omega$ as continuous.
As in the time domain, the PSD is an exponentially distributed random variable, and therefore specified entirely by its mean,
\begin{equation}\begin{aligned}
\langle S_{a}(\omega) \rangle = \frac{\pi \rho_a}{\bar{\omega}} \frac{p(\omega)}{\omega}\,.
\label{eq:Sa-rel}
\end{aligned}\end{equation}
Once more, this result reduces to the correct dark-matter expression in the non-relativistic limit.
The energy of the non-relativistic wave is specified by its speed, drawn from a distribution $f(v)$.
Changing variables to $v_{\omega} = \sqrt{2\omega/m_a-2}$, in the non-relativistic limit, we have
\begin{equation}\begin{aligned}
\langle S_{a}^{\DM}(\omega) \rangle = \frac{\pi \rho_{\DM}}{m_a^3} \frac{f(v_{\omega})}{v_{\omega}}\,.
\label{eq:Sa-DM}
\end{aligned}\end{equation}
This agrees with the dark-matter case in~\cite{Foster:2017hbq}, demonstrating that the general expression in \eqref{eq:Sa-rel} contains the non-relativistic limit as a special case.

Nevertheless, the scaling in \eqref{eq:Sa-rel} is misleading.
While it quantifies the power distribution in fluctuations of the axion, experiments can only measure the induced fluctuations in SM fields derivatively coupled to the axion.
Accordingly, it is more appropriate to consider the power in $g_{a\SM} \partial a$, where $g_{a\SM}$ is the axion-SM coupling.
Taking $\partial a \sim \omega a$, we can determine the parametrics of the accessible power by approximating $p(\omega)$ as a uniform distribution over a range of width $\bar{\omega}/Q_a$.
Doing so, the power scales as
\begin{equation}
\langle S_{g\partial a}(\bar{\omega}) \rangle \sim \frac{g_{a\SM}^2 \rho_a Q_a}{\bar{\omega}}\,.
\label{eq:S-scaling}
\end{equation}

\subsection{Rough Sensitivity}
\label{sec:est-sens}

We will now use \eqref{eq:S-scaling} to determine the parametric sensitivity of dark-matter experiments to the C$a$B, leaving a detailed calculation of the sensitivities to the following subsections.
We begin with the following simple estimate: assume the C$a$B can be detected if the power it deposits at $\bar{\omega}$ matches the power produced by the dark-matter axion at the sensitivity threshold.
For the moment, we assume that experiments extract power from a relativistic and non-relativistic axion wave identically, although we will later justify this assumption up to $\mathcal{O}(1)$ factors.
In order to compute the power matching, we assume optimistically that the coupling saturates the existing bounds, $g_{a\SM} = g_{a\SM}^{\rm SE}$, so that for a fixed $Q_a$ and $\bar{\omega}$ we can constrain $\rho_a$.
For the dark-matter power, we need the dark-matter equivalent of \eqref{eq:S-scaling}, which is obtained by setting $\rho_a = \rho_{\DM}$, $Q_a = Q_a^{\DM} \sim 10^6$, $\bar{\omega} \sim m_a$, and fixing the coupling to an existing sensitivity threshold, denoted by $g_{a\SM}^{\rm lim}$.
Equating the powers at the frequency $\bar{\omega}=m_a$, we expect sensitivity to an axion background that constitutes the following fraction of the CMB energy density,
\begin{equation} 
\frac{\rho_a}{\rho_{\gamma}} = \frac{\rho_{\DM}}{\rho_{\gamma}} \left( \frac{g_{a\SM}^{\rm lim}}{g_{a\SM}^{\rm SE}} \right)^2 \frac{Q_a^{\DM}}{Q_a^{{\text{C$a$B}}}} \hspace{0.5cm} {\rm (naive)}.
\label{eq:est-lim-naive}
\end{equation}

This scaling is overly pessimistic.
The C$a$B will deposit its power over a much wider range than dark matter, so there is more information than can be gleaned by comparing power at a single frequency.
In principle, our sensitivity depends on how this additional information is obtained, either through a broadband or resonant readout strategy, and so we will consider the two cases separately.
As we do so, however, we emphasize a fundamental challenge: the broad nature of the signal will make it harder to distinguish from backgrounds.
There are handles, for instance as we will show for the axion-photon coupling, the signal power will continue to scale quadratically with the magnetic field, and further for the case of the C$a$B from dark-matter decay, there can be a unique daily-modulation signal.
Beyond such remarks, we will not attempt to determine the optimal analysis for a relativistic signal here, although we note it will likely require a more accurate characterization of the background than in dark-matter searches.
Indeed, both ADMX and HAYSTAC usually remove features much broader than that expected of dark matter, see for example~\cite{Brubaker:2017rna,Du:2018uak}, which raises the possibility that a signal of the C$a$B may already be hiding in existing data, albeit in the most optimistic scenarios.

For a broadband readout of the axion power, we integrate over a range of energies, and therefore at the level of the integrated signal power the distribution $p(\omega)$ would seem irrelevant.
Yet even when the entire spectrum is resolved, the width of $p(\omega)$ still determines an important physical property of the axion field: the coherence time.
The coherence time has a straightforward interpretation in the frequency domain.
Recall that the measurement time $T$ determines the frequency resolution of the associated discrete Fourier transform, according to $\Delta \omega = 2\pi/T$.
For sufficiently small $T$, the entire signal will fit within a single bin, and the signal amplitude will be associated with one draw from the exponential distribution as outlined in Sec.~\ref{sec:prop-rel-ax}.
As $T$ is increased, eventually the resolution will be sufficient to resolve the structure in $p(\omega)$.
At this stage the signal will occupy multiple bins, each of which will have an independent exponential draw that then combine incoherently.
The transition between these two cases defines the coherence time, which we can quantify by $2\pi/\tau = \sigma_{\omega}$, with $\sigma_{\omega}$ the width of $p(\omega)$.
Parametrically, we expect $\sigma_{\omega} \sim \bar{\omega}/Q_a$, so that $\tau \sim 2 \pi Q_a/\bar{\omega}$, or numerically,
\begin{equation} 
\tau \sim Q_a \left( \frac{1~{\rm neV}}{\bar{\omega}} \right)~\mu{\rm s}\,.
\end{equation} 
Consequently, the coherence time of the C$a$B will generally be short on the timescale of experimental measurements,\footnote{Throughout we will always assume that we are operating in the $T > \tau$ regime so that the energy distributions can be resolved.} implying that we will operate in the regime $T > \tau$, where we expect the sensitivity to the signal power to be impeded by the increased background that enters when the signal is distributed over a broader range.
As the C$a$B has a coherence time that is smaller than for dark matter by a factor of $Q_a^{\DM}/Q_a^{\text{C$a$B}} \gg 1$, the background will be enhanced by this same scale, and subsequently there is a reduction in signal power sensitivity of $\sqrt{Q_a^{\DM}/Q_a^{\text{C$a$B}}}$.
This leads to a refined estimate for the broadband sensitivity of
\begin{equation} 
\frac{\rho_a}{\rho_{\gamma}} = \frac{\rho_{\DM}}{\rho_{\gamma}} \left( \frac{g_{a\SM}^{\rm lim}}{g_{a\SM}^{\rm SE}} \right)^2 \sqrt{\frac{Q_a^{\DM}}{Q_a^{\text{C$a$B}}}} \hspace{0.5cm} {\rm (broadband)}.
\label{eq:est-lim-broadband}
\end{equation}

Turning to a resonant detection strategy, the estimate in \eqref{eq:est-lim-naive} will be modified by the experimental quality factor $Q$, associated with the cavity or readout circuit of the instrument, in three ways.
Firstly, the power recorded by the resonator is controlled by ${\rm min}(Q,Q_a)$, so that for $Q < Q_a^{\DM}$, we have overestimated the deposited dark-matter power.
For the moment, we will assume that we are in this limit, for instance ADMX and HAYSTAC currently operate with $Q \sim 10^5$ and $Q \sim 10^4$, respectively.
We also expect that $Q_a^{\text{C$a$B}} \ll Q$, so that for equal couplings and density, we expect the C$a$B power to be suppressed by a factor of $Q/Q_a^{\text{C$a$B}}$, rather than $Q_a^{\DM}/Q_a^{\text{C$a$B}}$ as assumed in \eqref{eq:est-lim-naive}.
The experimental quality factor will enter a second time in defining the instrumental bandwidth of $\omega_0/Q$, where $\omega_0$ is the resonant frequency.
The bandwidth conventionally dictates the range over which the signal can be analyzed.
For dark matter, the signal is narrower than the bandwidth by a factor of $Q_a^{\DM}/Q$.
This implies that when searching for dark matter, the background can be restricted to a smaller range, suppressing its contribution by $Q_a^{\DM}/Q$.
For the C$a$B there is no such suppression -- the signal extends over the full bandwidth -- so using the Dicke radiometer equation~\cite{Dicke:1946glx}, our sensitivity will suffer due to the increased background by a further factor of $\sqrt{Q_a^{\DM}/Q}$, similar to the broadband consideration.
The third consideration is in the C$a$B's favor.
Its broad nature implies that the signal will deposit power over many bandwidths collected during a dark-matter search.
At most the number of bins can be $Q/Q_a^{\text{C$a$B}}$, producing an enhancement in the sensitivity of $\sqrt{Q/Q_a^{\text{C$a$B}}}$.\footnote{$N$ additional measurements can be thought of as scaling the experimental measurement time $T \to NT$.
Assuming $T > \tau$, our sensitivity to the power will scale as $\sqrt{T} \to \sqrt{N T}$~\cite{Budker:2013hfa}.}
Taken together, these three factors modify \eqref{eq:est-lim-naive} to
\begin{equation}
\frac{\rho_a}{\rho_{\gamma}} = \frac{\rho_{\DM}}{\rho_{\gamma}} \left( \frac{g_{a\SM}^{\rm lim}}{g_{a\SM}^{\rm SE}} \right)^2 \sqrt{\frac{Q_a^{\DM}}{Q_a^{\text{C$a$B}}}} \hspace{0.5cm} {\rm (resonant)}.
\label{eq:est-lim-resonant}
\end{equation} 
As $Q$ has dropped out, we are left with the same parametric scaling as for broadband detection.

Given an experimental limit on or sensitivity for dark matter, we can estimate our sensitivity using either \eqref{eq:est-lim-broadband} or \eqref{eq:est-lim-resonant}. For several specific instruments, we have already shown the results in Fig.~\ref{fig:CaB}.
We can also consider the expected reach more generally.
Taking $Q_a^{\DM} = 10^6$ and $Q_a^{\text{C$a$B}} = 1$, all that remains is to fix the couplings.
Focussing on the axion-photon coupling, we have $g_{a\SM} = \ga \sim \alpha/(2\pi f_a)$.
For the C$a$B, we take the optimistic value of $g_{a\SM}^{\rm SE} = \ga^{\rm SE} = 0.66 \times 10^{-10}~{\rm GeV}^{-1}$, whereas for dark matter, we exploit the fact that a broad goal of the axion dark-matter program is to probe $\ga$ values at the scale of the QCD axion, which satisfies $m_a f_a \simeq m_{\pi} f_{\pi}$.
Assuming this goal is achieved across a wide range of masses, then the corresponding sensitivity to a relativistic population is given by
\begin{equation} 
\frac{\rho_a}{\rho_{\gamma}} \simeq \left( \frac{\bar{\omega}}{1~\mu{\rm eV}} \right)^2\,,
\label{eq:rhoreach}
\end{equation} 
so that for energies below $1~\mu{\rm eV}$, we could be sensitive to a C$a$B with an energy density below that of the CMB.
In what follows we will refine this estimate.

\subsection{Relativistic Axion E\&M}
\label{sec:ra-em}

We now specialize our discussion to detecting the C$a$B through a coupling to electromagnetism.
As is well known, the coupling in \eqref{eq:Lga} leads to the following classical equations of motion~\cite{Sikivie:1983ip}
\begin{equation}\begin{aligned}
\nabla \cdot \bE &= \rho - \ga \bB \cdot \nabla a\,, \\
\nabla \cdot \bB &= 0\,, \\
\nabla \times \bE &= - \partial_t \bB\,, \\
\nabla \times \bB &= \partial_t \bE + \bJ + \ga \left( \bB\,\partial_t a- \bE \times \nabla a \right), \\
(\Box + m_a^2) a &= \ga \bE \cdot \bB\,,
\label{eq:axionEM}
\end{aligned}\end{equation}
with $\rho$ and $\bJ$ the charge and current densities, respectively.
For non-relativistic dark-matter axions the momentum is parametrically smaller than the energy ($\nabla a \sim \bk \ll \omega \sim \partial_t a$) and this justifies neglecting the two terms involving $\nabla a$.
This leaves a single modification to the Amp\'ere-Maxwell equation, which by analogy enters as an effective current $\bJ_{\rm eff} = \ga \bB\,\partial_t a$.
The axion field thereby converts magnetic field lines into oscillating currents, identifying large magnetic fields as a central ingredient in the detection of axion dark matter.

For the C$a$B spatial gradients cannot be neglected, yielding two additional sources in \eqref{eq:axionEM}.
The first of these is the generation of an additional effective current, $\bJ_{\rm eff} = - \ga \bE \times \nabla a$.
As our focus is on searching for the C$a$B with existing axion dark-matter detectors, which rely on large magnetic fields, this term will not be relevant. 
The second gradient term, which provides a contribution to Gauss' law cannot be immediately discarded.
In detail, a relativistic axion field generates an effective charge density $\rho_{\rm eff} = - \ga \bB \cdot \nabla a$; again by analogy, the axion converts magnetic field lines into oscillating lines of charge.
This effect is proportional to $\bB \cdot \nabla a \sim \bB \cdot \bk a$, and is therefore dependent on the incident direction of the axion relative to the experimentally established magnetic field.
Nevertheless, we will see that for all the experiments we consider the effective charge does not significantly contribute to the signal.
Yet the incident direction of the axion remains detectable: the relativistic field can undergo appreciable spatial oscillations over the instrument, leading to an interference pattern that depends on the incoming angle of the axion wave.
We will show explicitly how this effect arises for resonant cavity instruments.
For a true cosmological relic, the signal, like the CMB, will be almost completely isotropic (up to $\sim 10^{-3}$ variations associated with our peculiar velocity with respect to the Hubble flow), resulting in an effectively time independent signal.
In the scenario where the C$a$B arises from dark-matter decay, the galactic component will be far from isotropic, instead pointing preferentially towards the Galactic Center.
Given that decaying dark-matter will emerge as a case that can be probed already by existing datasets (as it can generate $\rho_a \gg \rho_{\gamma}$), this modulation will be an important fingerprint of a genuine C$a$B signal.

The final equation in \eqref{eq:axionEM} allows for backreaction of the electromagnetic fields on the axion itself.
To determine when this is effective, consider a particularly simple experimental configuration with $\bB \cdot \bk = |\bE| = 0$, but with a DC magnetic field of strength $B_0$.
From Amp\'ere-Maxwell, the axion will induce an AC electric field oscillating parallel to the magnetic field, and with amplitude $E \sim \ga B_0 a$, generating a backreaction of $\ga \bE \cdot \bB \sim \ga^2 B_0^2 a$.
For a relativistic field we take $(\Box + m_a^2) a \simeq \Box a$, and so the condition for backreaction to be irrelevant is parametrically $\ga^2 B_0^2/\bar{\omega}^2 \ll 1$, with
\begin{equation} 
\frac{\ga B_0}{\bar{\omega}} \sim 10^{-8} \bigg( \frac{\ga}{\ga^{\rm SE}} \bigg) \bigg( \frac{1~{\rm neV}}{\bar{\omega}} \bigg) \bigg( \frac{B_0}{1~{\rm T}} \bigg)\,.
\end{equation} 
The size of the backreaction is negligible for the parameter space considered in this work, though might be of phenomenological interest for experiments looking for significantly lower frequency axions.

More generally, effects subleading in $\ga$ can be neglected. This motivates studying the fields in \eqref{eq:axionEM} in powers of $\ga$ (see also \cite{Ouellet:2018nfr}),
\begin{equation}\begin{aligned}
\bE & = \bE_0 + \bE_a + \mathcal{O}(\ga^2) \,,\\
\bB & = \bB_0 + \bB_a + \mathcal{O}(\ga^2) \,.
\end{aligned}\end{equation}
Fields carrying a subscript 0 are the dominant fields generated by the experiment, for example a large static magnetic field in ADMX or HAYSTAC, whereas a subscript $a$ denotes axion-induced effects, which are $\mathcal{O}(\ga)$.
To simplify the discussion, we will assume the large fields are DC (i.e. static), as is the case for many axion dark-matter proposals, although not all, see e.g.~\cite{Berlin:2019ahk,Lasenby:2019prg,Berlin:2020vrk}. Under this assumption, the equations for the DC fields reduce to those of electro- and magneto-statics, so that all the physics of interest is contained in the equations for the axion-induced AC fields,\footnote{Here and throughout, we will neglect all couplings to the detector and readout circuit for simplicity of the discussion. This assumption will not qualitatively impact our results, however, we note that in detail these contributions can be important, see for instance Ref.~\cite{Lasenby:2019hfz}. We leave a detailed treatment of the C$a$B response including the full matter effects to future work, and thank Robert Lasenby for emphasizing the importance of this.}
\begin{equation}\begin{aligned}
\nabla \cdot \bE_a &= - \ga \bB_0 \cdot \nabla a\,, \\
\nabla \cdot \bB_a &= 0\,, \\
\nabla \times \bE_a &= - \partial_t \bB_a\,, \\
\nabla \times \bB_a &= \partial_t \bE_a + \ga \left( \bB_0\,\partial_t a- \bE_0 \times \nabla a \right).
\end{aligned}\end{equation}
We can separate the equations as follows,
\begin{align}
(\nabla^2 - \partial_t^2) \bE_a 
=\; &\ga \bB_0 \partial_t^2 a - \ga (\bB_0 \cdot \nabla) \nabla a \nonumber \\
-\;& \ga \bE_0 \times \nabla (\partial_t a) - \ga (\nabla a \cdot \nabla ) \bB_0 \nonumber \\
-\;& \ga \nabla a \times (\nabla \times \bB_0), \label{eq:ACfull} \\
(\nabla^2 - \partial_t^2) \bB_a
=\;&
\ga \bE_0 \nabla^2 a 
- \ga (\bE_0 \cdot \nabla) \nabla a \nonumber \\
+\;& \ga \bB_0 \times \nabla (\partial_t a)
+ \ga (\nabla a \cdot \nabla) \bE_0 \nonumber \\
-\;& \ga (\partial_t a) (\nabla \times \bB_0 )
- \ga (\nabla \cdot \bE_0) \nabla a\,. \nonumber
\end{align}

In the following subsections we will solve the equations in \eqref{eq:ACfull} for different experimental configurations.
Before doing so, we consider the equations parametrically.
Firstly, in the non-relativistic limit, we can drop all axion gradients, and the equations reduce significantly,
\begin{equation}\begin{aligned}
(\nabla^2 - \partial_t^2) \bE_a 
= \;&\ga \bB_0 \partial_t^2 a\,, \\
(\nabla^2 - \partial_t^2) \bB_a
=\;&- \ga (\partial_t a) (\nabla \times \bB_0 ).
\label{eq:ACnonrel}
\end{aligned}\end{equation}
There are two relevant spatial scales in the problem.
The first is the experimental size $L$, which through the boundary conditions dictates the scale over which the primary fields vary.
The second is the Compton wavelength of the axion field, $\lambda_a \sim 1/\bar{\omega}$, or in the non-relativistic case $\lambda_a \sim 1/m_a$.\footnote{Even in the non-relativistic case, the relevant spatial scale is the Compton wavelength, and not the distance over which the phase of axion field itself varies, which is set by the coherence length.
The rationale is that the axion field will drive oscillations in the electromagnetic fields, which having a lightlike dispersion will vary over a spatial scale set by the timescale of their oscillations.
}
Accordingly, in order to understand the relevance of various terms in \eqref{eq:ACnonrel} qualitatively we can substitute $\nabla \to 1/L$ and $\partial_t \to 1/\lambda_a$, and then determine the relevance of each term for specific experiments. A more careful discussion of these scalings is provided in Ref.~\cite{Ouellet:2018nfr}.

To begin with, resonant cavity instruments are designed with a principle that $\lambda_a \sim L$, and therefore all terms are relevant.
Experiments searching for lighter dark matter, such as ABRACADABRA or DMRadio, have $\lambda_a \gg L$, suppressing time derivatives with respect to spatial gradients.
In particular, \eqref{eq:ACnonrel} then implies that $E_a \sim (L/\lambda_a) B_a$, so that the induced electric fields are parametrically suppressed with respect to the magnetic fields, a point that has been widely discussed~\cite{Goryachev:2018vjt,Ouellet:2018nfr,Kim:2018sci,Beutter:2018xfx,Lasenby:2019hfz}.
In the high mass regime considered by, for instance MADMAX~\cite{TheMADMAXWorkingGroup:2016hpc,Brun:2019lyf}, where $\lambda_a \ll L$, we instead neglect the spatial gradients.
Accordingly, $B_a \sim (\lambda_a/L) E_a$, so that the dominant effect is now the induced electric fields.

A similar analysis can be performed in the relativistic case.
We first reinstate the terms containing gradients of the axion field, and then replace those gradients with their parametric scaling of $1/\lambda_a$.
For simplicity, we consider DC field configurations that are purely magnetic.
Then, \eqref{eq:ACfull} reduces to
\begin{align}
(\nabla^2 - \partial_t^2) \bE_a 
= \;&\ga \bB_0 \partial_t^2 a - \ga (\bB_0 \cdot \nabla) \nabla a \label{eq:EMcavity} \\
- \;&\ga (\nabla a \cdot \nabla ) \bB_0- \ga \nabla a \times (\nabla \times \bB_0), \nonumber \\
(\nabla^2 - \partial_t^2) \bB_a
=\;&\ga \bB_0 \times \nabla (\partial_t a) - \ga (\partial_t a) (\nabla \times \bB_0 ). \nonumber
\end{align}
Compared to \eqref{eq:ACnonrel}, we see for $\bB_a$ there is a single additional term that depends on the relative angle between $\bB_0$ and $\bk$.
For a resonant cavity, once more all terms are in principle relevant.
In the low-frequency limit ($\lambda_a \gg L$), we have
\begin{align}
\nabla^2 \bE_a 
\simeq \;&-\ga (\nabla a \cdot \nabla ) \bB_0- \ga \nabla a \times (\nabla \times \bB_0), \nonumber \\
\nabla^2 \bB_a
\simeq\;&- \ga (\partial_t a) (\nabla \times \bB_0 ).
\label{eq:EMlowmass}
\end{align}
We now see that $E_a \sim B_a$, so that the induced electric field is no longer parametrically suppressed.
Nevertheless, the origin of the two effects is different.
The AC magnetic field is generated by $\bJ_{\rm eff}$, whereas the AC electric field originates from $\rho_{\rm eff}$.
An identical analysis in the high-frequency regime ($\lambda_a \ll L$), results in
\begin{equation}\begin{aligned}
\partial_t^2 \bE_a 
\simeq \;&\ga (\bB_0 \cdot \nabla) \nabla a - \ga \bB_0 \partial_t^2 a\,, \\
\partial_t^2 \bB_a
\simeq\;& -\ga \bB_0 \times \nabla (\partial_t a),
\end{aligned}\end{equation}
and again, neither field is suppressed.

Going forward, we will specialize to two specific scenarios from which we can largely infer how the C$a$B could appear in experiments designed to search for dark matter.
In particular, we will consider broadband and resonant detection for $\lambda_a \gg L$, and resonant detection in the cavity regime $\lambda \sim L$.
We will not consider the regime where $\lambda_a \ll L$, relevant for experiments such as MADMAX.
For a C$a$B, such experiments will not be able to reach axion energy densities relevant for cosmic sources, given the scaling in \eqref{eq:rhoreach} (see also Fig.~\ref{fig:CaB}), though they may have promise in looking for dark-matter decay.
Nonetheless, this is the parameter range relevant for the thermal C$a$B, and therefore it may be interesting to consider dedicated experiments searching for such a background.
We will not pursue this direction here.

\subsection{Low-Frequency Detection ($\lambda_a \gg L$)}

Armed with the expressions for the induced electric and magnetic fields, we now compute the C$a$B sensitivity of axion dark-matter instruments focusing on the frequencies well below a $ \mu$eV ($\lambda_a \gg 1~{\rm m}$).
For the C$a$B, our estimated sensitivity in \eqref{eq:rhoreach} suggests that such experiments are ideal for probing cosmic relics, for which measurements of $\Delta N_{\rm eff} $ bound $\rho_a < \rho_{\gamma}$.
While existing instruments only have sensitivity for dark-matter axions with a coupling comparable to the star-emission bounds~\cite{Ouellet:2018beu,Ouellet:2019tlz,Gramolin:2020ict,Crisosto:2019fcj}, these results are paving the way for future experiments that will probe the couplings predicted for the QCD axion, such as DMRadio~\cite{SnowmassOuellet,SnowmassChaudhuri}.
In this mass range, both broadband and resonant search strategies have been proposed.
As such, in this section we will consider both types of detection, and to be concrete envision a large scale realization of the DMRadio (or equivalently ABRACADABRA) instrument.\footnote{For the specific case of DMRadio, it will likely only be realized on large scales as a resonant instrument given the advantages of a resonant approach for dark-matter searches~\cite{Chaudhuri:2018rqn,Chaudhuri:2019ntz}.}

Our starting point is the geometry of DMRadio and ABRACADABRA: a large toroidal magnet with field strength $B_0$.
To understand the effects generated by the C$a$B on such an instrument, we can use the equations of axion electrodynamics in the $\lambda_a \gg L$ limit, as stated in \eqref{eq:EMlowmass}.
From the second equation, we see that the axion field will convert the DC magnetic field into an oscillating toroidal current, which will then induce an AC magnetic field in the center of the torus.
In the presence of an axion field, a pickup loop placed in the center of the torus would see a varying magnetic flux in a region where conventionally there should be none.
This detection principle is identical to the conventional strategy for detecting axion dark matter with such an instrument; that the effect would be the same is clear from the fact the equation for $B_a$ in \eqref{eq:EMlowmass} does not involve any gradients of the axion field.
Even though the C$a$B modes have a significantly smaller de Broglie wavelength than dark matter, there is no issue of this leading to an incoherent effect across the instrument, as that would only occur for $\lambda_a \lesssim L$, outside the range considered by these instruments.

There are, however, differences for the C$a$B detection from the conventional dark-matter axion search.
Firstly, the range of frequencies over which the AC magnetic field will be excited are significantly larger than for dark matter, as we have emphasized many times already.
A second difference is that unlike in the non-relativistic case, there is now an unsuppressed electric field generated from the effective charge.
Recalling that the relativistic axion converts magnetic field lines into oscillating charge lines, the instrument would behave like a torus of oscillating charge, inducing an axial AC electric field near the center of the torus.
Supposing the pickup loop used to search for $B_a$ is perfectly perpendicular to this field, the above detection scheme is unaffected.
Nevertheless, the oscillating electric field is present, and its detection could provide a confirmation of any magnetic field excess.

Turning to the actual detector response, integrating the effective current over the torus, the flux induced in the pickup loop will be~\cite{Kahn:2016aff}
\begin{equation} 
\Phi_{\rm pickup}(t) = \ga B_0 V_B \partial_t a(t)\,,
\end{equation} 
where $B_0$ is the magnetic field at the inner radius of the torus and $V_B$ is the magnetic field volume.
In ABRACADABRA this flux is read out by inductively coupling the pickup loop to a SQUID, which will observe a flux $\Phi_a$ proportional to $\Phi_{\rm pickup}$, with a proportionality constant $\beta$.
For a 100 m$^3$ instrument, $\beta \simeq 0.5\%$~\cite{Kahn:2016aff}.
As discussed in Sec.~\ref{sec:prop-rel-ax}, this signal can be searched for by collecting a time series data set of the SQUID flux, taking the discrete Fourier transform of the measurements, and finally forming the PSD, $S_{\Phi}(\omega)$.
Going through these steps and recalling that the PSD of the C$a$B is exponentially distributed, with mean given in \eqref{eq:Sa-rel}, the result is that the power will be an exponentially distributed quantity, with mean
\begin{equation}\begin{aligned}
\langle S_{\Phi}(\omega) \rangle &  = \ga^2  \beta^2  B_0^2 V_B^2  \left\langle S _{ \partial _t a } \right\rangle  + \lambda_B(\omega) \\ 
& = \ga^2 \rho_a \beta^2 B_0^2 V_B^2\frac{\pi \omega p(\omega)}{\bar{\omega}} + \lambda_B(\omega)\,,
\label{eq:SABRA}
\end{aligned}\end{equation}
where we have introduced the time derivative of \eqref{eq:Sa-rel}, $ \left\langle S _{ \partial _t a } \right\rangle =  \omega ^2 \left\langle S _{ a} \right\rangle $. 
Here $\lambda_B(\omega)$ is the contribution to the flux from background sources, which in general will be frequency dependent.
For a broadband strategy, $\lambda_B$ is limited by noise within the SQUID, numerically given by $\lambda_B \simeq 1.6 \times 10^5~{\rm eV}^{-1}$.
The steps leading to this result parallel closely those in the non-relativistic case considered in \cite{Foster:2017hbq}, and we refer there for additional details.
In particular, recalling that in the non-relativistic limit $p(\omega) = f(v_{\omega})/(m_a v_{\omega})$, \eqref{eq:SABRA} contains the dark-matter result as a special case.

Importantly, the exponential nature of the PSD implies that we can exploit the full likelihood framework of~\cite{Foster:2017hbq}.
We can then analytically determine the expected sensitivity to our signal through the use of the Asimov data set~\cite{Cowan:2010js}, where instead of considering the distribution of our sensitivity on a set of simulated data, we instead replace the data with its asymptotic expectation.
Doing so, for a given $\ga$, our sensitivity to the C$a$B density is given by
\begin{equation}\begin{aligned}
\frac{\rho_a}{\rho_{\gamma}} = &\frac{1}{\ga^2 \rho_{\gamma}} \frac{1}{\beta^2 B_0^2 V_B^2} \sqrt{\frac{2 {\rm TS}}{T \pi}} \\
\times &\left[ \int d\omega\, \left( \frac{\omega p(\omega)}{\bar{\omega} \lambda_B} \right)^2 \right]^{-1/2}\,.
\label{eq:broadband-sens}
\end{aligned}\end{equation}
Here $T$ is the data collection time, and ${\rm TS}$ is the test statistic associated with the sensitivity threshold (for 95\% expected limits ${\rm TS} \simeq 2.71$, whereas for an $n$-$\sigma$ discovery, ${\rm TS} = n^2$ up to the look elsewhere effect).
To arrive at this result we assumed that $T \gg \tau$, such that $p(\omega)$ is well resolved, and also that we are in the background dominated regime.
Further, we have left the terminals on the frequency integration unspecified, but these can at largest be the range over which the experiment has sensitivity.

The result in \eqref{eq:broadband-sens} is consistent with the simple estimate claimed in Sec.~\ref{sec:est-sens}.
To see this, we model the broad C$a$B as a log-uniform distribution $p(\omega) = Q_a^{\text{C$a$B}}/\omega$ defined over a range $\bar{\omega}/Q_a$.
Then, treating $\lambda_B$ as independent of frequency, our sensitivity can be written (dropping $\rho_{\gamma}$)
\begin{equation}\begin{aligned}
\rho_a = &\sqrt{\frac{2{\rm TS}}{\pi}} \left(\frac{\sqrt{\bar{\omega}/Q_a^{\text{C$a$B}}}}{\ga^2} \right) \left( \frac{\lambda_B}{\beta^2 B_0^2 V_B^2 \sqrt{T}}\right)\,.
\label{eq:broadband-sens-est}
\end{aligned}\end{equation}
For dark matter, we instead approximate the distribution as a uniform $p(\omega) = Q_a^{\DM}/m_a$ over a narrow range, which results in an identical expression but with $Q_a^{\text{C$a$B}} \to Q_a^{\DM}$ and $\bar{\omega} \to m_a$.
Taking $m_a = \bar{\omega}$, the ratio of the two expressions yields \eqref{eq:est-lim-resonant}.

The sensitivity in \eqref{eq:broadband-sens} represents the quantitative result for broadband sensitivity, however to provide additional intuition -- especially for why the width of $p(\omega)$ plays a fundamental role -- we can rederive the result qualitatively from a signal-to-noise ratio~\cite{Budker:2013hfa,Kahn:2016aff}.
The signal strength is simply the average signal flux, which is given by (using \eqref{eq:asq-mean})
\begin{equation}
|\Phi_a| = \ga |\partial_t a|  \beta B_0 V_B
\sim \ga \sqrt{\rho_a}  \beta B_0 V_B\,,
\end{equation}
whereas the flux noise is given by $\sqrt{\lambda_B}$.
Given the signal and background flux magnitudes, if we perform a measurement for a time $T$, the naive expectation is our sensitivity will grow as $\sqrt{T}$, and indeed for a time it will.
Nevertheless, as described already, the C$a$B has a finite coherence time $\tau \sim 2\pi Q_a/\bar{\omega}$, and for $T > \tau$ the signal will no longer combine coherently, leading to the signal-to-noise only scaling with the parametrically reduced $(T \tau)^{1/4}$~\cite{Budker:2013hfa}.
Assuming $T > \tau$, then our sensitivity is given by $|\Phi_a| (T \tau)^{1/4}/\sqrt{\lambda_B}=1$.
Together, these scalings provide
\begin{equation} 
\rho_a = \frac{1}{\sqrt{2\pi}} \left( \frac{\sqrt{\bar{\omega}/Q_a}}{\ga^2} \right) \left( \frac{\lambda_B}{\beta^2 B_0^2 V_B^2 \sqrt{T}} \right)\,,
\end{equation} 
which is parametrically identical to \eqref{eq:broadband-sens-est}, and demonstrates that the appearance of the width of $p(\omega)$ is a consequence of measuring the C$a$B over times longer than the field is coherent.

The alternative qualitatively different readout strategy proposed in this frequency range is resonant detection.
For the toroidal geometry described above, this can be achieved by reading out the pickup loop through a resonant circuit, which for our purposes can be characterized by four parameters: the quality factor $Q$, resonant frequency $\omega_0$, total circuit inductance $L_T$, and thermal noise temperature $T_0$.
As in the broadband case, we can generalize the known dark-matter result (see e.g. Ref.~\cite{Foster:2017hbq}) to the C$a$B as follows,
\begin{equation}\begin{aligned} 
\frac{\rho_a}{\rho_{\gamma}} &= \frac{1}{\ga^2 \rho_{\gamma}} \frac{2 L_T T_0}{Q \omega_0 B_0^2 V_B^2} \sqrt{\frac{2 {\rm TS}}{T \pi}} \\
&\times \left[ \int d\omega \left( \frac{\omega p(\omega)}{\bar{\omega}} \right)^2 \right]^{-1/2},
\end{aligned}\end{equation}
which, up to experimental factors, is identical to \eqref{eq:broadband-sens}.

This expression determines the expected C$a$B energy density sensitivity for a given set of experimental parameters.
Yet we can recast this result in the spirit of Sec.~\ref{sec:est-sens}, and forecast C$a$B sensitivity in terms of the expected dark-matter reach.
Indeed \eqref{eq:broadband-sens} holds equally well for dark matter, taking $\rho_a = \rho_{\DM}$, and evaluating
\begin{equation} 
\int d\omega \left(\frac{\omega p(\omega)}{\bar{\omega}} \right)^2 = \frac{1}{m_a} \int dv \frac{f(v)^2}{v} \simeq \frac{2}{3} \frac{Q_a^{\DM}}{m_a}\,,
\end{equation} 
where in the final step we assumed $f(v)$ follows the canonical standard halo model.
For a resonant instrument, the frequency range is commonly restricted to a single bandwidth of size $\Delta \omega = \omega_0/Q$ around $\omega_0$.
For $Q < Q_a^{\DM}$ this detail is irrelevant in the dark-matter computation, as for $m_a = \omega_0$, then the dark-matter distribution is contained entirely within the bandwidth.
For the C$a$B however, we will generically have $Q_a^{\text{C$a$B}} \ll Q$, and therefore expect $p(\omega)$ to be constant over the range of integration, leading to
\begin{equation} 
\int d\omega \left(\frac{\omega p(\omega)}{\bar{\omega}} \right)^2 \simeq \frac{\omega_0}{Q} \left( \frac{\omega_0 p(\omega_0)}{\bar{\omega}} \right)^2\,.
\end{equation} 
Having computed the result for both dark matter and relativistic axions, we can take the ratio to determine the C$a$B sensitivity as a function of the experimentally achieved dark-matter coupling, $\ga^{\rm lim}$, to be
\begin{equation} 
\frac{\rho_a}{\rho_{\gamma}} = \sqrt{\frac{2}{3}} \frac{\rho_{\DM}}{\rho_{\gamma}} \left( \frac{\ga^{\rm lim}}{\ga^{\rm SE}} \right)^2 \left( \frac{\bar{\omega}}{m_a} \right) \left( \frac{\sqrt{Q_a^{\DM} Q}}{m_a p(m_a)} \right)\,.
\label{eq:resonant-sens}
\end{equation} 
If we approximate the energy distribution as log flat, so $p(m_a) = Q_a/m_a$, take $\bar{\omega}=m_a$, and recall that the C$a$B sensitivity can in principle be enhanced across at most $N \sim Q/Q_a^{\text{C$a$B}}$ bandwidths, then this result reproduces the parametric scaling given in \eqref{eq:est-lim-resonant}.

\subsection{Resonant Cavity Detection ($\lambda_a \sim L$)}

We next consider detection with a physical resonator, as pursued by both ADMX and HAYSTAC, where a microwave cavity is constructed in order to resonantly enhance power produced at a frequency tuned to the cavity dimension, $\omega_0 \sim 1/L$.
The design principle for these instruments is to optimize the search for the narrow spectral feature dark matter predicts when its mass is such that $1/m_a \sim L \sim {\rm m}$.
In particular, in the presence of a large static magnetic field $\bB_0$, an axion background will source oscillating electromagnetic fields, thereby generating potentially detectable power in the cavity.
As we will show in this section, this statement is true for both dark matter and the C$a$B.
Importantly, the existing reach of ADMX is already sufficient to probe open parameter space, although a reanalysis of the data would be required, and the same will soon be true of HAYSTAC.

Parametrically our final sensitivity will be identical to the resonant circuit expression \eqref{eq:resonant-sens}.
There will, however, be important differences.
According to the expressions in \eqref{eq:EMcavity}, the C$a$B will source AC electric and magnetic fields in the presence of a large $\bB_0$, and if we work in a regime where the DC field is spatially uniform, we have
\begin{equation} 
(\nabla^2 - \partial_t^2) \bE_a 
= \ga \bB_0 \partial_t^2 a\,,
\label{eq:EMcavity-uniform}
\end{equation} 
whilst $\bB_a$ can be determined from Faraday's law.
This result is identical to the case of dark matter: we have dropped the contribution from the effective charge as it cannot excite resonant cavity modes.\footnote{We thank Asher Berlin and Kevin Zhou for teaching us this point. Further discussion can be found in, for example, Ref.~\cite{Berlin:2022mia}.}
Unlike for a non-relativistic axion, when $\lambda_a \sim L$ the axion wave can undergo ${\cal O}(1)$ spatial oscillations across the detector, which will induce a new daily modulation effect we will explore.
Taking a simplified picture of the C$a$B where $a(t,\bx) \propto \cos (\omega t - \bk \cdot \bx)$, we will find an explicit dependence on the direction of $\bk$.
In particular, when we compute the cavity form factor -- which captures the overlap between the axion source and the relevant mode of the cavity being measured -- a dependence on the direction of $\bk$ and hence the C$a$B will appear.
For a cosmic relic, the C$a$B will be incident approximately isotropically on the detector, and the effect will take on a sky-averaged, and effectively time-independent, value.\footnote{If a detection is made, future experiments could further look for axion spatial correlations analogous to those present in the CMB.}
However, for local sources of relativistic axions, the distribution can be anisotropic.
This will certainly be the case for dark-matter decays in the Milky Way, where the flux predominantly originates from the Galactic Center.
In the rest frame of the Earth, where the orientation of the cavity is time invariant, the direction of the Galactic Center will vary throughout the day, leading to a daily variation in the cavity form factor.
Accordingly, in general the C$a$B power, which is proportional to the form factor, will undergo $\mathcal{O}(1)$ variations throughout the day.
These daily modulations provide a novel handle that can be used to distinguish a C$a$B with a local origin from potential backgrounds, although in the present work we will not quantitatively calculate this effect.

Our focus is instead on determining the C$a$B sensitivity of these instruments, which will require a determination of the power a relativistic axion field deposits in a cylindrical cavity as used by both ADMX and HAYSTAC.
We do so by repeating the analogous non-relativistic computation~\cite{Krauss:1985ub,Sikivie:1985yu} (see~\cite{Brubaker:2018ebj} for a recent review) while accounting for the three important modifications in the relativistic case.
Two of these we have already discussed: the additional gradient term in \eqref{eq:EMcavity-uniform} and the fact that the C$a$B carries power over a much broader frequency range.
The third relativistic novelty arises from the fact the axion field need not be spatially coherent across the instrument when $|\bk| = \omega \sim 1/L$, which can suppress the integrated power deposited over the cavity.

We begin with \eqref{eq:EMcavity-uniform}.
We will solve this equation within a cylindrical cavity, where a DC magnetic field $\bB_0 = B_0 \hz$ has been established along the cavity axis, and we will assume the field is spatially uniform.
The cavity will have a set of normal modes for the electric fields it can support, $\bbe_{\ell m n}(\bx)$, indexed by three integers $(\ell,m,n)$, and satisfying
\begin{equation} 
(\nabla^2 + \omega_{\ell m n}^2) \bbe_{\ell m n}(\bx) = 0\,,
\end{equation} 
with $\omega_{\ell m n}$ the resonant frequency of the mode.
On dimensional grounds we expect $\omega_{\ell m n} \sim 1/L$, however the exact value will be determined by the spatial variation of the modes.
As these basis modes must satisfy the electric boundary conditions, $\mathcal{O}(1)$ changes to the resonant frequency can be achieved by modifying these boundary conditions, as ADMX achieves by mechanically varying the location of tuning rods within the instrument.
We can normalize the basis modes as follows,
\begin{equation} 
\int d^3 \bx\, \bbe_{\ell m n} \cdot \bbe_{\ell' m' n'}^* = \delta_{\ell \ell'} \delta_{m m'} \delta_{nn'}\,.
\end{equation} 
The integral is performed over the cavity volume $V$, which reveals that the modes carry dimension $\bbe \sim V^{-1/2}$.

As the orthonormal modes are a complete basis for the electric field in the cavity, we can write $\bE_a = \sum \alpha_{\ell mn} \bbe_{\ell mn}$ and solve for the coefficients.
Using this expansion in \eqref{eq:EMcavity-uniform}, and transforming to the frequency domain, we obtain
\begin{equation}\begin{aligned}
&\sum_{\ell, m,n} (\omega^2 - \omega_{\ell m n}^2) \alpha_{\ell mn} \bbe_{\ell mn} \\
= &-\omega^2 B_0 \ga a(\omega) {\rm cos} (\bk \cdot \bx) \hz\,,
\end{aligned}\end{equation} 
where we have assumed a simple form for the axions spatial dependence, $a(\omega,\bx) = a(\omega) \cos( \bk \cdot \bx)$.
Using the orthonormality condition, we can then isolate the electric field coefficients as,
\begin{equation}\begin{aligned}
\alpha_{\ell mn} = &-\frac{\ga B_0\, a(\omega)\,\omega^2}{\omega^2 - \omega_{\ell m n}^2} \\
\times &\int d^3 \bx \cos (\bk \cdot \bx) \hz \cdot \bbe_{\ell mn}^*\,.
\label{eq:ACEcoeffs}
\end{aligned}\end{equation} 
The first line of this result identifies $\omega_{\ell m n}$ as the resonant frequencies, with the apparent divergence a remnant of our idealized treatment of the problem.
In particular we have neglected the fact that the electric field will penetrate the walls of the cavity and dissipate energy due to the finite resistance of the material, which is the origin of the cavity quality factor, $Q$.
This dissipation can be accounted for as in the standard dark-matter calculation.
More novel is the second line of \eqref{eq:ACEcoeffs}, and we define
\begin{equation} 
\kappa_{\ell m n} = \int d^3 \bx \cos (\bk \cdot \bx) \hz \cdot \bbe_{\ell mn}^*\,,
\end{equation} 
which we will use to define a generalized cavity form factor shortly, and note dimensionally $\kappa_{\ell m n} \sim V^{1/2}$.

The energy density in the AC cavity fields for this mode is given by $U_{\ell m n} = (|\bE_a|^2 + |\bB_a|^2)/2 \simeq |\bE_a|^2$, as for frequencies near the resonant frequency we will have $|\bB_a| \simeq |\bE_a|$ from Faraday's law.
Collecting our expressions above, the energy density has the form
\begin{equation} 
U_{\ell m n} = \frac{1}{4} \ga^2 B_0^2 V C_{\ell m n} \omega_{\ell m n}^2 |a(\omega)|^2 \mathcal{T}(\omega)\,.
\label{eq:cavityenergydensity}
\end{equation} 
Here we have defined a transfer function, $\mathcal{T}(\omega)$, which accounts for the resonant response of the circuit when dissipation is included,
\begin{equation} 
\mathcal{T}(\omega) = \frac{1}{(\omega-\omega_{\ell m n})^2 + (\omega_{\ell m n}/2Q)^2}\,,
\label{eq:Transfer}
\end{equation} 
which is sharply peaked around its maximum, $\mathcal{T}(\omega_{\ell m n}) = 4Q^2/\omega_{\ell m n}^2$.
The energy density is further expressed in terms of a cavity form factor $C_{\ell m n} = |\kappa_{\ell m n}|^2/V$, where cavity volume has been introduced to ensure this is an intensive quantity.
In detail, we define a relativistic cavity form factor,
\begin{equation} 
\hspace{-0.05cm}C^{\text{C$a$B}}_{\ell m n} = \frac{1}{V}\left| \int d^3 \bx \cos (\bk \cdot \bx) \hz \cdot \bbe_{\ell mn}^* \right|^2\hspace{-0.05cm},
\label{eq:cavityrel}
\end{equation} 
which can be contrasted with the conventional result used for dark matter
\begin{equation} 
C^{\DM}_{\ell m n} = \frac{1}{V}\left| \int d^3 \bx\, \hz \cdot \bbe_{\ell mn}^* \right|^2\hspace{-0.05cm}.
\end{equation} 

Let us discuss several details of these form factors, which parameterize the overlap between the induced $\bE_a$ and the static $\bB_0$.
In both cases, numerically we have $C < 1$.
Further, we expect $C$ to be $\mathcal{O}(1)$ for the lowest lying mode -- higher modes correspond to basis functions of shorter wavelength, which generically suppress the integral.
In the non-relativistic case, contributions to $\bbe$ perpendicular to $\hz$ do not contribute, which identifies the cavity transverse magnetic (TM) modes as relevant.
We have yet to specify $\bbe$, however for a cylindrical cavity they are well known (and our results are qualitatively similar for other geometries).
Taking the cylinder to have height $L$, and radius $R \sim L$, we have
\begin{equation} 
(\bbe_{\ell00})_z = \frac{1}{\sqrt{V}}\frac{J_0(\omega_{\ell00} r)}{J_1(\omega_{\ell00} R)}\,,
\label{eq:cavitymodes}
\end{equation} 
where $J_n$ are Bessel functions of the first kind, and the resonant frequency is given by $\omega_{\ell00} = j_{0\ell}/R$, with $j_{0\ell}$ the $\ell$th zero of $J_0$.
Explicitly evaluating the cavity form factor, we obtain
\begin{equation} 
C^{\DM}_{\ell 0 0} = \frac{4}{j_{0\ell}^2}\,.
\end{equation} 
Numerically, $C_{100} \simeq  0.69$, and $C_{\ell00} \sim C_{100}/\ell^2$, so that the response of higher order modes is rapidly suppressed.
Accordingly, it is common to focus on the lowest lying mode, defining $\omega_0 = \omega_{100}$ and $C = C_{100}$.

\begin{figure}[!t]
\vspace{-0.1cm}
\centering
\includegraphics[width=.465\textwidth]{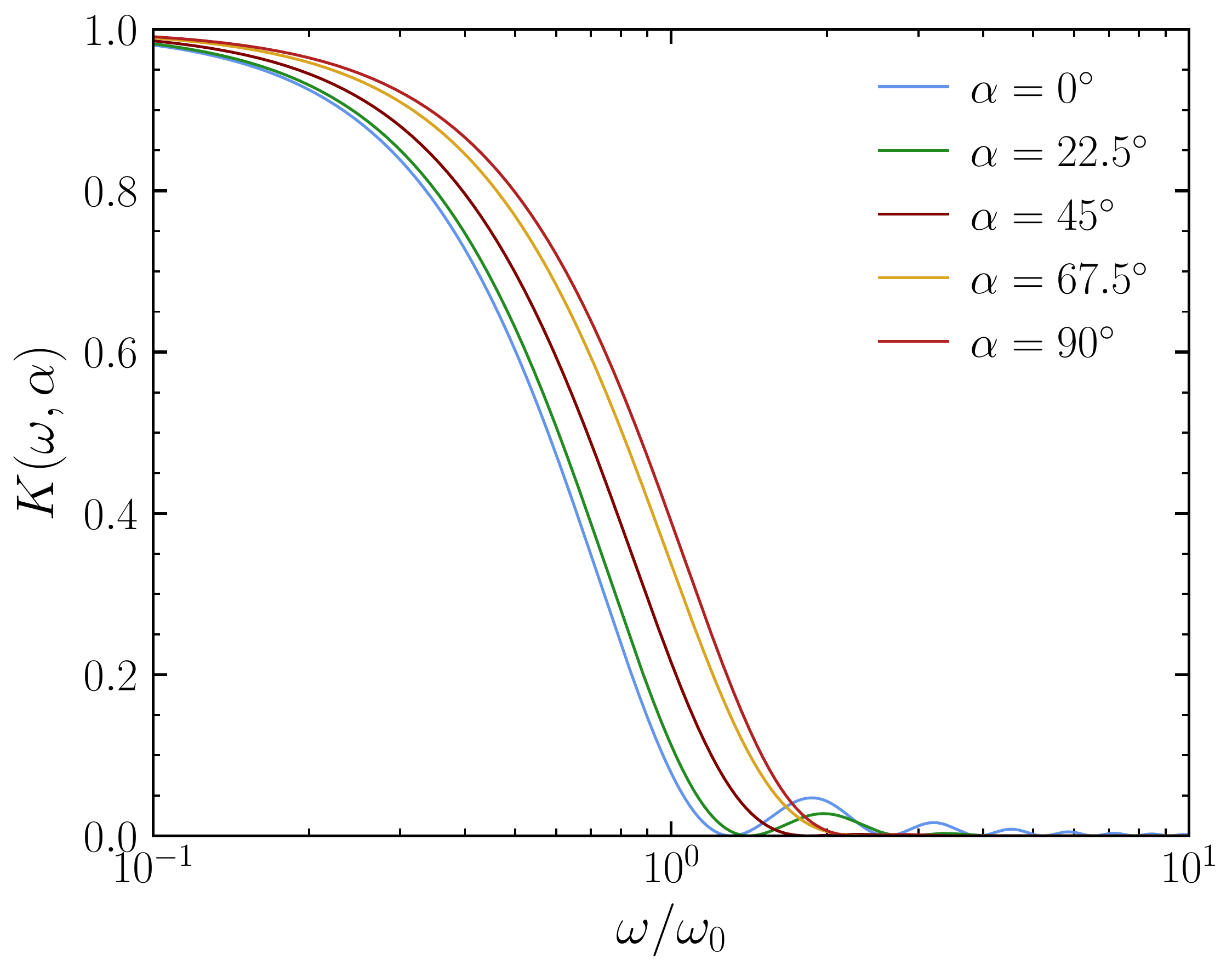}
\vspace{-0.3cm}
\caption{The suppression of relativistic axion power deposited in resonant cavity instruments, as encoded in $K$, defined in \eqref{eq:Crel}.
We fix the cavity radius to be equal to the height, $R=L$, and demonstrate the approximate insensitivity of the result to the relative angle of the incident axions and the detector magnetic field, $\alpha = \arccos(\hn \cdot \hz)$.
For $R \neq L$, a large dependence on $\alpha$ can arise, generating a potential daily modulation in the C$a$B signal.
Note that this factor enters in the relativistic case in addition to the transfer function in \eqref{eq:Transfer}, which is sharply peaked at $\omega_0$.
}
\label{fig:ADMXSuppression}
\end{figure}

The relativistic cavity factor is complicated by a dependence on the incident angle of the axion through $\hn$ (where $\bk = \omega \hn$) and a frequency dependence through $\bk$.
With a view to searching for the C$a$B through a repurposed dark-matter search, we will only consider the response induced for the lowest TM mode, although we note from \eqref{eq:cavityrel} that in this case transverse electric modes could also contribute.
Combining \eqref{eq:cavityrel} and \eqref{eq:cavitymodes} for $\ell=0$,
\begin{equation}\begin{aligned}
C^{\text{C$a$B}} &= C \left[ \frac{\sin(\omega L c_{\alpha})}{\omega L c_{\alpha}} \frac{J_0(\omega R s_{\alpha})}{1-(\omega R s_{\alpha}/j_{01})^2} \right]^2 \\
&\equiv C K(\omega,\alpha)\,,
\label{eq:Crel}
\end{aligned}\end{equation}
where we employ a shorthand $s_{\alpha} = \sin \alpha$ and $c_{\alpha} = \cos \alpha$, with $\alpha = {\rm arccos} (\hn \cdot \hz)$.
The additional relativistic novelty is $K(\omega,\alpha)$, which accounts for the incoherence of the axion field over the experimental volume, and is shown in Fig.~\ref{fig:ADMXSuppression}.
As shown there, for $\omega \ll \omega_0$ we have $K(\omega,\alpha) \to 1$, corresponding to the limit where the axion is spatially coherent over the instrument, whereas for $\omega \gg \omega_0$ instead $K(\omega,\alpha) \to 0$ as a result of destructive interference across the cavity, and for $L \sim R$ the result is only weakly dependent on $\alpha$.\footnote{Both ADMX and HAYSTAC have $L \sim 5 R$, and we will discuss this more realistic case shortly.}
This factor effectively removes the contribution of frequency with $\omega > \omega_0$, and we will approximate it by a step function, $K(\omega,\alpha) \simeq \Theta(\omega_0 - \omega)$.

Having determined the modified form factor, we can return to determining the measurable signal power, which is related to the cavity energy density by $P = \omega_0 U/Q$. Using \eqref{eq:cavityenergydensity}, we have
\begin{equation} 
P_a(\omega) = \ga^2 \frac{B_0^2 V C \omega_0^3}{4Q} |a(\omega)|^2 \mathcal{T}(\omega) \Theta(\omega_0 - \omega)\,.
\end{equation} 
The form of $|a(\omega)|^2$ has already been extensively discussed, it will be an exponentially distributed variable with mean given in \eqref{eq:Sa-rel}. Accordingly, the power will also be exponentially distributed, however the total power on average will be given by
\begin{equation}\begin{aligned}
P_a^{\text{C$a$B}} =\, &\frac{\ga^2\pi \rho_a}{\bar{\omega}} \frac{B_0^2 V C \omega_0^3}{4Q} \int_0^{\omega_0} \frac{d\omega}{2\pi} \frac{p(\omega)}{\omega} \mathcal{T}(\omega) \\
\simeq\, &\frac{\pi}{8} \ga^2 \rho_a p(\omega_0) \frac{\omega_0}{\bar{\omega}} B_0^2 V C\,,
\label{eq:relpower-cavity}
\end{aligned}\end{equation}
where in the final step we assumed that $Q_a^{\text{C$a$B}} \ll Q$, so that $p(\omega)$ only varies slowly over the range where the transfer function has appreciable support.

In \eqref{eq:relpower-cavity} we have an expression for the signal power the relativistic axion will deposit in the cavity when analyzing a single resonant frequency. Combined with an expected background contribution and a set of experimental parameters, this result is sufficient to forecast the C$a$B sensitivity. Here, however, we will instead use a matched power approach to obtain the projected reach. In particular, existing ADMX limits are a combination of individual experimental runs, so rather than combining these run-by-run, we will simply recast the combined dark-matter $\ga$ limits.
To do so, we require the dark-matter analogue of \eqref{eq:relpower-cavity}, which is obtained by integrating over the full frequency range, and taking a mean $|a(\omega)|^2$ as given in \eqref{eq:Sa-DM}.
Re-evaluating the integral, we find
\begin{equation} 
P_a^{\DM} \simeq \frac{1}{2} \ga^2 B_0^2 Q V C \frac{\rho_{\DM}}{m_a}\,.
\label{eq:nonrelpower-cavity}
\end{equation} 

We cannot directly match C$a$B and dark-matter powers above, as comparing the signal strength will only determine the sensitivity assuming the background is equal in the two cases.
However, as already discussed in Sec.~\ref{sec:est-sens}, it is not, and the larger background for the C$a$B will reduce its sensitivity by $\sqrt{Q_a^{\DM}/Q}$.
Once this is accounted for, we can combine \eqref{eq:relpower-cavity} and \eqref{eq:nonrelpower-cavity} (evaluated at $\omega_0=m_a$), to obtain our estimated sensitivity in this regime of
\begin{equation} 
\frac{\rho_a}{\rho_{\gamma}} = \frac{4}{\pi} \frac{\rho_{\DM}}{\rho_{\gamma}} \left( \frac{\ga^{\rm lim}}{\ga^{\rm SE}} \right)^2 \left( \frac{\bar{\omega}}{m_a} \right) \left( \frac{\sqrt{Q_a^{\DM} Q}}{m_a p(m_a)} \right)\,,
\label{eq:cavity-sens}
\end{equation} 
which, as promised, is identical to \eqref{eq:resonant-sens} up to a numerical prefactor.

Several aspects of the above discussion are overly idealized.\footnote{We thank Jonathan Ouellet for this observation.}
In particular, we considered a configuration where the scale at which the relativistic power suppression occurs -- as encoded in $K(\omega,\alpha)$ -- matches the resonant frequency scale, both set by the characteristic size of the cavity.
In practice, however, microwave cavity instruments introduce tuning rods in order to vary the resonant frequency, which will shift $\omega_0$ by a factor of a few above the characteristic value $\sim 1/L$.
Yet as the relativistic suppression occurs for $\omega \gtrsim 1/L$, when the axion field is no longer spatially coherent across the cavity, $K(\omega,\alpha)$ as it appears in Fig.~\ref{fig:ADMXSuppression} will remain qualitatively unchanged.
Combining the two effects, naively this would suggest a significant reduction in sensitivity as the modes that are resonantly enhanced also experience the incoherent suppression.
Nonetheless, in an actual instrument, the geometry is such that $L \neq R$, and once that is accounted for $K(\omega,\alpha)$ now varies considerably with angle, and for $\alpha=90^{\circ}$ the suppression is postponed till higher frequencies, so that the power can still be absorbed, and with the appearance of a significant daily-modulation effect that can be exploited.

\section{Projected Limits}
\label{sec:lim}

Having outlined various forms the C$a$B can take, and having determined the experimental sensitivity to it, in this section we combine these results to sketch projected sensitivities.
As we will show, detecting the C$a$B will not necessarily require dedicated instruments, instead the rapid progress in the search for axion dark matter will simultaneously open enormous swaths of relativistic axion parameter space.
The present discussion will not be exhaustive.
Instead we will show the estimated reach in three cases to demonstrate various aspects of the detection schemes we have proposed, and the interplay with specific C$a$B candidates.
Firstly, we will discuss the reach of the existing resonant cavity instruments HAYSTAC and ADMX for a simple Gaussian $p(\omega)$ as predicted in the parametric resonance scenario, showing that an order of magnitude improvement in the $\ga$ sensitivity of ADMX would translate to sensitivity to $\rho_a < \rho_{\gamma}$, although this would still be short of the prediction of parametric-resonance production discussed in Sec.~\ref{sec:parametric-resonance}.
Secondly, we demonstrate that a large scale broadband ABRACADABRA style instrument could probe relativistic axions originating from cosmic strings in the parameter space where they could help alleviate the Hubble tension.
Finally, we will consider the case of most immediate interest: indirect detection of dark matter decaying to axions.
As this scenario allows $\rho_a > \rho_{\gamma}$, we will see that ADMX is already sensitive to unexplored parameter space, and the situation will improve dramatically with future instruments.

Recall that the C$a$B signal is determined by three quantities: $\rho_a$, $\ga$, and $p(\omega)$ (equivalently, the form of $ \Omega _a ( \omega ) $ fixes $ \rho _a $ and $ p ( \omega ) $).
As already discussed, here we will take the approach of fixing $\ga$ to $\ga^{\rm SE}$, the largest value consistent with star-emission constraints.
When considering detection with frequencies $\omega \gtrsim 10^{-9}$ eV, we take $\ga^{\rm SE} = 0.66 \times 10^{-10}~{\rm GeV}^{-1}$ for consistency with CAST~\cite{Anastassopoulos:2017ftl} and Horizontal Branch~\cite{Ayala:2014pea,Carenza:2020zil} constraints.
In the future, this limit may be tightened by IAXO~\cite{Vogel:2013bta} (see also Refs.~\cite{Mukherjee:2018oeb,Mukherjee:2019dsu} for future searches using the CMB). Should these experiments detect an axion signal, the same axion could be produced in the early Universe, and would strongly motivate further searches for the C$a$B.
At lower frequencies, the bounds strengthen further, and we will adopt $\ga^{\rm SE} = 3.6 \times 10^{-12}~{\rm GeV}^{-1}$ as determined from super star clusters~\cite{Dessert:2020lil}.

\subsection{Gaussian}

To begin with, we consider searching for a Gaussian energy distribution in existing resonant cavity instruments. Such a distribution can be motivated by the parametric-resonant production mechanism reviewed in Sec.~\ref{sec:parametric-resonance}, however here we can also envision it as providing an opportunity to explore our formalism. In detail, we consider a positive definite Gaussian with mean $\mu = \bar{\omega}$, and variable width $\sigma = \kappa \bar{\omega}$, where we will explore several values of $\kappa$. In detail, we take
\begin{equation} 
p(\omega) = \frac{\sqrt{2/\pi}\, e^{- (\omega/\bar{\omega}-1)^2/2\kappa^2}}{\kappa \bar{\omega} [1-{\rm erf}[-1/\sqrt{2} \kappa])} \Theta[\omega]\,.
\label{eq:dist-Gauss}
\end{equation} 
Fixing $\ga$, we can then determine a limit on $\rho_a$ for a given $\kappa$.

To do so, we will recast existing bounds on axion dark matter collected by the ADMX and HAYSTAC instruments.
ADMX is already probing the couplings predicted for the QCD axion for $m_a \sim 2-3~\mu{\rm eV}$~\cite{Asztalos:2003px,Du:2018uak,Braine:2019fqb}.
Given \eqref{eq:rhoreach}, we would therefore expect the instrument to be on the verge of $\rho_a \sim \rho_{\gamma}$ sensitivities for the C$a$B.
HAYSTAC, on the other hand, is already within a factor of a few from the QCD prediction for $m_a \sim 23-24~\mu{\rm eV}$~\cite{Zhong:2018rsr}.
Here we take these existing limits and recast them using \eqref{eq:cavity-sens}.

\begin{figure}[!t]
\vspace{-0.1cm}
\centering
\includegraphics[width=.465\textwidth]{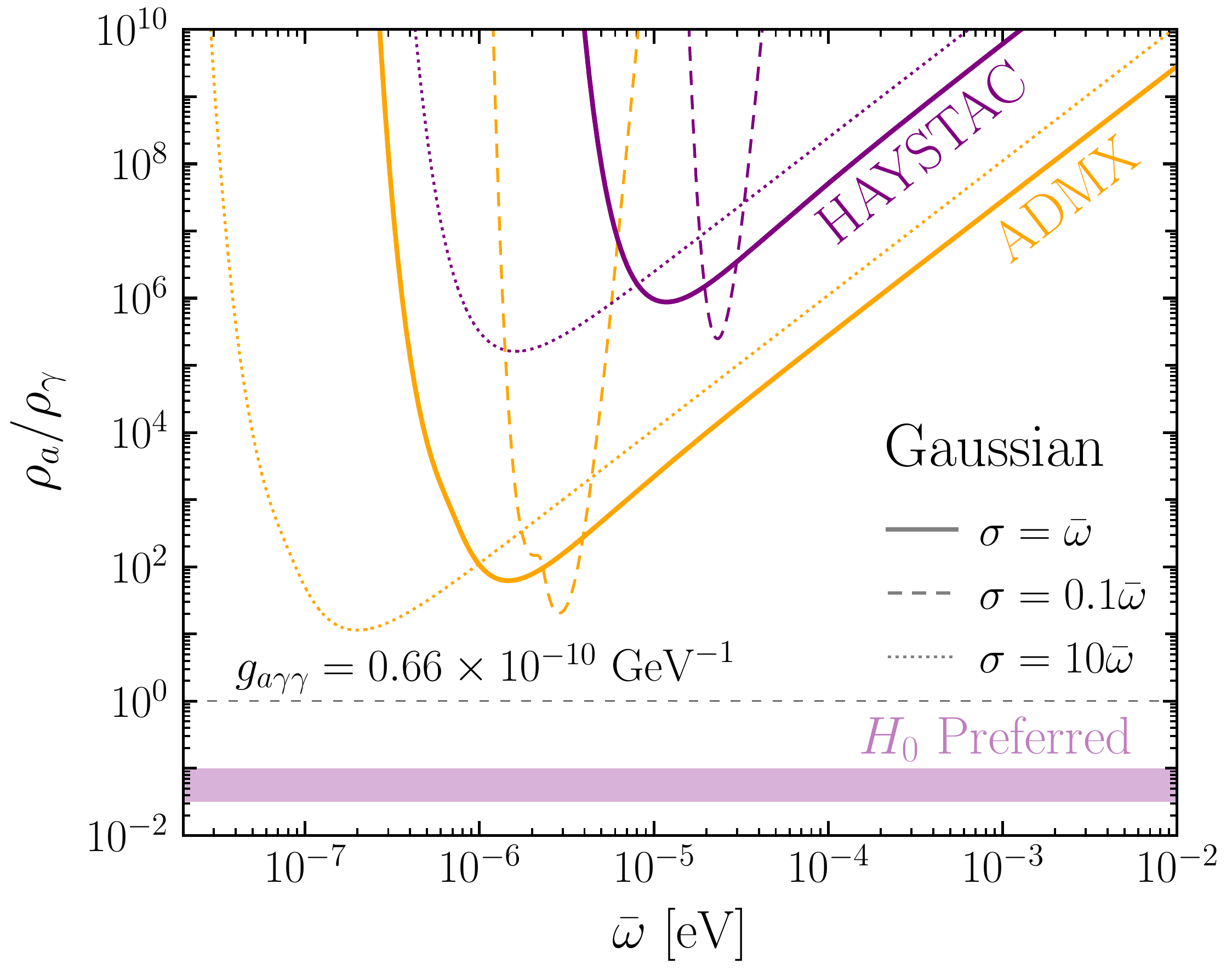}
\vspace{-0.3cm}
\caption{Sensitivity for a Gaussian C$a$B, as predicted by parametric-resonance scenario, that can be obtained by searching for a relativistic axion in data already collected by ADMX and HAYSTAC in the search for axion dark matter.
In particular we take $p(\omega)$ as in \eqref{eq:dist-Gauss}, and consider three values of the width $\kappa$.
If the C$a$B is a cosmological relic, then it must have $\rho_a < \rho_{\gamma}$, which ADMX would reach with an order of magnitude improvement in its $\ga$ reach.
}
\label{fig:GaussLim}
\end{figure}

Our forecast sensitivity is provided in Fig.~\ref{fig:GaussLim}.
In determining the plotted sensitivities, we combined the single bandwidth sensitivities in \eqref{eq:cavity-sens} across multiple bins, accounting for the spread of $p(\omega)$.
In doing so we assumed the frequency range scanned by the instruments was divided into bins of width $\omega_0/Q$, taking $Q= 10^5$ for ADMX and $10^4$ for HAYSTAC.
In detail, we compute a limit $\rho_{a,i}$ in each bin, indexed by $i$, and then determine the combined limit as $\rho_a^{-2} = \sum \rho_{a,i}^{-2}$.
Note that in the event of the limit being identical across $N$ bins, this returns the expected $\rho_a = \rho_{a,i}/\sqrt{N}$.

As the figure demonstrates, at present neither instrument is sensitive to the $\rho_a < \rho_{\gamma}$ required for a cosmological relic.
Nonetheless, ADMX is within two orders of magnitude of the relevant parameter space, which from \eqref{eq:cavity-sens}, would be achieved with an order of magnitude improvement in their sensitivity to the dark matter $\ga$.
A factor of 30 improvement would allow ADMX to access the parameter space required to reduce the $H_0$ tension.
The situation is more challenging for HAYSTAC, which would require at least three orders of magnitude improvement in their coupling sensitivity to reach $\rho_a \sim \rho_{\gamma}$, although widening the range of masses considered by a factor $\alpha$ would also enhance there sensitivity by $\sqrt{\alpha}$.

Note that in Fig.~\ref{fig:GaussLim}, the peak sensitivity depends on the width of the distribution, $\kappa$, and is not always located within the range of masses directly probed by ADMX and HAYSTAC for $\kappa > 1$.
To understand this, recall from \eqref{eq:S-scaling} that the signal power is determined by $\rho_a/\bar{\omega} = n_a$, the number density.
For a fixed $\rho_a$, we can increase $n_a$ by decreasing $\bar{\omega}$, and still obtain a constraint as long as $p(\omega)$ has support.

\subsection{Cosmic Strings}

Next we consider sensitivity to a cosmic-string origin of the C$a$B as discussed in Sec.~\ref{sec:cosmic-string}.
In this work we will use the specific results provided in Refs.~\cite{Gorghetto:2018myk,Gorghetto:2020qws}, as already discussed, however we note that further improvements in the predicted string spectrum will impact the sensitivities we present.
Regardless, as demonstrated in Fig.~\ref{fig:drhodomega_string}, the cosmic-string spectrum is expected to be especially broad.
As such, we will use it as an example to forecast sensitivity with a futuristic broadband instrument operating in the low-frequency regime, defined by $\lambda_a \gg L$.

The spectrum and energy density of cosmic-string axions is determined determined by the symmetry breaking scale, $f_a$, and subsequent temperature at which the string network enters the scaling regime, $T_d \leq f_a$.
In terms of their impact on the predicted C$a$B spectrum, $T_d < f_a$ provides an effective cutoff on the spectrum at higher frequencies, whereas the energy density in the spectrum is controlled by $\sim f_a^2$.
Accordingly, for a given $T_d$, we can construct our sensitivity to $f_a$ by determining where a detectable axion power is produced.
Recall our broadband sensitivity to $\rho_a$ was given in \eqref{eq:broadband-sens} for an ABRACADABRA type instrument.
Assuming a frequency independent background, we can rearrange that result to obtain
\begin{equation} 
\int d\omega\, \left( \omega \frac{dn_a}{d\omega} \right)^2 = \frac{2 {\rm TS}}{T \pi} \left(\frac{\lambda_B}{\beta^2 B_0^2 V_B^2 (\ga^{\rm SE})^2} \right)^2\,.
\label{eq:CosmicStringLim}
\end{equation} 
For a fixed $\ga^{\rm SE}$, the left hand side determines the signal strength, so that the result determines our sensitivity.
The differential energy density can be determined from \eqref{eq:OmegaString} as
\begin{equation}\begin{aligned}
\frac{dn_a}{d\omega} = \,&\frac{\rho_c}{\omega^2} \Omega_a(\omega) \\
= \,&\frac{8}{3 M_{\rm Pl}^2 \omega}
\int_{a_d}^{1} da\,a^2 \frac{\xi \mu_{\rm eff}}{H} 
\rho_{\SM}(a) F(\omega/Ha)\,.
\end{aligned}\end{equation}
For a given $T_d$ and $f_a$, this provides all the ingredients for the signal prediction, and what remains is to set the experimental parameters.
For this purpose we use the parameters adopted in the most optimistic scenario provided in the original ABRACADABRA proposal~\cite{Kahn:2016aff}, which involved a 100 m$^3$ volume instrument with a 5 T magnetic field operating for a year.
In order to determine the expected 95\% sensitivity we further take ${\rm TS}=2.71$.
Lastly, we need to specify the frequency range of the search, which will enter in the terminals for the signal integral in \eqref{eq:CosmicStringLim}.
Here we take $2 \times 10^{-13}~{\rm eV} < \omega < 10^{-7}~{\rm eV}$, where the lower limit is set at 50 Hz, where the $1/f$ noise is expected to begin dominating, and the upper limit is determined by the physical size of the instrument.
Going to such low frequencies requires us to take the enhanced star emission bounds of $\ga^{\rm SE} = 3.6 \times 10^{-12}~{\rm GeV}^{-1}$.

\begin{figure}[!t]
\vspace{-0.1cm}
\centering
\includegraphics[width=.465\textwidth]{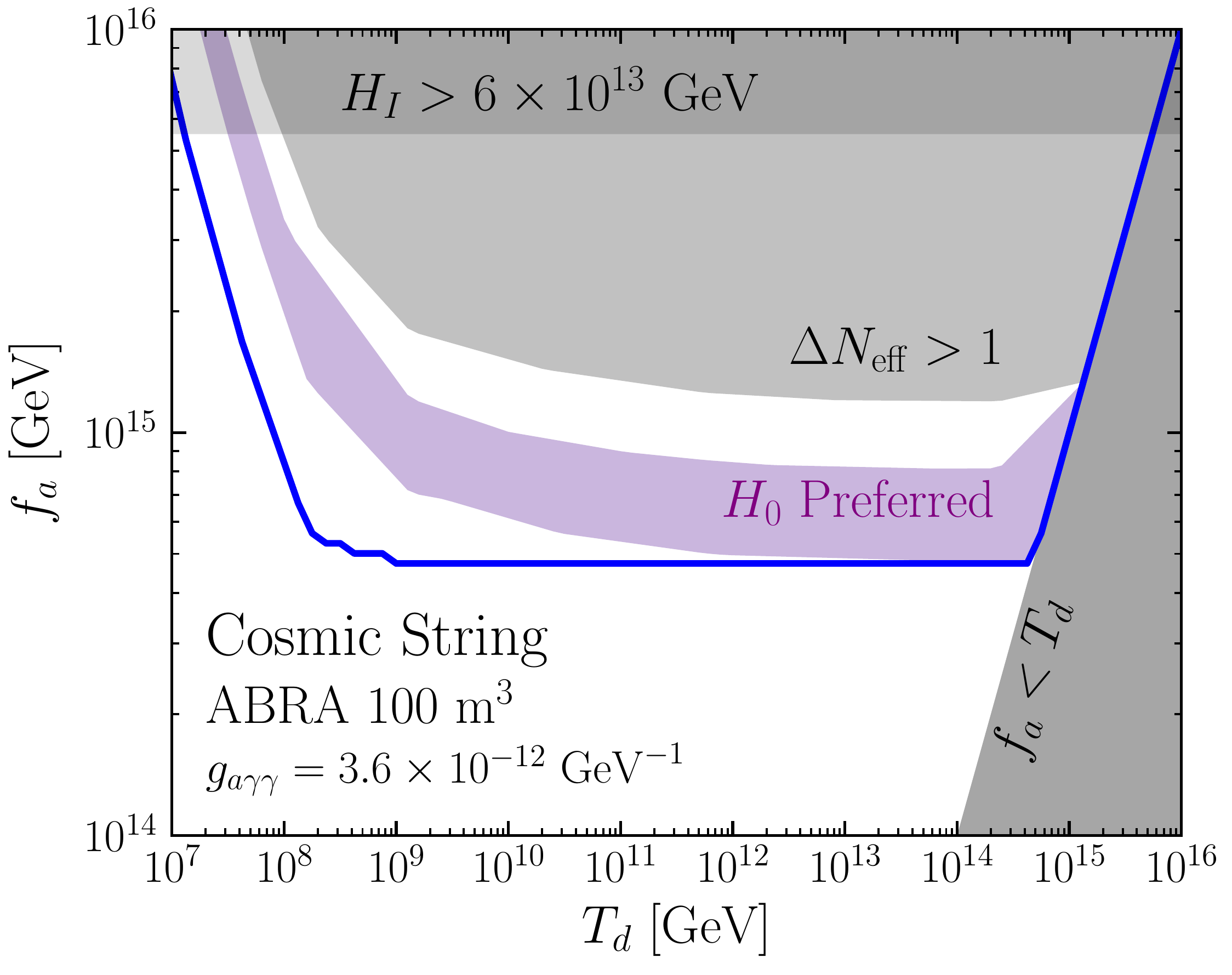}
\vspace{-0.3cm}
\caption{Sensitivity to the cosmic-string C$a$B at a future large scale broadband instrument, as a function of the axion decay constant $f_a$ and decoupling temperature $T_d$. In particular, we consider a 100 m$^3$ volume ABRACADABRA type instrument, operating with a 5 T field for one year.
}
\label{fig:StringLim}
\end{figure}

The end result of the above discussion is the forecast sensitivity shown in Fig.~\ref{fig:StringLim}.
In the same figure we also depict the parameter space where the cosmic-string spectrum over this frequency range would be in tension with $\Delta N_{\rm eff}$ bounds, and once more where the model could reduce the Hubble tension. This entire parameter space would be covered by the large scale broadband instrument. The sensitivity is dominated by the contribution at the low-frequency end of the instrument, and the flattening of the sensitivity at $T_d \sim 10^9$ GeV arises when the decoupling induced cutoff in the spectrum occurs within the experimental frequency range.

\subsection{Dark-Matter Indirect Detection}

\begin{figure*}[!t]
\vspace{-0.1cm}
\centering
\includegraphics[width=.465\textwidth]{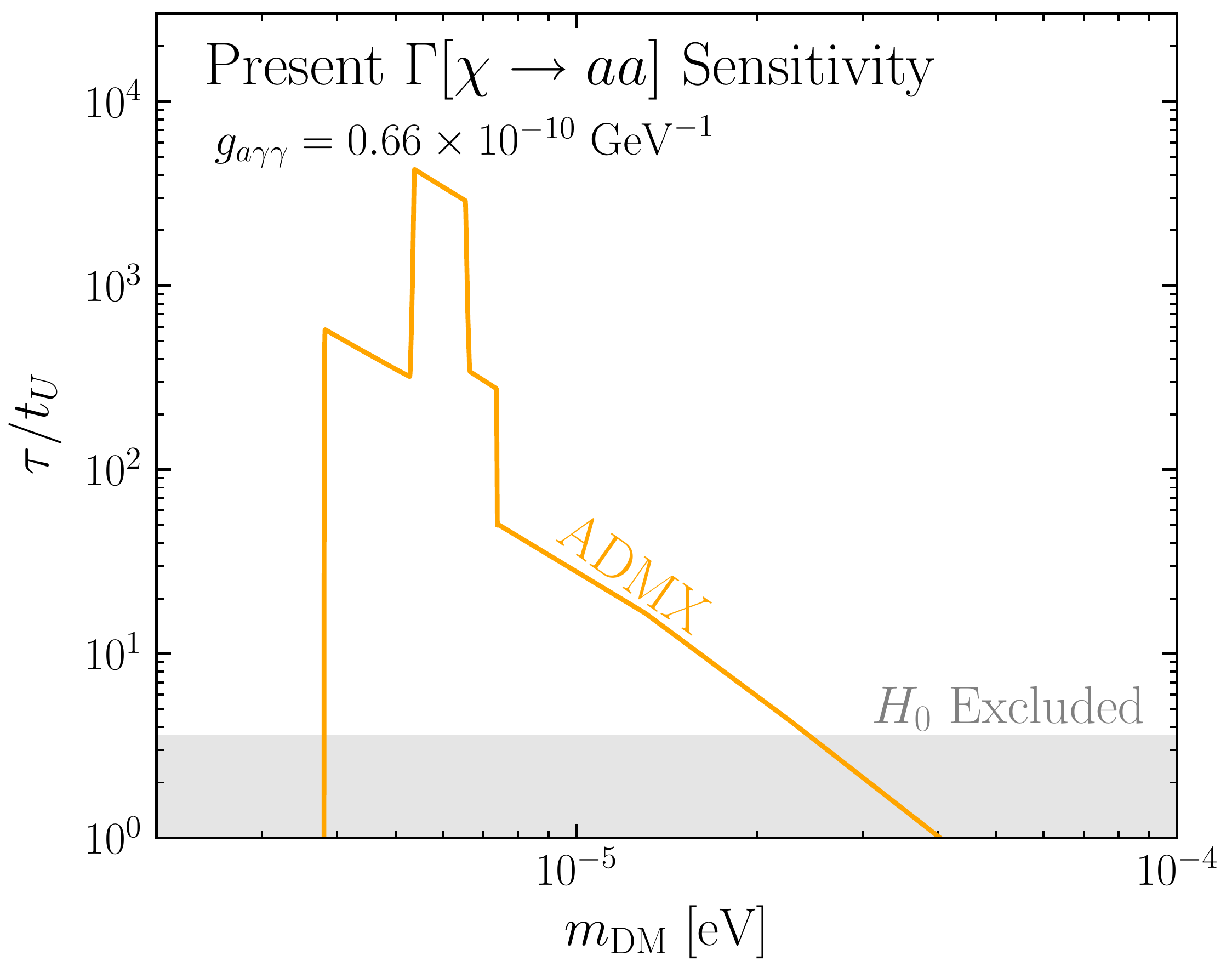}
\hspace{0.5cm}
\includegraphics[width=.47\textwidth]{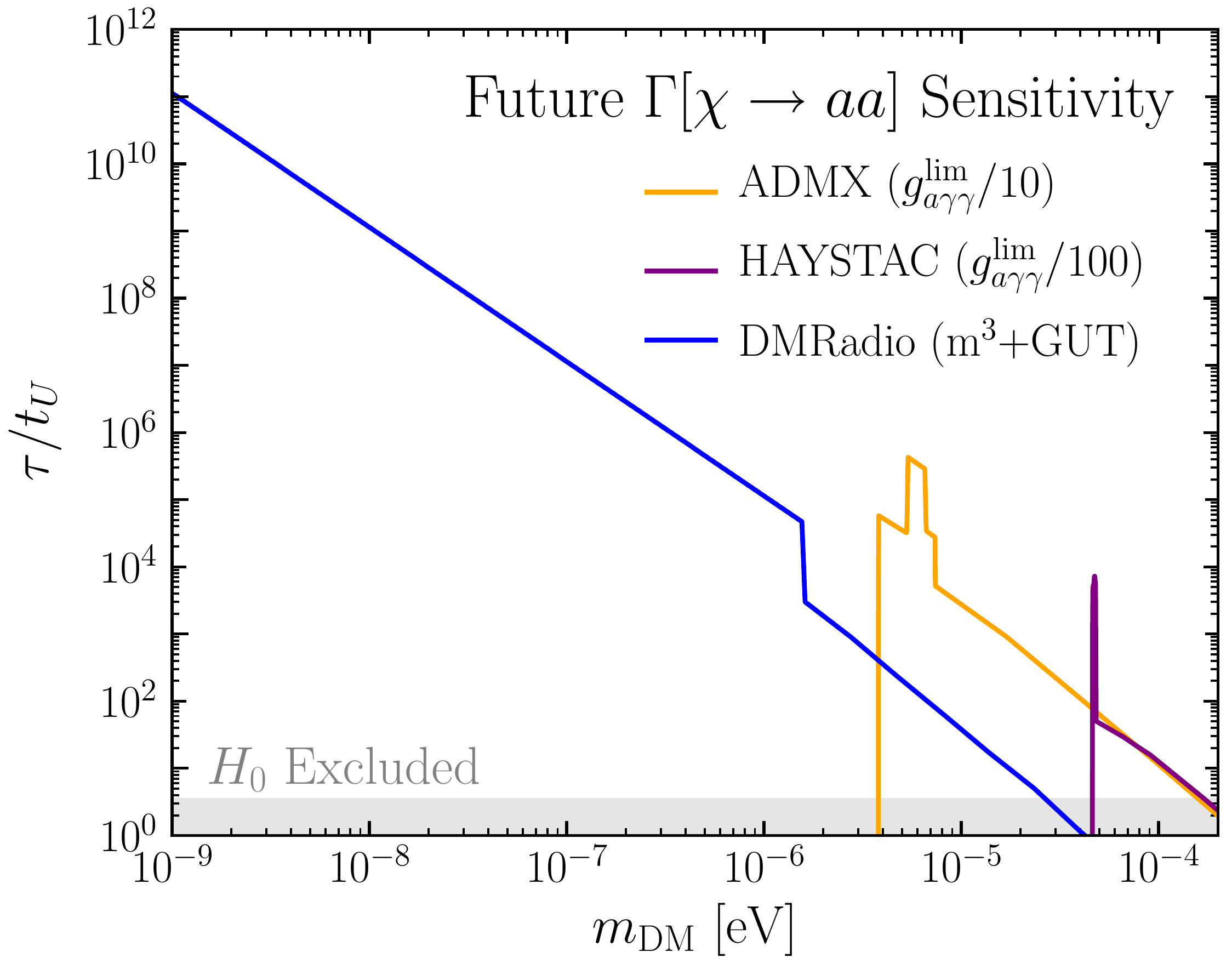}
\vspace{-0.3cm}
\caption{Sensitivity to relativistic axions resulting from dark-matter decay using existing (left) and future (right) resonant instruments.  We emphasize that the forecast sensitivities are likely to be obtained on significantly different timescales for the various instruments. The $H_0$ exclusion region is a result of the decay of dark matter to a relativistic species leading to an observable modification of cosmology, and we show the constraints $\tau \gtrsim 3.6\,t_U$ obtained by the DES collaboration~\cite{Chen:2020iwm}.
}
\label{fig:DecayLim}
\end{figure*}

In the examples thus far, the relevant C$a$B parameter space will only be reached in future instruments. This in a consequence of bounds from observational cosmology limiting $\rho_a \lesssim \rho_\gamma $. However, if the C$a$B is produced in the late Universe, $\rho_a > \rho_{\gamma}$ is allowed, and in principle may already be detectable.
Axions produced from dark-matter decay, discussed in Sec.~\ref{sec:DMDecay}, is one such scenario.
Searching for these axions would open up a new channel in the broader search for dark-matter indirect detection.

Recall that relativistic axions produced from dark-matter decay will receive a contribution from both decays in the local Milky Way halo and extragalactic dark matter.
The latter will arrive approximately isotropically at the Earth, whereas the former will preferentially originate from the Galactic Center given its higher dark-matter density.
For detection at resonant cavity instruments, we reiterate that the local decays will be associated with an observable daily modulation, as the signal is proportional to $\sin^4 \alpha$, with $\alpha$ the relative angle of the incident axions and the cavity magnetic field.
Effectively this search uses the resonant cavities as dark-matter telescopes, although with peak sensitivity obtained when the instrument is perpendicular to the source.
Whilst this effect can be used as a fingerprint of a genuine signal, for the sensitivity estimates we present here we will simply take the sky averaged value.

Doing so, our sensitivity for resonant cavity instruments was provided in \eqref{eq:cavity-sens}, which we can re-express as
\begin{equation} 
\frac{dn_a}{d\omega}(m_a) = \frac{4}{\pi} \rho_{\DM} \left( \frac{\ga^{\rm lim}}{\ga^{\rm SE}} \right)^2 \left( \frac{\sqrt{Q_a^{\DM} Q}}{m_a^2} \right)\,,
\end{equation} 
where the signal has been written in terms of the differential number density evaluated at $\omega=m_a$.
This number density is a combination of the local and extragalactic densities given in \eqref{eq:dndw-localdecay} and \eqref{eq:dndw-egldecay}, respectively. As discussed, the local distribution will be broadened by the Doppler shifts arising from both the finite velocity distribution of the dark matter, and also the Earth's motion relative to the halo, although here we will simply model the distribution as a Gaussian with relative width $10^{-3}$.
With these distributions, for a given $m_{\DM}$, the flux is dictated by the dark-matter lifetime $\tau$.

While a search for indirect detection with axions has not yet been performed, there are constraints on dark matter decaying to a relativistic species from cosmology, which roughly limit the lifetime to be longer than the age of the Universe~\cite{Poulin:2016nat,Haridasu:2020xaa,Chen:2020iwm}.
Repurposed axion dark-matter searches will be able to do considerably better.
To begin with, by repeating the approach of combining experimental bandwidths as we did for the Gaussian case above, we determine that ADMX is already sensitive to open parameter space, as we show on the left of Fig.~\ref{fig:DecayLim}.
The strongest sensitivity is obtained when $2m_{\DM}$ falls within the ADMX search window of $2-3~\mu{\rm eV}$, as this corresponds to detecting the local population of axions, which have energies peaking at $m_{\DM}/2$.
For higher dark-matter masses, the decay can still be detected thanks to the redshifted extragalactic spectrum.
We emphasize once more that this figure shows the sensitivity ADMX can obtain to this scenario using existing data.
Even for local decays, the signal is significantly broader than that expected of dark matter, and thus in existing searches the effect would have been discarded as background.

The future prospects for this search are considerable, as we demonstrate on the right of Fig.~\ref{fig:DecayLim}.
Performing a naive forecast of ADMX and HAYSTAC by assuming they improve their reach for $\ga$ by one and two orders of magnitude, respectively, they will both be able to probe open parameter space.
At lower masses, were a future instrument such as DMRadio able to reach the projected sensitivity of the QCD axion $\ga$ prediction from 0.5 neV to 0.8 $\mu$eV~\cite{SnowmassOuellet,SnowmassChaudhuri}, their reach would be considerable.
In particular, repurposing the low-frequency resonant result in~\eqref{eq:resonant-sens}, and assuming DMRadio operates with $Q=10^6$ (although their actual readout strategy will be more complex), we see that the instrument will be sensitive to an enormous range of parameters.
We note that the future projections shown here may have different timelines and do not necessarily represent a fair comparison between instruments.

\section{Discussion}
\label{sec:conc}

In this work, we considered the possibility of detecting a Cosmic {\em axion} Background, an ultrarelativistic background of relic axions. Being naturally light, axions produced in the early Universe are typically relativistic and may exist over a large range of possible energies. Its detection would have imprints of the history of the Universe in its energy spectrum, in close analogy with the program searching for a stochastic gravitational wave background. In particular, we discuss sources sensitive to the reheat temperature of the Universe (thermal production), the origin of dark matter (dark-matter decay), inflation (parametric resonance), and early Universe phase transitions (cosmic-string emission). Most of these production mechanisms predict axions with energies in the range detectable by current and future axion dark-matter experiments. The exception is the well-motivated thermal C$a$B, where new ideas will be required to probe the predicted energies and densities. For relativistic axion production before recombination, probes of the expansion of the Universe limit the axion energy density to be below that of the CMB. Moreover, recent measurements have found a persistent discrepancy between the early and late measurements. An additional source of radiation is the simplest solution to partially alleviate the tension, motivating a value for the C$a$B energy density and we discuss the production of axions in light of this target. If an axion is discovered from star emission searches such as hinted by the recent Xenon-1T excess~\cite{Aprile:2020tmw} or by future experiments such as IAXO~\cite{Vogel:2013bta}, there would additionally be a clear target coupling for the axion, greatly increasing the urgency of conducting C$a$B searches. 

For detection, we have focused on axions coupled to photons. Relativistic axion relics contribute terms to Maxwell's equations which are negligible for axion dark matter. The influence of the new terms depends on the experimental design and we study both resonant cavity experiments (e.g., ADMX, HAYSTAC) and lumped-circuit readout instruments (e.g., ABRACADABRA and DMRadio). For resonant cavities, the power deposited in the cavity becomes sensitive to the axion propagation direction and will exhibit a daily modulation for non-isotropic sources (such as dark-matter decay). For oscillating magnetic field searches, we show that the power deposited is independent of the incoming axion direction, however oscillating electric fields develop within the experiment that might be useful in confirming a positive signal. In either case, we develop simple relations to estimate the prospective sensitivity of axion experiments to a C$a$B. While present searches for dark-matter axions can have sensitivity to a C$a$B, this typically requires a dedicated search due to the presence of a broad energy spectra. In particular, with a search for a relatively broad axion energy distribution, current data from the ADMX experiment may be able to discover dark matter decaying into axions, thereby turning the instrument into an indirect-detection axion telescope. In the future, axion detection experiments will be able to discover ambient axion densities well below that of the CMB and probe a wide range of motivated sources.

Throughout this work, we have largely neglected the influence of a prospective axion mass, $m_a$. If, for a given axion source, $m_a$ becomes comparable to the axion energy as the Universe expands, then axions would cluster in galaxies. This would increase the local density, analogous to the C$\nu$B. If axions become highly non-relativistic they make up a form of dark matter as currently searched for by axion haloscopes. Alternatively, if the axion is still mildly relativistic, it can have a relatively wide energy distribution and still be overlooked by present analyses. The detection prospects of axions with intermediate masses is an interesting question we leave for future work. Experiments could also look for C$a$B with other interaction terms, such as the axion-nucleon coupling. Since the projections of experiments such as CASPEr~\cite{Budker:2013hfa} have sensitives well below star-emission bounds, they will also be able to probe cosmological sources of relativistic axions. In addition, one could consider other relativistic bosons such as dark photons. We leave the study of their prospective sources and detection capabilities for future work.

\section*{Acknowledgments}
We thank Yang Bai, Karl van Bibber, Keisuke Harigaya, and Alexander Leder for useful discussions.
We further acknowledge significant and insightful feedback on a draft provided by Anson Hook, Yonatan Kahn, Robert Lasenby, and Jonathan Ouellet.
Finally, we thank Asher Berlin and Kevin Zhou for pointing out our misuse of the effective charge in an earlier version of this work.
JD is supported in part by the DOE under contract DE-AC02-05CH11231 and in part by the NSF CAREER grant PHY-1915852.
HM is supported by the Director, Office of Science, Office of High Energy Physics of the US Department of Energy under the Contract No. DE-AC02-05CH11231, NSF grant PHY-1915314, the JSPS Grant-in-Aid for Scientific Research JP17K05409, MEXT Grant-in-Aid for Scientific Research on Innovative Areas JP15H05887, JP15K21733, by WPI, MEXT,
Japan, and Hamamatsu Photonics, KK.
NLR is supported by the Miller Institute for Basic Research in Science at the University of California, Berkeley.
This work made use of resources provided by the National Energy Research Scientific Computing Center, a US Department of Energy Office of Science User Facility supported by Contract No. DE-AC02-05CH11231.

\appendix 
\section{Breaking the Axion-Photon Coupling Relation with Kinetic Mixing}
\label{app:mixing}

The axion-photon coupling has a natural value related to its decay constant.
If the interactions arises from integrating out charged fermions, $f$, of charge $Q_f$, then the coupling generated is,
\begin{equation}
\ga = \sum_f \frac{\alpha Q^2}{2\pi f_a } \,.
\end{equation}  
In the absence of parametrically small charges, the expectation is that $\ga \sim \alpha / 2\pi f_a$.
However, it is conceptually simple to envision scenarios where the coupling is well below this scale either through axion-axion~\cite{Babu:1994id} or photon-dark photon~\cite{Daido:2018dmu} kinetic mixing (see also Refs.~\cite{Agrawal:2017cmd,Dror:2020zru} for related discussions).

To see this we first consider a kinetic mixing between an axion, $a$, and a second axion, $b$.
We take the second axion to have a decay constant $f_b$ that couples to electromagnetism, whereas $a$ does not couple to any charged fields, so that it has no direct photon coupling.
In detail, we take the following Lagrangian,
\begin{equation}\begin{aligned}
{\cal L} \supset\, &\frac{1}{2} ( \partial_\mu a )^2 + \frac{1}{2} ( \partial_\mu b )^2 +  \varepsilon  (\partial_\mu a) (\partial^\mu b) \\
&+ \frac{\alpha}{8\pi f_b} b F_{\mu \nu} \tilde{F}^{\mu\nu} \,.
\end{aligned}\end{equation} 
To diagonalize the axion kinetic terms to leading order in $\varepsilon$, we send $a \to a$, $ b\to b - \varepsilon a$. Once axion masses are introduced, this transformation leaves the rest of the Lagrangian approximately unchanged only if $ m _a \gg m _b $. With this shift the Lagrangian for $a$ can now be written as,
\begin{equation} 
{\cal L} \supset  \frac{1}{2} ( \partial _\mu a ) ^2    -  \frac{ \alpha \varepsilon }{ 8\pi f _b  } a F _{\mu \nu } \tilde{F} ^{\mu\nu}  \,.
\end{equation} 
Importantly, the coupling between $a$ and electromagnetism is determined by the free parameters $\varepsilon$ and $f_b$ --- it can be small even if $f_a$ is well below the weak scale.
While axion-axion mixing breaks the relationship between $\ga$ and $f_a$, it also requires an additional light axion that may influence the phenomenology.

Alternatively, one can consider a kinetic mixing between the photon and a dark photon.
Consider the Lagrangian,
\begin{equation}\begin{aligned}
\mathcal{L}\supset\,& \frac{\alpha'}{8\pi f_a} a F'_{\mu\nu} \tilde{F}^{\prime \mu \nu} -\frac{1}{4} F_{\mu\nu} F^{\mu\nu} \\
&-\frac{1}{4}F^\prime _{\mu\nu}  F^{\prime \mu\nu } -\frac{\epsilon}{4}F_{\mu\nu} F^{\prime \mu\nu}\,,
\end{aligned}\end{equation}
where $\alpha'$ is the dark gauge coupling and the photon-dark photon mixing is set by $\epsilon$.
If the dark photon, $A'$, has a mass below the photon plasma mass, then the photon-dark photon mixing can be eliminated with the transformation $A' \to A'$, $A \to A - \epsilon A'$.
This leaves the dark photon approximately massless, but generates an axion-photon coupling,
\begin{equation} 
\mathcal{L} \supset \frac{\epsilon^2 \alpha'}{8\pi f_a} a F  \tilde{F}\,.
\end{equation} 
The coefficient can be arbitrarily small even for $f_a$ below the electroweak scale, breaking its conventional relationship to $\ga$.
Interestingly, there may also be an opportunity to discover additional light states in this scenario.
If $f_a \lesssim 1~{\rm TeV}$ then in the limit that the dark-photon mass is negligible, there are millicharged particles in the spectrum.
These particles can have a mass of, at most, $\sim 4\pi f_a$.
For $f_a \lesssim ~100~{\rm eV}$ solar cooling bounds on millicharged particles would require $\epsilon \lesssim 10^{-13}$~\cite{Vogel:2013raa} and may place a bound on the parameter space. 

\bibliographystyle{JHEP}
\bibliography{CaB}

\providecommand{\href}[2]{#2}\begingroup\raggedright\begin{thebibliography}{100}

\bibitem{Peccei:1977hh}
R.~Peccei and H.~R. Quinn, {\it {CP Conservation in the Presence of
  Instantons}},  {\em Phys. Rev. Lett.} {\bf 38} (1977) 1440--1443.

\bibitem{Peccei:1977ur}
R.~Peccei and H.~R. Quinn, {\it {Constraints Imposed by CP Conservation in the
  Presence of Instantons}},  {\em Phys. Rev. D} {\bf 16} (1977) 1791--1797.

\bibitem{Weinberg:1977ma}
S.~Weinberg, {\it {A New Light Boson?}},  {\em Phys. Rev. Lett.} {\bf 40}
  (1978) 223--226.

\bibitem{Wilczek:1977pj}
F.~Wilczek, {\it {Problem of Strong $P$ and $T$ Invariance in the Presence of
  Instantons}},  {\em Phys. Rev. Lett.} {\bf 40} (1978) 279--282.

\bibitem{Svrcek:2006yi}
P.~Svrcek and E.~Witten, {\it {Axions In String Theory}},  {\em JHEP} {\bf 06}
  (2006) 051, [\href{http://arxiv.org/abs/hep-th/0605206}{{\tt
  hep-th/0605206}}].

\bibitem{Arvanitaki:2009fg}
A.~Arvanitaki, S.~Dimopoulos, S.~Dubovsky, N.~Kaloper, and J.~March-Russell,
  {\it {String Axiverse}},  {\em Phys. Rev. D} {\bf 81} (2010) 123530,
  [\href{http://arxiv.org/abs/0905.4720}{{\tt arXiv:0905.4720}}].

\bibitem{Halverson:2019cmy}
J.~Halverson, C.~Long, B.~Nelson, and G.~Salinas, {\it {Towards string theory
  expectations for photon couplings to axionlike particles}},  {\em Phys. Rev.
  D} {\bf 100} (2019), no.~10 106010,
  [\href{http://arxiv.org/abs/1909.05257}{{\tt arXiv:1909.05257}}].

\bibitem{Preskill:1982cy}
J.~Preskill, M.~B. Wise, and F.~Wilczek, {\it {Cosmology of the Invisible
  Axion}},  {\em Phys. Lett. B} {\bf 120} (1983) 127--132.

\bibitem{Abbott:1982af}
L.~Abbott and P.~Sikivie, {\it {A Cosmological Bound on the Invisible Axion}},
  {\em Phys. Lett. B} {\bf 120} (1983) 133--136.

\bibitem{Dine:1982ah}
M.~Dine and W.~Fischler, {\it {The Not So Harmless Axion}},  {\em Phys. Lett.
  B} {\bf 120} (1983) 137--141.

\bibitem{Baumann:2016wac}
D.~Baumann, D.~Green, and B.~Wallisch, {\it {New Target for Cosmic Axion
  Searches}},  {\em Phys. Rev. Lett.} {\bf 117} (2016), no.~17 171301,
  [\href{http://arxiv.org/abs/1604.08614}{{\tt arXiv:1604.08614}}].

\bibitem{Conlon:2013isa}
J.~P. Conlon and M.~C.~D. Marsh, {\it {The Cosmophenomenology of Axionic Dark
  Radiation}},  {\em JHEP} {\bf 10} (2013) 214,
  [\href{http://arxiv.org/abs/1304.1804}{{\tt arXiv:1304.1804}}].

\bibitem{Conlon:2013txa}
J.~P. Conlon and M.~D. Marsh, {\it {Excess Astrophysical Photons from a
  0.1--1 keV Cosmic Axion Background}},  {\em Phys. Rev. Lett.} {\bf 111}
  (2013), no.~15 151301, [\href{http://arxiv.org/abs/1305.3603}{{\tt
  arXiv:1305.3603}}].

\bibitem{Cicoli:2014bfa}
M.~Cicoli, J.~P. Conlon, M.~C.~D. Marsh, and M.~Rummel, {\it {3.55 keV photon
  line and its morphology from a 3.55 keV axionlike particle line}},  {\em
  Phys. Rev. D} {\bf 90} (2014) 023540,
  [\href{http://arxiv.org/abs/1403.2370}{{\tt arXiv:1403.2370}}].

\bibitem{Cui:2017ytb}
Y.~Cui, M.~Pospelov, and J.~Pradler, {\it {Signatures of Dark Radiation in
  Neutrino and Dark Matter Detectors}},  {\em Phys. Rev. D} {\bf 97} (2018),
  no.~10 103004, [\href{http://arxiv.org/abs/1711.04531}{{\tt
  arXiv:1711.04531}}].

\bibitem{Higaki:2013qka}
T.~Higaki, K.~Nakayama, and F.~Takahashi, {\it {Cosmological constraints on
  axionic dark radiation from axion-photon conversion in the early Universe}},
  {\em JCAP} {\bf 09} (2013) 030, [\href{http://arxiv.org/abs/1306.6518}{{\tt
  arXiv:1306.6518}}].

\bibitem{Evoli:2016zhj}
C.~Evoli, M.~Leo, A.~Mirizzi, and D.~Montanino, {\it {Reionization during the
  dark ages from a cosmic axion background}},  {\em JCAP} {\bf 05} (2016) 006,
  [\href{http://arxiv.org/abs/1602.08433}{{\tt arXiv:1602.08433}}].

\bibitem{Verde:2019ivm}
L.~Verde, T.~Treu, and A.~Riess, {\it {Tensions between the Early and the Late
  Universe}},  {\em Nature Astron.} {\bf 3} (7, 2019) 891,
  [\href{http://arxiv.org/abs/1907.10625}{{\tt arXiv:1907.10625}}].

\bibitem{Aghanim:2018eyx}
{\bf Planck} Collaboration, N.~Aghanim et~al., {\it {Planck 2018 results. VI.
  Cosmological parameters}},  {\em Astron. Astrophys.} {\bf 641} (2020) A6,
  [\href{http://arxiv.org/abs/1807.06209}{{\tt arXiv:1807.06209}}].

\bibitem{Graham:2015ouw}
P.~W. Graham, I.~G. Irastorza, S.~K. Lamoreaux, A.~Lindner, and K.~A. van
  Bibber, {\it {Experimental Searches for the Axion and Axion-Like Particles}},
   {\em Ann. Rev. Nucl. Part. Sci.} {\bf 65} (2015) 485--514,
  [\href{http://arxiv.org/abs/1602.00039}{{\tt arXiv:1602.00039}}].

\bibitem{Irastorza:2018dyq}
I.~G. Irastorza and J.~Redondo, {\it {New experimental approaches in the search
  for axion-like particles}},  {\em Prog. Part. Nucl. Phys.} {\bf 102} (2018)
  89--159, [\href{http://arxiv.org/abs/1801.08127}{{\tt arXiv:1801.08127}}].

\bibitem{Raffelt:1996wa}
G.~Raffelt, {\em {Stars as laboratories for fundamental physics}: {The
  astrophysics of neutrinos, axions, and other weakly interacting particles}}.
\newblock 5, 1996.

\bibitem{Sikivie:1983ip}
P.~Sikivie, {\it {Experimental Tests of the Invisible Axion}},  {\em Phys. Rev.
  Lett.} {\bf 51} (1983) 1415--1417. [Erratum: Phys.Rev.Lett. 52, 695 (1984)].

\bibitem{Moriyama:1995bz}
S.~Moriyama, {\it {A Proposal to search for a monochromatic component of solar
  axions using Fe-57}},  {\em Phys. Rev. Lett.} {\bf 75} (1995) 3222--3225,
  [\href{http://arxiv.org/abs/hep-ph/9504318}{{\tt hep-ph/9504318}}].

\bibitem{Anastassopoulos:2017ftl}
{\bf CAST} Collaboration, V.~Anastassopoulos et~al., {\it {New CAST Limit on
  the Axion-Photon Interaction}},  {\em Nature Phys.} {\bf 13} (2017) 584--590,
  [\href{http://arxiv.org/abs/1705.02290}{{\tt arXiv:1705.02290}}].

\bibitem{Ayala:2014pea}
A.~Ayala, I.~Domínguez, M.~Giannotti, A.~Mirizzi, and O.~Straniero, {\it
  {Revisiting the bound on axion-photon coupling from Globular Clusters}},
  {\em Phys. Rev. Lett.} {\bf 113} (2014), no.~19 191302,
  [\href{http://arxiv.org/abs/1406.6053}{{\tt arXiv:1406.6053}}].

\bibitem{Carenza:2020zil}
P.~Carenza, O.~Straniero, B.~D\"obrich, M.~Giannotti, G.~Lucente, and
  A.~Mirizzi, {\it {Constraints on the coupling with photons of heavy
  axion-like-particles from Globular Clusters}},  {\em Phys. Lett. B} {\bf 809}
  (2020) 135709, [\href{http://arxiv.org/abs/2004.08399}{{\tt
  arXiv:2004.08399}}].

\bibitem{Payez:2014xsa}
A.~Payez, C.~Evoli, T.~Fischer, M.~Giannotti, A.~Mirizzi, and A.~Ringwald, {\it
  {Revisiting the SN1987A gamma-ray limit on ultralight axion-like particles}},
   {\em JCAP} {\bf 02} (2015) 006, [\href{http://arxiv.org/abs/1410.3747}{{\tt
  arXiv:1410.3747}}].

\bibitem{Bar:2019ifz}
N.~Bar, K.~Blum, and G.~D'Amico, {\it {Is there a supernova bound on axions?}},
   {\em Phys. Rev. D} {\bf 101} (2020), no.~12 123025,
  [\href{http://arxiv.org/abs/1907.05020}{{\tt arXiv:1907.05020}}].

\bibitem{Reynolds:2019uqt}
C.~S. Reynolds, M.~D. Marsh, H.~R. Russell, A.~C. Fabian, R.~Smith, F.~Tombesi,
  and S.~Veilleux, {\it {Astrophysical limits on very light axion-like
  particles from Chandra grating spectroscopy of NGC 1275}},
  \href{http://arxiv.org/abs/1907.05475}{{\tt arXiv:1907.05475}}.

\bibitem{Dessert:2020lil}
C.~Dessert, J.~W. Foster, and B.~R. Safdi, {\it {X-ray Searches for Axions from
  Super Star Clusters}},  \href{http://arxiv.org/abs/2008.03305}{{\tt
  arXiv:2008.03305}}.

\bibitem{Krauss:1985ub}
L.~Krauss, J.~Moody, F.~Wilczek, and D.~E. Morris, {\it {Calculations for
  Cosmic Axion Detection}},  {\em Phys. Rev. Lett.} {\bf 55} (1985) 1797.

\bibitem{Sikivie:1985yu}
P.~Sikivie, {\it {Detection Rates for 'Invisible' Axion Searches}},  {\em Phys.
  Rev. D} {\bf 32} (1985) 2988. [Erratum: Phys.Rev.D 36, 974 (1987)].

\bibitem{Asztalos:2003px}
{\bf ADMX} Collaboration, S.~Asztalos et~al., {\it {An Improved RF cavity
  search for halo axions}},  {\em Phys. Rev. D} {\bf 69} (2004) 011101,
  [\href{http://arxiv.org/abs/astro-ph/0310042}{{\tt astro-ph/0310042}}].

\bibitem{Du:2018uak}
{\bf ADMX} Collaboration, N.~Du et~al., {\it {A Search for Invisible Axion Dark
  Matter with the Axion Dark Matter Experiment}},  {\em Phys. Rev. Lett.} {\bf
  120} (2018), no.~15 151301, [\href{http://arxiv.org/abs/1804.05750}{{\tt
  arXiv:1804.05750}}].

\bibitem{Braine:2019fqb}
{\bf ADMX} Collaboration, T.~Braine et~al., {\it {Extended Search for the
  Invisible Axion with the Axion Dark Matter Experiment}},  {\em Phys. Rev.
  Lett.} {\bf 124} (2020), no.~10 101303,
  [\href{http://arxiv.org/abs/1910.08638}{{\tt arXiv:1910.08638}}].

\bibitem{Zhong:2018rsr}
{\bf HAYSTAC} Collaboration, L.~Zhong et~al., {\it {Results from phase 1 of the
  HAYSTAC microwave cavity axion experiment}},  {\em Phys. Rev. D} {\bf 97}
  (2018), no.~9 092001, [\href{http://arxiv.org/abs/1803.03690}{{\tt
  arXiv:1803.03690}}].

\bibitem{Lee:2020cfj}
S.~Lee, S.~Ahn, J.~Choi, B.~Ko, and Y.~Semertzidis, {\it {Axion Dark Matter
  Search around 6.7 $\mu$eV}},  {\em Phys. Rev. Lett.} {\bf 124} (2020), no.~10
  101802, [\href{http://arxiv.org/abs/2001.05102}{{\tt arXiv:2001.05102}}].

\bibitem{Sikivie:2013laa}
P.~Sikivie, N.~Sullivan, and D.~Tanner, {\it {Proposal for Axion Dark Matter
  Detection Using an LC Circuit}},  {\em Phys. Rev. Lett.} {\bf 112} (2014),
  no.~13 131301, [\href{http://arxiv.org/abs/1310.8545}{{\tt
  arXiv:1310.8545}}].

\bibitem{Chaudhuri:2014dla}
S.~Chaudhuri, P.~W. Graham, K.~Irwin, J.~Mardon, S.~Rajendran, and Y.~Zhao,
  {\it {Radio for hidden-photon dark matter detection}},  {\em Phys. Rev. D}
  {\bf 92} (2015), no.~7 075012, [\href{http://arxiv.org/abs/1411.7382}{{\tt
  arXiv:1411.7382}}].

\bibitem{Kahn:2016aff}
Y.~Kahn, B.~R. Safdi, and J.~Thaler, {\it {Broadband and Resonant Approaches to
  Axion Dark Matter Detection}},  {\em Phys. Rev. Lett.} {\bf 117} (2016),
  no.~14 141801, [\href{http://arxiv.org/abs/1602.01086}{{\tt
  arXiv:1602.01086}}].

\bibitem{Silva-Feaver:2016qhh}
M.~Silva-Feaver et~al., {\it {Design Overview of DM Radio Pathfinder
  Experiment}},  {\em IEEE Trans. Appl. Supercond.} {\bf 27} (2017), no.~4
  1400204, [\href{http://arxiv.org/abs/1610.09344}{{\tt arXiv:1610.09344}}].

\bibitem{Ouellet:2018beu}
J.~L. Ouellet et~al., {\it {First Results from ABRACADABRA-10 cm: A Search for
  Sub-$\mu$eV Axion Dark Matter}},  {\em Phys. Rev. Lett.} {\bf 122} (2019),
  no.~12 121802, [\href{http://arxiv.org/abs/1810.12257}{{\tt
  arXiv:1810.12257}}].

\bibitem{Ouellet:2019tlz}
J.~L. Ouellet et~al., {\it {Design and implementation of the ABRACADABRA-10 cm
  axion dark matter search}},  {\em Phys. Rev. D} {\bf 99} (2019), no.~5
  052012, [\href{http://arxiv.org/abs/1901.10652}{{\tt arXiv:1901.10652}}].

\bibitem{Gramolin:2020ict}
A.~V. Gramolin, D.~Aybas, D.~Johnson, J.~Adam, and A.~O. Sushkov, {\it {Search
  for axion-like dark matter with ferromagnets}},
  \href{http://arxiv.org/abs/2003.03348}{{\tt arXiv:2003.03348}}.

\bibitem{SnowmassOuellet}
J.~L. Ouellet et~al., {\it {Probing the QCD Axion with DMRadio-m$^3$}},  {\em
  Snowmass 2021 Letter of Interest} {\bf CF2} (2020), no.~217. Available at
  \url{https://www.snowmass21.org/docs/files/summaries/CF/SNOWMASS21-CF2_CF0-IF1_IF0_Ouellet-217.pdf}.

\bibitem{SnowmassChaudhuri}
S.~Chaudhuri et~al., {\it {DMRadio-GUT: Probing GUT-scale QCD Axion Dark
  Matter}},  {\em Snowmass 2021 Letter of Interest} {\bf CF2} (2020), no.~219.
  Available at
  \url{https://www.snowmass21.org/docs/files/summaries/CF/SNOWMASS21-CF2_CF0-IF1_IF0_Saptarshi_Chaudhuri-219.pdf}.

\bibitem{TheMADMAXWorkingGroup:2016hpc}
{\bf MADMAX Working Group} Collaboration, A.~Caldwell, G.~Dvali, B.~Majorovits,
  A.~Millar, G.~Raffelt, J.~Redondo, O.~Reimann, F.~Simon, and F.~Steffen, {\it
  {Dielectric Haloscopes: A New Way to Detect Axion Dark Matter}},  {\em Phys.
  Rev. Lett.} {\bf 118} (2017), no.~9 091801,
  [\href{http://arxiv.org/abs/1611.05865}{{\tt arXiv:1611.05865}}].

\bibitem{Millar:2016cjp}
A.~J. Millar, G.~G. Raffelt, J.~Redondo, and F.~D. Steffen, {\it {Dielectric
  Haloscopes to Search for Axion Dark Matter: Theoretical Foundations}},  {\em
  JCAP} {\bf 01} (2017) 061, [\href{http://arxiv.org/abs/1612.07057}{{\tt
  arXiv:1612.07057}}].

\bibitem{Ioannisian:2017srr}
A.~N. Ioannisian, N.~Kazarian, A.~J. Millar, and G.~G. Raffelt, {\it
  {Axion-photon conversion caused by dielectric interfaces: quantum field
  calculation}},  {\em JCAP} {\bf 09} (2017) 005,
  [\href{http://arxiv.org/abs/1707.00701}{{\tt arXiv:1707.00701}}].

\bibitem{Berlin:2019ahk}
A.~Berlin, R.~T. D'Agnolo, S.~A. Ellis, C.~Nantista, J.~Neilson, P.~Schuster,
  S.~Tantawi, N.~Toro, and K.~Zhou, {\it {Axion Dark Matter Detection by
  Superconducting Resonant Frequency Conversion}},
  \href{http://arxiv.org/abs/1912.11048}{{\tt arXiv:1912.11048}}.

\bibitem{Lasenby:2019prg}
R.~Lasenby, {\it {Microwave cavity searches for low-frequency axion dark
  matter}},  {\em Phys. Rev. D} {\bf 102} (2020), no.~1 015008,
  [\href{http://arxiv.org/abs/1912.11056}{{\tt arXiv:1912.11056}}].

\bibitem{Liu:2018icu}
H.~Liu, B.~D. Elwood, M.~Evans, and J.~Thaler, {\it {Searching for Axion Dark
  Matter with Birefringent Cavities}},  {\em Phys. Rev. D} {\bf 100} (2019),
  no.~2 023548, [\href{http://arxiv.org/abs/1809.01656}{{\tt
  arXiv:1809.01656}}].

\bibitem{Obata:2018vvr}
I.~Obata, T.~Fujita, and Y.~Michimura, {\it {Optical Ring Cavity Search for
  Axion Dark Matter}},  {\em Phys. Rev. Lett.} {\bf 121} (2018), no.~16 161301,
  [\href{http://arxiv.org/abs/1805.11753}{{\tt arXiv:1805.11753}}].

\bibitem{Berlin:2020vrk}
A.~Berlin, R.~T. D'Agnolo, S.~A. Ellis, and K.~Zhou, {\it {Heterodyne Broadband
  Detection of Axion Dark Matter}},
  \href{http://arxiv.org/abs/2007.15656}{{\tt arXiv:2007.15656}}.

\bibitem{Marsh:2018dlj}
D.~J. Marsh, K.-C. Fong, E.~W. Lentz, L.~r. Smejkal, and M.~N. Ali, {\it
  {Proposal to Detect Dark Matter using Axionic Topological Antiferromagnets}},
   {\em Phys. Rev. Lett.} {\bf 123} (2019), no.~12 121601,
  [\href{http://arxiv.org/abs/1807.08810}{{\tt arXiv:1807.08810}}].

\bibitem{Trickle:2019ovy}
T.~Trickle, Z.~Zhang, and K.~M. Zurek, {\it {Detecting Light Dark Matter with
  Magnons}},  {\em Phys. Rev. Lett.} {\bf 124} (2020), no.~20 201801,
  [\href{http://arxiv.org/abs/1905.13744}{{\tt arXiv:1905.13744}}].

\bibitem{Lawson:2019brd}
M.~Lawson, A.~J. Millar, M.~Pancaldi, E.~Vitagliano, and F.~Wilczek, {\it
  {Tunable axion plasma haloscopes}},  {\em Phys. Rev. Lett.} {\bf 123} (2019),
  no.~14 141802, [\href{http://arxiv.org/abs/1904.11872}{{\tt
  arXiv:1904.11872}}].

\bibitem{Gelmini:2020kcu}
G.~B. Gelmini, A.~J. Millar, V.~Takhistov, and E.~Vitagliano, {\it {Probing
  dark photons with plasma haloscopes}},  {\em Phys. Rev. D} {\bf 102} (2020),
  no.~4 043003, [\href{http://arxiv.org/abs/2006.06836}{{\tt
  arXiv:2006.06836}}].

\bibitem{deSalas:2020hbh}
P.~F. de~Salas and A.~Widmark, {\it {Dark matter local density determination:
  recent observations and future prospects}},
  \href{http://arxiv.org/abs/2012.11477}{{\tt arXiv:2012.11477}}.

\bibitem{Lentati:2015qwp}
L.~Lentati et~al., {\it {European Pulsar Timing Array Limits On An Isotropic
  Stochastic Gravitational-Wave Background}},  {\em Mon. Not. Roy. Astron.
  Soc.} {\bf 453} (2015), no.~3 2576--2598,
  [\href{http://arxiv.org/abs/1504.03692}{{\tt arXiv:1504.03692}}].

\bibitem{Shannon:2015ect}
R.~Shannon et~al., {\it {Gravitational waves from binary supermassive black
  holes missing in pulsar observations}},  {\em Science} {\bf 349} (2015),
  no.~6255 1522--1525, [\href{http://arxiv.org/abs/1509.07320}{{\tt
  arXiv:1509.07320}}].

\bibitem{Arzoumanian:2018saf}
{\bf NANOGRAV} Collaboration, Z.~Arzoumanian et~al., {\it {The NANOGrav 11-year
  Data Set: Pulsar-timing Constraints On The Stochastic Gravitational-wave
  Background}},  {\em Astrophys. J.} {\bf 859} (2018), no.~1 47,
  [\href{http://arxiv.org/abs/1801.02617}{{\tt arXiv:1801.02617}}].

\bibitem{LIGOScientific:2019vic}
{\bf LIGO Scientific, Virgo} Collaboration, B.~Abbott et~al., {\it {Search for
  the isotropic stochastic background using data from Advanced
  LIGO\textquoteright{}s second observing run}},  {\em Phys. Rev. D} {\bf 100}
  (2019), no.~6 061101, [\href{http://arxiv.org/abs/1903.02886}{{\tt
  arXiv:1903.02886}}].

\bibitem{Turner:1986tb}
M.~S. Turner, {\it {Thermal Production of Not SO Invisible Axions in the Early
  Universe}},  {\em Phys. Rev. Lett.} {\bf 59} (1987) 2489. [Erratum:
  Phys.Rev.Lett. 60, 1101 (1988)].

\bibitem{Chang:1993gm}
S.~Chang and K.~Choi, {\it {Hadronic axion window and the big bang
  nucleosynthesis}},  {\em Phys. Lett. B} {\bf 316} (1993) 51--56,
  [\href{http://arxiv.org/abs/hep-ph/9306216}{{\tt hep-ph/9306216}}].

\bibitem{Masso:2002np}
E.~Masso, F.~Rota, and G.~Zsembinszki, {\it {On axion thermalization in the
  early universe}},  {\em Phys. Rev. D} {\bf 66} (2002) 023004,
  [\href{http://arxiv.org/abs/hep-ph/0203221}{{\tt hep-ph/0203221}}].

\bibitem{Hannestad:2005df}
S.~Hannestad, A.~Mirizzi, and G.~Raffelt, {\it {New cosmological mass limit on
  thermal relic axions}},  {\em JCAP} {\bf 07} (2005) 002,
  [\href{http://arxiv.org/abs/hep-ph/0504059}{{\tt hep-ph/0504059}}].

\bibitem{Graf:2010tv}
P.~Graf and F.~D. Steffen, {\it {Thermal axion production in the primordial
  quark-gluon plasma}},  {\em Phys. Rev. D} {\bf 83} (2011) 075011,
  [\href{http://arxiv.org/abs/1008.4528}{{\tt arXiv:1008.4528}}].

\bibitem{Salvio:2013iaa}
A.~Salvio, A.~Strumia, and W.~Xue, {\it {Thermal axion production}},  {\em
  JCAP} {\bf 01} (2014) 011, [\href{http://arxiv.org/abs/1310.6982}{{\tt
  arXiv:1310.6982}}].

\bibitem{Ferreira:2018vjj}
R.~Z. Ferreira and A.~Notari, {\it {Observable Windows for the QCD Axion
  Through the Number of Relativistic Species}},  {\em Phys. Rev. Lett.} {\bf
  120} (2018), no.~19 191301, [\href{http://arxiv.org/abs/1801.06090}{{\tt
  arXiv:1801.06090}}].

\bibitem{Arias-Aragon:2020shv}
F.~Arias-Aragon, F.~D'Eramo, R.~Z. Ferreira, L.~Merlo, and A.~Notari, {\it
  {Production of Thermal Axions across the ElectroWeak Phase Transition}},
  \href{http://arxiv.org/abs/2012.04736}{{\tt arXiv:2012.04736}}.

\bibitem{DEramo:2018vss}
F.~D'Eramo, R.~Z. Ferreira, A.~Notari, and J.~L. Bernal, {\it {Hot Axions and
  the $H_0$ tension}},  {\em JCAP} {\bf 1811} (2018) 014,
  [\href{http://arxiv.org/abs/1808.07430}{{\tt arXiv:1808.07430}}].

\bibitem{Gong:2008gi}
Y.~Gong and X.~Chen, {\it {Cosmological Constraints on Invisible Decay of Dark
  Matter}},  {\em Phys. Rev. D} {\bf 77} (2008) 103511,
  [\href{http://arxiv.org/abs/0802.2296}{{\tt arXiv:0802.2296}}].

\bibitem{Poulin:2016nat}
V.~Poulin, P.~D. Serpico, and J.~Lesgourgues, {\it {A fresh look at linear
  cosmological constraints on a decaying dark matter component}},  {\em JCAP}
  {\bf 08} (2016) 036, [\href{http://arxiv.org/abs/1606.02073}{{\tt
  arXiv:1606.02073}}].

\bibitem{Vattis:2019efj}
K.~Vattis, S.~M. Koushiappas, and A.~Loeb, {\it {Dark matter decaying in the
  late Universe can relieve the H0 tension}},  {\em Phys. Rev. D} {\bf 99}
  (2019), no.~12 121302, [\href{http://arxiv.org/abs/1903.06220}{{\tt
  arXiv:1903.06220}}].

\bibitem{Haridasu:2020xaa}
B.~S. Haridasu and M.~Viel, {\it {Late-time decaying dark matter: constraints
  and implications for the $H_0$-tension}},  {\em Mon. Not. Roy. Astron. Soc.}
  {\bf 497} (2020), no.~2 1757--1764,
  [\href{http://arxiv.org/abs/2004.07709}{{\tt arXiv:2004.07709}}].

\bibitem{Chen:2020iwm}
{\bf DES} Collaboration, A.~Chen et~al., {\it {Constraints on Decaying Dark
  Matter with DES-Y1 and external data}},
  \href{http://arxiv.org/abs/2011.04606}{{\tt arXiv:2011.04606}}.

\bibitem{Lisanti:2017qoz}
M.~Lisanti, S.~Mishra-Sharma, N.~L. Rodd, B.~R. Safdi, and R.~H. Wechsler, {\it
  {Mapping Extragalactic Dark Matter Annihilation with Galaxy Surveys: A
  Systematic Study of Stacked Group Searches}},  {\em Phys. Rev. D} {\bf 97}
  (2018), no.~6 063005, [\href{http://arxiv.org/abs/1709.00416}{{\tt
  arXiv:1709.00416}}].

\bibitem{Navarro:1995iw}
J.~F. Navarro, C.~S. Frenk, and S.~D. White, {\it {The Structure of cold dark
  matter halos}},  {\em Astrophys. J.} {\bf 462} (1996) 563--575,
  [\href{http://arxiv.org/abs/astro-ph/9508025}{{\tt astro-ph/9508025}}].

\bibitem{Navarro:1996gj}
J.~F. Navarro, C.~S. Frenk, and S.~D. White, {\it {A Universal density profile
  from hierarchical clustering}},  {\em Astrophys. J.} {\bf 490} (1997)
  493--508, [\href{http://arxiv.org/abs/astro-ph/9611107}{{\tt
  astro-ph/9611107}}].

\bibitem{Abuter:2018drb}
{\bf GRAVITY} Collaboration, R.~Abuter et~al., {\it {Detection of the
  gravitational redshift in the orbit of the star S2 near the Galactic centre
  massive black hole}},  {\em Astron. Astrophys.} {\bf 615} (2018) L15,
  [\href{http://arxiv.org/abs/1807.09409}{{\tt arXiv:1807.09409}}].

\bibitem{Speckhard:2015eva}
E.~G. Speckhard, K.~C. Ng, J.~F. Beacom, and R.~Laha, {\it {Dark Matter
  Velocity Spectroscopy}},  {\em Phys. Rev. Lett.} {\bf 116} (2016), no.~3
  031301, [\href{http://arxiv.org/abs/1507.04744}{{\tt arXiv:1507.04744}}].

\bibitem{Gelmini:1980re}
G.~B. Gelmini and M.~Roncadelli, {\it {Left-Handed Neutrino Mass Scale and
  Spontaneously Broken Lepton Number}},  {\em Phys. Lett.} {\bf 99B} (1981)
  411--415.

\bibitem{Feng:1997tn}
J.~L. Feng, T.~Moroi, H.~Murayama, and E.~Schnapka, {\it {Third generation
  familons, b factories, and neutrino cosmology}},  {\em Phys. Rev. D} {\bf 57}
  (1998) 5875--5892, [\href{http://arxiv.org/abs/hep-ph/9709411}{{\tt
  hep-ph/9709411}}].

\bibitem{Dror:2020fbh}
J.~A. Dror, {\it {Discovering leptonic forces using nonconserved currents}},
  {\em Phys. Rev. D} {\bf 101} (2020), no.~9 095013,
  [\href{http://arxiv.org/abs/2004.04750}{{\tt arXiv:2004.04750}}].

\bibitem{Hannestad:2004qu}
S.~Hannestad, {\it {Structure formation with strongly interacting neutrinos -
  Implications for the cosmological neutrino mass bound}},  {\em JCAP} {\bf 02}
  (2005) 011, [\href{http://arxiv.org/abs/astro-ph/0411475}{{\tt
  astro-ph/0411475}}].

\bibitem{Barenboim:2020vrr}
G.~Barenboim, J.~Z. Chen, S.~Hannestad, I.~M. Oldengott, T.~Tram, and Y.~Y.
  Wong, {\it {Invisible neutrino decay in precision cosmology}},
  \href{http://arxiv.org/abs/2011.01502}{{\tt arXiv:2011.01502}}.

\bibitem{Dolgov:1989us}
A.~Dolgov and D.~Kirilova, {\it {On particle creation by a time dependent
  scalar field}},  {\em Sov. J. Nucl. Phys.} {\bf 51} (1990) 172--177.

\bibitem{PhysRevD.42.2491}
J.~H. Traschen and R.~H. Brandenberger, {\it Particle production during
  out-of-equilibrium phase transitions},  {\em Phys. Rev. D} {\bf 42} (Oct,
  1990) 2491--2504.

\bibitem{Kofman:1994rk}
L.~Kofman, A.~D. Linde, and A.~A. Starobinsky, {\it {Reheating after
  inflation}},  {\em Phys. Rev. Lett.} {\bf 73} (1994) 3195--3198,
  [\href{http://arxiv.org/abs/hep-th/9405187}{{\tt hep-th/9405187}}].

\bibitem{Kofman:1997yn}
L.~Kofman, A.~D. Linde, and A.~A. Starobinsky, {\it {Towards the theory of
  reheating after inflation}},  {\em Phys. Rev. D} {\bf 56} (1997) 3258--3295,
  [\href{http://arxiv.org/abs/hep-ph/9704452}{{\tt hep-ph/9704452}}].

\bibitem{Ema:2017krp}
Y.~Ema and K.~Nakayama, {\it {Explosive Axion Production from Saxion}},  {\em
  Phys. Lett. B} {\bf 776} (2018) 174--181,
  [\href{http://arxiv.org/abs/1710.02461}{{\tt arXiv:1710.02461}}].

\bibitem{Co:2017mop}
R.~T. Co, L.~J. Hall, and K.~Harigaya, {\it {QCD Axion Dark Matter with a Small
  Decay Constant}},  {\em Phys. Rev. Lett.} {\bf 120} (2018), no.~21 211602,
  [\href{http://arxiv.org/abs/1711.10486}{{\tt arXiv:1711.10486}}].

\bibitem{Dror:2018pdh}
J.~A. Dror, K.~Harigaya, and V.~Narayan, {\it {Parametric Resonance Production
  of Ultralight Vector Dark Matter}},  {\em Phys. Rev. D} {\bf 99} (2019),
  no.~3 035036, [\href{http://arxiv.org/abs/1810.07195}{{\tt
  arXiv:1810.07195}}].

\bibitem{Micha:2004bv}
R.~Micha and I.~I. Tkachev, {\it {Turbulent thermalization}},  {\em Phys. Rev.
  D} {\bf 70} (2004) 043538, [\href{http://arxiv.org/abs/hep-ph/0403101}{{\tt
  hep-ph/0403101}}].

\bibitem{Akrami:2018odb}
{\bf Planck} Collaboration, Y.~Akrami et~al., {\it {Planck 2018 results. X.
  Constraints on inflation}},  \href{http://arxiv.org/abs/1807.06211}{{\tt
  arXiv:1807.06211}}.

\bibitem{Gorghetto:2018myk}
M.~Gorghetto, E.~Hardy, and G.~Villadoro, {\it {Axions from Strings: the
  Attractive Solution}},  {\em JHEP} {\bf 07} (2018) 151,
  [\href{http://arxiv.org/abs/1806.04677}{{\tt arXiv:1806.04677}}].

\bibitem{Gorghetto:2020qws}
M.~Gorghetto, E.~Hardy, and G.~Villadoro, {\it {More Axions from Strings}},
  \href{http://arxiv.org/abs/2007.04990}{{\tt arXiv:2007.04990}}.

\bibitem{Buschmann:2019icd}
M.~Buschmann, J.~W. Foster, and B.~R. Safdi, {\it {Early-Universe Simulations
  of the Cosmological Axion}},  {\em Phys. Rev. Lett.} {\bf 124} (2020), no.~16
  161103, [\href{http://arxiv.org/abs/1906.00967}{{\tt arXiv:1906.00967}}].

\bibitem{Dine:2020pds}
M.~Dine, N.~Fernandez, A.~Ghalsasi, and H.~H. Patel, {\it {Comments on Axions,
  Domain Walls, and Cosmic Strings}},
  \href{http://arxiv.org/abs/2012.13065}{{\tt arXiv:2012.13065}}.

\bibitem{Baumann:2009ds}
D.~Baumann, {\it {Inflation}},  in {\em {Theoretical Advanced Study Institute
  in Elementary Particle Physics}: {Physics of the Large and the Small}},
  pp.~523--686, 2011.
\newblock \href{http://arxiv.org/abs/0907.5424}{{\tt arXiv:0907.5424}}.

\bibitem{Charnock:2016nzm}
T.~Charnock, A.~Avgoustidis, E.~J. Copeland, and A.~Moss, {\it {CMB constraints
  on cosmic strings and superstrings}},  {\em Phys. Rev. D} {\bf 93} (2016),
  no.~12 123503, [\href{http://arxiv.org/abs/1603.01275}{{\tt
  arXiv:1603.01275}}].

\bibitem{Choi:2015fiu}
K.~Choi and S.~H. Im, {\it {Realizing the relaxion from multiple axions and its
  UV completion with high scale supersymmetry}},  {\em JHEP} {\bf 01} (2016)
  149, [\href{http://arxiv.org/abs/1511.00132}{{\tt arXiv:1511.00132}}].

\bibitem{Agrawal:2017cmd}
P.~Agrawal, J.~Fan, M.~Reece, and L.-T. Wang, {\it {Experimental Targets for
  Photon Couplings of the QCD Axion}},  {\em JHEP} {\bf 02} (2018) 006,
  [\href{http://arxiv.org/abs/1709.06085}{{\tt arXiv:1709.06085}}].

\bibitem{Dror:2020zru}
J.~A. Dror and J.~M. Leedom, {\it {The Cosmological Tension of Ultralight Axion
  Dark Matter and its Solutions}},  \href{http://arxiv.org/abs/2008.02279}{{\tt
  arXiv:2008.02279}}.

\bibitem{Foster:2017hbq}
J.~W. Foster, N.~L. Rodd, and B.~R. Safdi, {\it {Revealing the Dark Matter Halo
  with Axion Direct Detection}},  {\em Phys. Rev. D} {\bf 97} (2018), no.~12
  123006, [\href{http://arxiv.org/abs/1711.10489}{{\tt arXiv:1711.10489}}].

\bibitem{Foster:2020fln}
J.~W. Foster, Y.~Kahn, R.~Nguyen, N.~L. Rodd, and B.~R. Safdi, {\it {Dark
  Matter Interferometry}},  \href{http://arxiv.org/abs/2009.14201}{{\tt
  arXiv:2009.14201}}.

\bibitem{Brubaker:2017rna}
B.~Brubaker, L.~Zhong, S.~Lamoreaux, K.~Lehnert, and K.~van Bibber, {\it
  {HAYSTAC axion search analysis procedure}},  {\em Phys. Rev. D} {\bf 96}
  (2017), no.~12 123008, [\href{http://arxiv.org/abs/1706.08388}{{\tt
  arXiv:1706.08388}}].

\bibitem{Dicke:1946glx}
R.~Dicke, {\it {The Measurement of Thermal Radiation at Microwave
  Frequencies}},  {\em Rev. Sci. Instrum.} {\bf 17} (1946), no.~7 268--275.

\bibitem{Budker:2013hfa}
D.~Budker, P.~W. Graham, M.~Ledbetter, S.~Rajendran, and A.~Sushkov, {\it
  {Proposal for a Cosmic Axion Spin Precession Experiment (CASPEr)}},  {\em
  Phys. Rev. X} {\bf 4} (2014), no.~2 021030,
  [\href{http://arxiv.org/abs/1306.6089}{{\tt arXiv:1306.6089}}].

\bibitem{Ouellet:2018nfr}
J.~Ouellet and Z.~Bogorad, {\it {Solutions to Axion Electrodynamics in Various
  Geometries}},  {\em Phys. Rev. D} {\bf 99} (2019), no.~5 055010,
  [\href{http://arxiv.org/abs/1809.10709}{{\tt arXiv:1809.10709}}].

\bibitem{Lasenby:2019hfz}
R.~Lasenby, {\it {Parametrics of electromagnetic searches for axion dark
  matter}},  \href{http://arxiv.org/abs/1912.11467}{{\tt arXiv:1912.11467}}.

\bibitem{Goryachev:2018vjt}
M.~Goryachev, B.~Mcallister, and M.~E. Tobar, {\it {Axion Detection with
  Precision Frequency Metrology}},  {\em Phys. Dark Univ.} {\bf 26} (2019)
  100345, [\href{http://arxiv.org/abs/1806.07141}{{\tt arXiv:1806.07141}}].

\bibitem{Kim:2018sci}
Y.~Kim, D.~Kim, J.~Jung, J.~Kim, Y.~C. Shin, and Y.~K. Semertzidis, {\it
  {Effective Approximation of Electromagnetism for Axion Haloscope Searches}},
  {\em Phys. Dark Univ.} {\bf 26} (2019) 100362,
  [\href{http://arxiv.org/abs/1810.02459}{{\tt arXiv:1810.02459}}].

\bibitem{Beutter:2018xfx}
M.~Beutter, A.~Pargner, T.~Schwetz, and E.~Todarello, {\it
  {Axion-electrodynamics: a quantum field calculation}},  {\em JCAP} {\bf 02}
  (2019) 026, [\href{http://arxiv.org/abs/1812.05487}{{\tt arXiv:1812.05487}}].

\bibitem{Brun:2019lyf}
{\bf MADMAX} Collaboration, P.~Brun et~al., {\it {A new experimental approach
  to probe QCD axion dark matter in the mass range above 40 $\mu$eV}},  {\em
  Eur. Phys. J. C} {\bf 79} (2019), no.~3 186,
  [\href{http://arxiv.org/abs/1901.07401}{{\tt arXiv:1901.07401}}].

\bibitem{Crisosto:2019fcj}
N.~Crisosto, G.~Rybka, P.~Sikivie, N.~Sullivan, D.~Tanner, and J.~Yang, {\it
  {ADMX SLIC: Results from a Superconducting LC Circuit Investigating Cold
  Axions}},  {\em Phys. Rev. Lett.} {\bf 124} (2020), no.~24 241101,
  [\href{http://arxiv.org/abs/1911.05772}{{\tt arXiv:1911.05772}}].

\bibitem{Chaudhuri:2018rqn}
S.~Chaudhuri, K.~Irwin, P.~W. Graham, and J.~Mardon, {\it {Fundamental Limits
  of Electromagnetic Axion and Hidden-Photon Dark Matter Searches: Part I - The
  Quantum Limit}},  \href{http://arxiv.org/abs/1803.01627}{{\tt
  arXiv:1803.01627}}.

\bibitem{Chaudhuri:2019ntz}
S.~Chaudhuri, K.~D. Irwin, P.~W. Graham, and J.~Mardon, {\it {Optimal
  Electromagnetic Searches for Axion and Hidden-Photon Dark Matter}},
  \href{http://arxiv.org/abs/1904.05806}{{\tt arXiv:1904.05806}}.

\bibitem{Cowan:2010js}
G.~Cowan, K.~Cranmer, E.~Gross, and O.~Vitells, {\it {Asymptotic formulae for
  likelihood-based tests of new physics}},  {\em Eur. Phys. J. C} {\bf 71}
  (2011) 1554, [\href{http://arxiv.org/abs/1007.1727}{{\tt arXiv:1007.1727}}].
  [Erratum: Eur.Phys.J.C 73, 2501 (2013)].

\bibitem{Brubaker:2018ebj}
B.~M. Brubaker, {\em {First results from the HAYSTAC axion search}}.
\newblock PhD thesis, Yale U., 2017.
\newblock \href{http://arxiv.org/abs/1801.00835}{{\tt arXiv:1801.00835}}.

\bibitem{Vogel:2013bta}
J.~Vogel et~al., {\it {IAXO - The International Axion Observatory}},  in {\em
  {8th Patras Workshop on Axions, WIMPs and WISPs}}, 2, 2013.
\newblock \href{http://arxiv.org/abs/1302.3273}{{\tt arXiv:1302.3273}}.

\bibitem{Mukherjee:2018oeb}
S.~Mukherjee, R.~Khatri, and B.~D. Wandelt, {\it {Polarized anisotropic
  spectral distortions of the CMB: Galactic and extragalactic constraints on
  photon-axion conversion}},  {\em JCAP} {\bf 04} (2018) 045,
  [\href{http://arxiv.org/abs/1801.09701}{{\tt arXiv:1801.09701}}].

\bibitem{Mukherjee:2019dsu}
S.~Mukherjee, D.~N. Spergel, R.~Khatri, and B.~D. Wandelt, {\it {A new probe of
  Axion-Like Particles: CMB polarization distortions due to cluster magnetic
  fields}},  {\em JCAP} {\bf 02} (2020) 032,
  [\href{http://arxiv.org/abs/1908.07534}{{\tt arXiv:1908.07534}}].

\bibitem{Aprile:2020tmw}
{\bf XENON} Collaboration, E.~Aprile et~al., {\it {Excess electronic recoil
  events in XENON1T}},  {\em Phys. Rev. D} {\bf 102} (2020), no.~7 072004,
  [\href{http://arxiv.org/abs/2006.09721}{{\tt arXiv:2006.09721}}].

\bibitem{Babu:1994id}
K.~Babu, S.~M. Barr, and D.~Seckel, {\it {Axion dissipation through the mixing
  of Goldstone bosons}},  {\em Phys. Lett. B} {\bf 336} (1994) 213--220,
  [\href{http://arxiv.org/abs/hep-ph/9406308}{{\tt hep-ph/9406308}}].

\bibitem{Daido:2018dmu}
R.~Daido, F.~Takahashi, and N.~Yokozaki, {\it {Enhanced
  axion\textendash{}photon coupling in GUT with hidden photon}},  {\em Phys.
  Lett. B} {\bf 780} (2018) 538--542,
  [\href{http://arxiv.org/abs/1801.10344}{{\tt arXiv:1801.10344}}].

\bibitem{Vogel:2013raa}
H.~Vogel and J.~Redondo, {\it {Dark Radiation constraints on minicharged
  particles in models with a hidden photon}},  {\em JCAP} {\bf 02} (2014) 029,
  [\href{http://arxiv.org/abs/1311.2600}{{\tt arXiv:1311.2600}}].

\end{thebibliography}\endgroup

\end{document}